\documentclass[10pt,sigconf,letterpaper]{acmart}
\newif\ifacmartsty
\acmartstytrue
\newif\ifcameraready
\camerareadyfalse

\ifacmartsty
\else
\usepackage[
  pass,%
]{geometry}
\fi

\usepackage{amsmath,amsfonts}
\usepackage[T1]{fontenc}
\usepackage[utf8]{inputenc}

\usepackage{upquote}

\usepackage{microtype}
\UseMicrotypeSet[protrusion]{basicmath} %

\usepackage{graphicx,grffile}

\usepackage{multirow}
\usepackage{xspace}
\usepackage{tabularx}
\usepackage{ragged2e}
\usepackage{booktabs}
\usepackage{paralist}
\usepackage[american]{babel}
\usepackage[shortlabels]{enumitem}
\usepackage{courier,color,wrapfig}
\usepackage{xcolor, soul}
\usepackage{xspace}
\usepackage{balance}
\usepackage{enumitem}
\usepackage{epstopdf}
\usepackage{multirow}
\usepackage{booktabs}
\usepackage{url}
\usepackage{pifont}
\usepackage{subfigure}
\usepackage{tikz}
\usepackage{mathtools}
\usepackage[normalem]{ulem}

\usepackage[algo2e]{algorithm2e}
\usepackage{algorithmic}
\usepackage{algorithm}
\usepackage{etoolbox}
\usepackage{ifthen}
\usepackage{sepfootnotes}
\usepackage[textsize=small]{todonotes}
\usepackage{setspace}
\usepackage{caption}
\usepackage{comment}

\makeatletter
\def\maxwidth{\ifdim\Gin@nat@width>\linewidth\linewidth\else\Gin@nat@width\fi}
\def\maxheight{\ifdim\Gin@nat@height>\textheight\textheight\else\Gin@nat@height\fi}
\makeatother
\setkeys{Gin}{width=\maxwidth,height=\maxheight,keepaspectratio}
\makeatletter
\g@addto@macro{\UrlBreaks}{\UrlOrds}
\makeatother

\ifacmartsty
\makeatletter
\let\origsection\section
\let\origsubsection\subsection

\renewcommand\section{\@ifstar{\starsection}{\nostarsection}}
\renewcommand\subsection{\@ifstar{\starsubsection}{\nostarsubsection}}

\newcommand\sectionprelude{\vspace{0ex}}
\newcommand\sectionpostlude{\vspace{0ex}}
\newcommand\subsectionprelude{\vspace{0ex}}
\newcommand\subsectionpostlude{\vspace{0ex}}

\newcommand\nostarsection[1]{\sectionprelude\origsection{#1}\sectionpostlude}
\newcommand\starsection[1]{\sectionprelude\origsection*{#1}\sectionpostlude}

\newcommand\nostarsubsection[1]{\subsectionprelude\origsubsection{#1}\subsectionpostlude}
\newcommand\starsubsection[1]{\subsectionprelude\origsubsection*{#1}\subsectionpostlude}

\makeatother
\fi

\newcommand\paraspace{\vspace*{1ex}}
\providecommand\parab[1]{\paraspace\noindent\textbf{#1.}}
\providecommand\parae[1]{\paraspace\textbf{\textit{#1.}}}

\ifacmartsty
\fi

\apptocmd\normalsize{%
\abovedisplayskip=5pt
\abovedisplayshortskip=5pt
\belowdisplayskip=5pt
\belowdisplayshortskip=5pt
}{}{}

\ifacmartsty
\renewcommand\footnotetextcopyrightpermission[1]{} %
\setcopyright{none}
\acmDOI{}
\acmISBN{}
\acmConference[Submitted for review]{}
\acmYear{2018}
\copyrightyear{}
\acmPrice{}
\fi

\newcommand{\sysname}{Gemini\xspace}
\newcommand{\gemini}{Gemini\xspace}

\newcommand{\set}[1]{\mathcal{#1}}

\newtheorem{theorem}{Theorem}

\newcommand{\ie}{\textit{i.e.,}\xspace}
\newcommand{\eg}{\textit{e.g.,}\xspace}
\newcommand{\vs}{\textit{vs.}\xspace}

\newcommand{\secref}[1]{\S\ref{#1}}
\newcommand{\figref}[1]{Fig.~\ref{#1}}

\newcommand{\algoref}[1]{Algorithm~\ref{#1}}
\newcommand{\thmref}[1]{Theorem~\ref{#1}}
\newcommand{\eqnref}[1]{Equation~(\ref{#1})}

\newcommand{\squishlist}{
 \begin{list}{$\bullet$}
 		{ \setlength{\itemsep}{0pt}
 			\setlength{\parsep}{1pt}
 			\setlength{\topsep}{1pt}
 			\setlength{\partopsep}{0pt}
 			\setlength{\leftmargin}{1.5em}
 			\setlength{\labelwidth}{1em}
 			\setlength{\labelsep}{0.5em} } }
\newcommand{\squishend}{
  \end{list}  }
  
\definecolor{mygray}{gray}{0.9}
\sethlcolor{mygray}

\begin{document}

\title{\sysname: Practical Reconfigurable Datacenter Networks with Topology and Traffic Engineering}
\author{Mingyang Zhang}
\affiliation{
    \institution{University of Southern California}
}
\author{Jianan Zhang}
\affiliation{
    \institution{Google}
}
\author{Rui Wang}
\affiliation{
    \institution{Google}
}

\author{Ramesh Govindan}
\affiliation{
    \institution{University of Southern California}
}
\author{Jeffrey C. Mogul}
\affiliation{
    \institution{Google}
}
\author{Amin Vahdat}
\affiliation{
    \institution{Google}
}
\date{}

\begin{abstract}

To reduce cost, datacenter network operators are exploring blocking network designs. An example of such a design is a "spine-free" form of a Fat-Tree, in which pods directly connect to each other, rather than via spine blocks. To maintain application-perceived performance in the face of dynamic workloads, these new designs must be able to reconfigure routing and the inter-pod topology. \gemini is a system designed to achieve these goals on commodity hardware while reconfiguring the network infrequently, rendering these blocking designs practical enough for deployment in the near future.

The key to \sysname is the joint optimization of topology and routing, using as input a robust estimation of future traffic derived from multiple historical traffic matrices. \sysname ``hedges'' against unpredicted bursts, by spreading these bursts across multiple paths, to minimize packet loss in exchange for a small increase in path lengths. It incorporates a robust decision algorithm to determine when to reconfigure, and whether to use hedging.

 Data from tens of production fabrics allows us to categorize these as either low-%
 or high-volatility; these categories seem stable.
 For the former, \sysname finds topologies and routing with near-optimal performance
 and cost.  For the latter, \sysname's use of multi-traffic-matrix optimization and hedging avoids
 the need for frequent topology reconfiguration, with only marginal increases in path length.
 As a result, \gemini can support existing workloads on these production fabrics using a spine-free topology that is half the cost of the existing topology on these fabrics.

\end{abstract}

\maketitle

\vspace{-0.16in}
\section{Introduction}
 \label{sec:intro}

\begin{figure*}
\centering
\begin{minipage}{0.6\columnwidth}
  \centering
  \includegraphics[width=1\columnwidth]{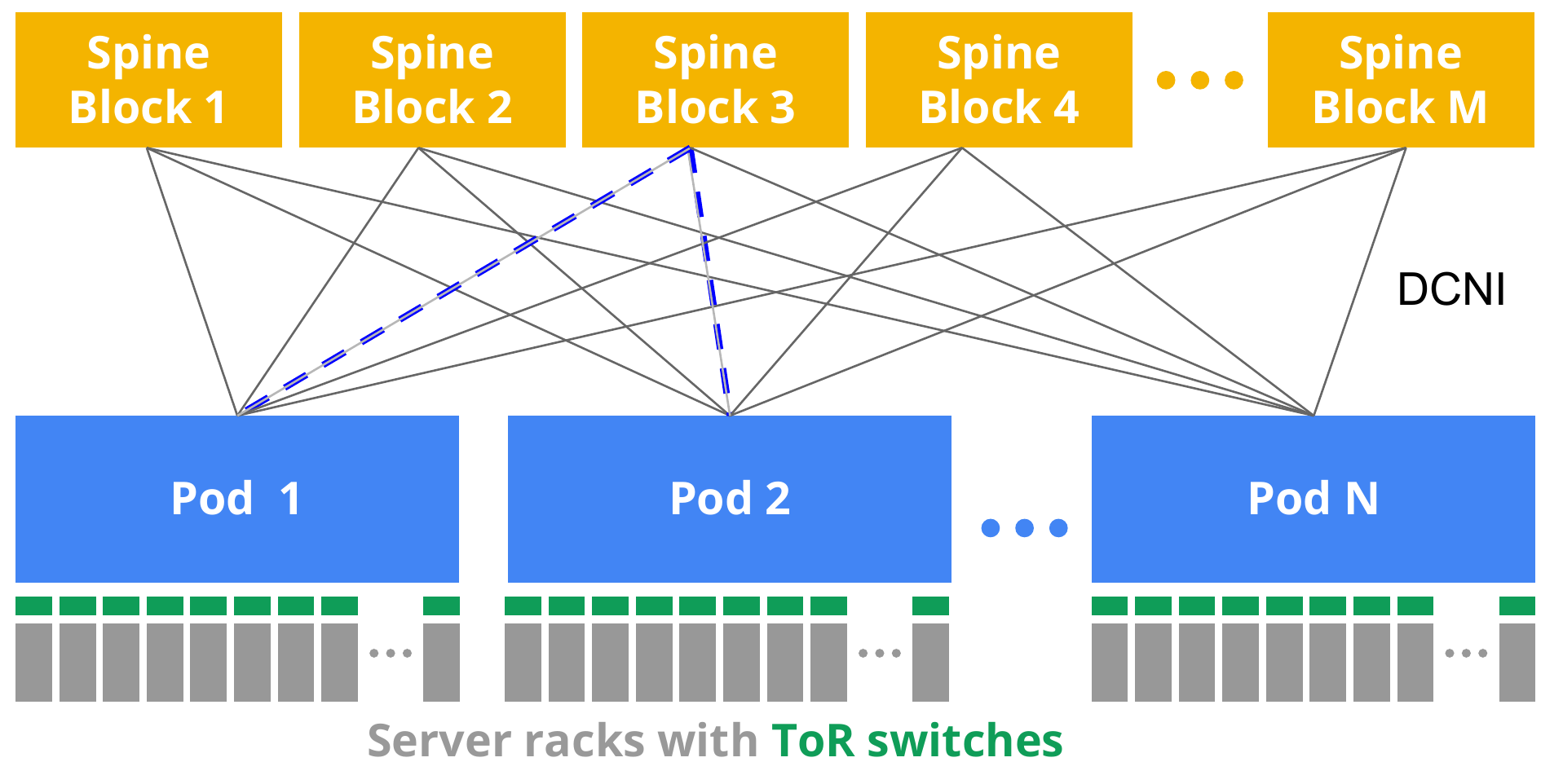}
  \setlength{\abovecaptionskip}{-12pt}
\setlength{\belowcaptionskip}{-12pt}
  \caption{\small FatTree: recursive Clos}
  \label{fig:fattree}
\end{minipage}
\begin{minipage}{0.66\columnwidth}
  \centering
  \includegraphics[width=0.9\columnwidth]{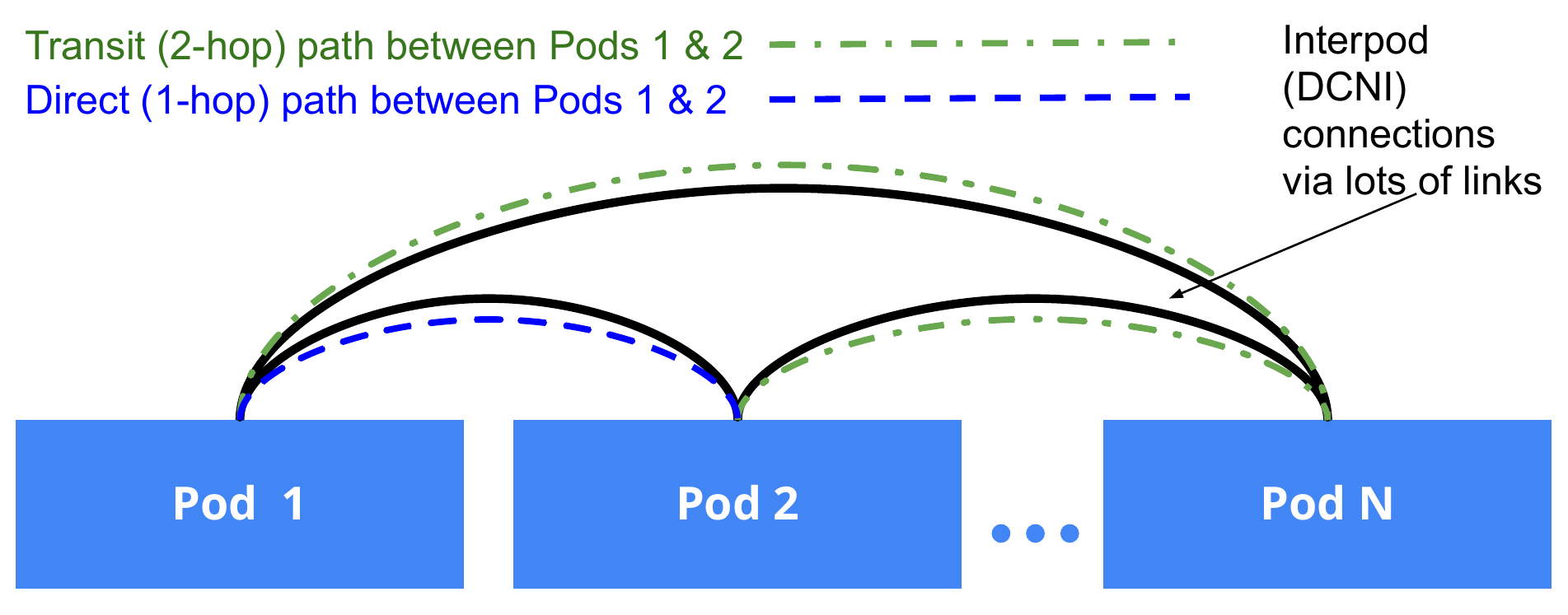}
  \setlength{\abovecaptionskip}{-2pt}
\setlength{\belowcaptionskip}{-4pt}
  \caption{\small Logical spine-free topology}
  \label{fig:logical-gemini}
\end{minipage}
\begin{minipage}{0.66\columnwidth}
  \centering
  \includegraphics[width=0.9\columnwidth]{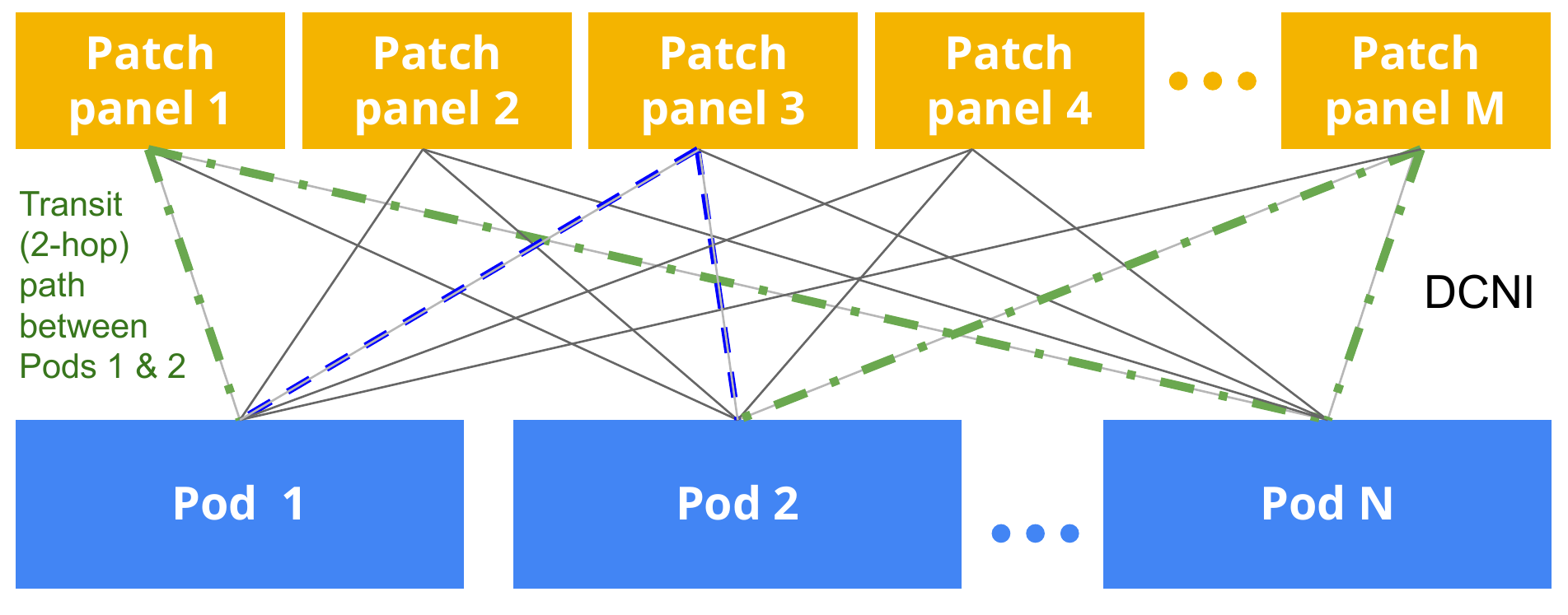}
  \setlength{\abovecaptionskip}{-2pt}
\setlength{\belowcaptionskip}{-4pt}
  \caption{\small Physical realization of spine-free topology, via patch panels}
  \label{fig:physical-gemini}
\end{minipage}
\end{figure*}

Datacenter topology designers must grapple with two competing objectives: cost and the need for reliable high
performance, in spite of dynamic workloads. Today's datacenter topologies use rearrangeably non-blocking
designs to support any admissible traffic matrix~\cite{fattree, jupiter, vl2, fb-dcn}.
These networks are hierarchical: sets of top-of-rack (\textit{ToR}) switches connect to
non-blocking Clos-based \textit{pods}, which are themselves connected, often by
Clos-based \textit{spines} -- see Fig.~\ref{fig:fattree}.

At larger scales (those which require an extra layer of switches), providing full bisection
bandwidth between the pods becomes expensive.
However, our observed inter-pod traffic
aggregates are both non-uniform and somewhat predictable (see \secref{sec:motivation}). This suggests we can
deploy a more efficient inter-pod network with non-uniform connectivity,
tuned to our predicted workloads rather than to the full-bisection worst case,
while matching the performance of non-blocking topologies \emph{on those workloads}.

If our traffic prediction, however, is imperfect, we risk load imbalance and high
packet loss.
To reduce that risk, we can reconfigure the inter-pod topology (the DataCenter
Network Interconnect, \textit{DCNI}) to match the traffic demand, or we can
reconfigure routing to rebalance the load to match the DCNI -- or we can do
both.

Prior research has explored reconfiguration at the top of the topology (the DCNI) 
through optical circuit switches (\textit{OCS}).
Helios~\cite{helios} used an OCS to establish dedicated pod-to-pod circuits
for long-lived elephant flows, 
and a separate spine network of electrical
packet switches to serve latency-sensitive flows. %
However, no commercially-available
OCS scales enough for Helios in large datacenters, and Helios required mechanisms to
detect and re-route elephant flows.

Other prior work uses reconfigurable ToR uplinks; as we discuss in 
\secref{sec:motivation}, practical considerations rule out these designs in today's
large datacenters.

\parab{Gemini: practical reconfigurability}
This paper describes \sysname, in which we replace the spine layer with
a passive \textit{restriping layer} 
that allows us to implement a reconfigurable DCNI;
we reconfigure both the DCNI topology
and routing at relatively long timescales; we use a novel approach to
predict future inter-pod demand; and we jointly optimize DCNI topology and routing based
on these predictions.   This allows us to oversubscribe the DCNI (w.r.t.
a worst-case traffic demand) without violating network-operator preferences for
minimizing and balancing link utilizations.\footnote{Readers should understand that we are \emph{not} proposing
a new fabric design with improved behavior on worst-case traffic; we are
describing a \emph{complete system}, including a control plane, that performs well
on traffic patterns we can observe. Therefore, our evaluations are based on link utilization metrics (see \secref{sec:metrics}), rather than theoretical properties (e.g., bisection bandwidth or oversubscription ratio) that ignore both the behavior of the control plane and the actual workload.}

Fig.~\ref{fig:logical-gemini} shows \sysname's \emph{spine-free} logical topology, where pods are directly connected.  
It also illustrates the
possibility of using two-hop \emph{transit} routing, to provide extra capacity
between pairs of pods.  Fig.~\ref{fig:physical-gemini} shows how one can use
a set of patch panels to create reconfigurable inter-pod links.

A spine-free network is significantly less expensive than the equivalent spine-based network,
as we discuss in \secref{sec:motivation}.

Because we only need to reconfigure the DCNI at long timescales, we can build the restriping layer
either out of patch panels (using humans to make changes) or relatively inexpensive,
commercially-available OCSs.   As the restriping technology improves,
\sysname can exploit faster DCNI reconfiguration to yield better results.
Our approach allows the restriping layer to be
split across multiple, independent, and relatively small patch panels or OCSs,
rather than requiring a single, impractically large switch or panel.

We must also tolerate long timescales for reconfiguring inter-pod routing. This can take several seconds~\cite{updateSchedule}, because (1) one TCAM rule update can take several mSec~\cite{updateSchedule}, (2) we might have to reconfigure hundreds
or thousands of rules, across hundreds of switches,
and (3) routing-table updates must be carefully sequenced across  switches, to minimize packet loss due to black holes and loops~\cite{zupdate, updateSchedule, consistentupdate}.

Given long reconfiguration timescales,
reconfigurability is practical \textit{only if the interval between reconfigurations is two
or three orders of magnitude larger than the reconfiguration time}.  Otherwise,
we risk violating overall fabric-availibility SLOs, because during restriping, the
fabric's total capacity is somewhat reduced, increasing the chance that a traffic
spike will cause packet loss.
Also, restriping is potentially error-prone, so
frequent restriping increases the risk of an SLO-impacting error.

\parab{Contributions} We show that it is both \textit{possible} and
\textit{useful} to reconfigure datacenter inter-pod topologies at infrequent intervals,
by jointly optimizing topology and routing for predicted, skewed workloads.
This allows us to eliminate 50\% of the expensive long-range transceivers, and a large fraction of the network
switches, without having to replace these with an expensive OCS.
Specific contributions include:
\squishlist

    \item \textbf{Traffic prediction}: It is widely assumed that datacenter traffic is
    unpredictable~\cite{vl2}.  We show, through production-network measurements, that
    while inter-pod traffic matrices (TMs) do change at short time-scales,
    these TMs have both predictable and unpredictable components,
    and the predictable components can be stable over days or weeks (\secref{sec:motivation}). 
    Inspired by
    prior work on robust routing~\cite{Applegate2004,Applegate2003, cope, criticalTM},
    we base our predictions on the \textit{convex hull} of a set of measured
    traffic matrices, over an \textit{aggregation window} of a few days or weeks (\secref{sec:design:traffic_model}).
    
    \item \textbf{Optimization}: Most robust optimization work~\cite{Applegate2003, Applegate2004, cope, criticalTM} only focuses on routing.
    \sysname\textit{jointly} optimizes topology and routing to
    find more opportunities to reduce both worst-case maximium link utilization (MLU) and average link utilization (ALU),
    across all traffic matrices at the extrema of a convex hull 
    (\secref{sec:design:joint_solver}).
    
    \item \textbf{Transit routing}: While the direct (optical-only) path between a pair
    of pods minimizes latency and link loading, our optimizer sometimes finds a better
    solution (w.r.t. MLU) using \textit{transit routing}, where packets 
    from Pod A to Pod B
    travel via Pod C, to reduce congestion in the case where the direct A-to-B links are overloaded (\secref{sec:design:joint_solver}).
    
    \item \textbf{Hedging}: Inter-pod demand can burst significantly at short timescales
    (\secref{sec:motivation}).
    Practical ``optimality''  requires good behavior both on average \emph{and} in the tail.
    To \textit{hedge} against the risk of unexpected bursts, the optimizer can %
    spread traffic across more paths than necessary for the expected case (\secref{sec:design:joint_solver}).
    Hedging gives us more robust
    handling of a wider range of workloads, in exchange for a little expected-case path stretch 
    due to more use of transit.
 
\squishend 
    
    Prior papers have often compared datacenter network designs
    based on flow-completion times (FCTs).   However, our traffic-matrix traces lack FCT data, and we use simulation-based evaluations that yield link utilizations 
    (we cannot accurately simulate FCTs, for a workload mixing thousands of applications, at scale). 
    \secref{sec:metrics} reports on the correlation between FCTs and link utilizations from actual production networks.
    These results support our use of utilization-based metrics in evaluating
    Gemini against other designs (\secref{sec:evaluation}).

    In particular, we show that a \sysname network outperforms a spine-free network constructed to have the same total DCNI cost, and often performs almost as well as 
     a higher-cost spine-based DCNI. 
     We also show how various aspects of \sysname, including topology reconfiguration, routing reconfiguration, hedging, and parameter choices contribute to the improvements.
    
\emph{This work does not raise any ethical issues.}

\vspace{-0.17in}
\section{Motivation} 
\label{sec:motivation}

\sysname's design is motivated by several interesting properties of today's DCNIs. We quantify these properties using measurements from 22 production fabrics.
Each property motivates a different aspect of \sysname's design.

\parab{Cost} For practical reasons, such as bursty traffic,
network operators avoid running networks at full utilization~\cite{jupiter,fbtraffic, microburst}. Therefore, a non-blocking fabric, such as a full-bisection bandwidth Clos, is too expensive, and
operators rely on over-subscription~\cite{fattree, fbtraffic, Benson-mstraffic} at different layers of the network.
The dominant cost of a high-speed network is in the optics\footnote{E.g., assuming 40G long-reach optics
at $\$7$ $Gb/s$~\cite{optical-handbook}, removing each $100G$ spine block, with $512$ ports, would save $360K$
in downlink optics  -- not counting switches, internal cabling, power, cooling, etc.}
and associated fiber~\cite{jupiter,rail}, and allowing
2:1 over-subscription of a Clos DNCI by removing half of the pod-to-spine links can remove almost half of
the cost of this layer.
Alternatively, by removing the spine blocks and directly connecting pods in \sysname, as shown in ~\figref{fig:physical-gemini}, we achieve similar cost reductions while preserving much of the performance
of a non-oversubscribed Clos.

Note that the simplest spine-free design, 
a uniform-mesh DCNI with a Valiant Load Balancing (VLB) routing scheme \cite{VLB}, offers little cost benefit over a spine-full DCNI, because it would require provisioning each pod with a DCNI link capacity of twice its DCNI demand -- an overprovisioning ratio of 2:1. This is required to serve \emph{all} traffic patterns (especially worst-case ones) with demand-oblivious routing.
This obliterates the cost savings from removing spines, by transferring that cost to the pods themselves.   \sysname makes a spine-free DCNI feasible without overprovisioning the pods.

\parab{Constraints on practical reconfigurability}
Many recent proposals ~\cite{mirrormirror, projecttor, firefly, rotornet, opera, sirius} have
proposed mechanisms for reconfiguring datacenter networks at the ToR level, using either
a reconfigurable fabric of optical cables, free-space optics, or high-capacity wireless networks. In response to a shift in traffic demand, these approaches can reconfigure a fabric on timescales of nsec to msec.
However, these proposals rely on custom hardware that is not yet commercially available at scale~\cite{rotornet, opera, sirius} or cannot be deployed in current production datacenters~\cite{beyondfattree, rotornet}, or require accurate flow size information for scheduling, which might not be always available~\cite{flowsize}.  Instead, \sysname yields good results with
relatively infrequent DCNI reconfiguration (\secref{sec:eval:sensitivity} discusses \sysname's sensitivity to
the reconfiguration interval.)

\begin{figure}[htb]	
{		 
    \centering	
    \includegraphics[width=0.95\columnwidth]{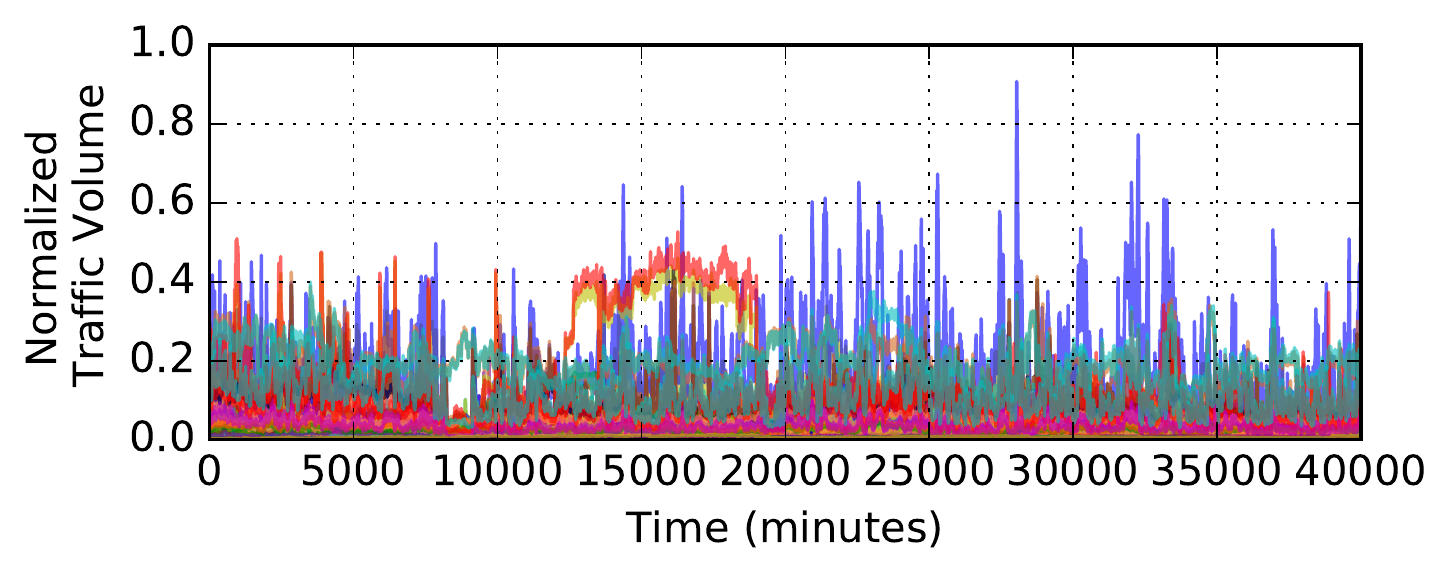}
    \vspace{-0.05in}
    \begin{spacing}{0.1}
    {\footnotesize Each color represents one pod-pair (``commodity'').}
    \end{spacing}
    \setlength{\belowcaptionskip}{-10pt}
    \caption{\small Normalized traffic vs. time: all pod-pairs, fabric F5 }
    \label{fig:traffic_time_series}
}
\end{figure}

\parab{Dynamic inter-pod traffic}
Even when highly aggregated, inter-pod level traffic is quite dynamic. \figref{fig:traffic_time_series} shows the inter-pod traffic of all pod pairs in an
illustrative fabric, F5;
traffic varies significantly at small time scales, with many traffic spikes.
This dynamism makes it hard to accurately predict traffic matrices in real time.
Because we need a robust, stable traffic model, %
we instead build models using the convex hull of past traffic matrices  (\secref{sec:design:traffic_model}).

\begin{figure}[htb]
  \centering
  \vspace{-0.15in}
  \includegraphics[width=0.95\columnwidth]{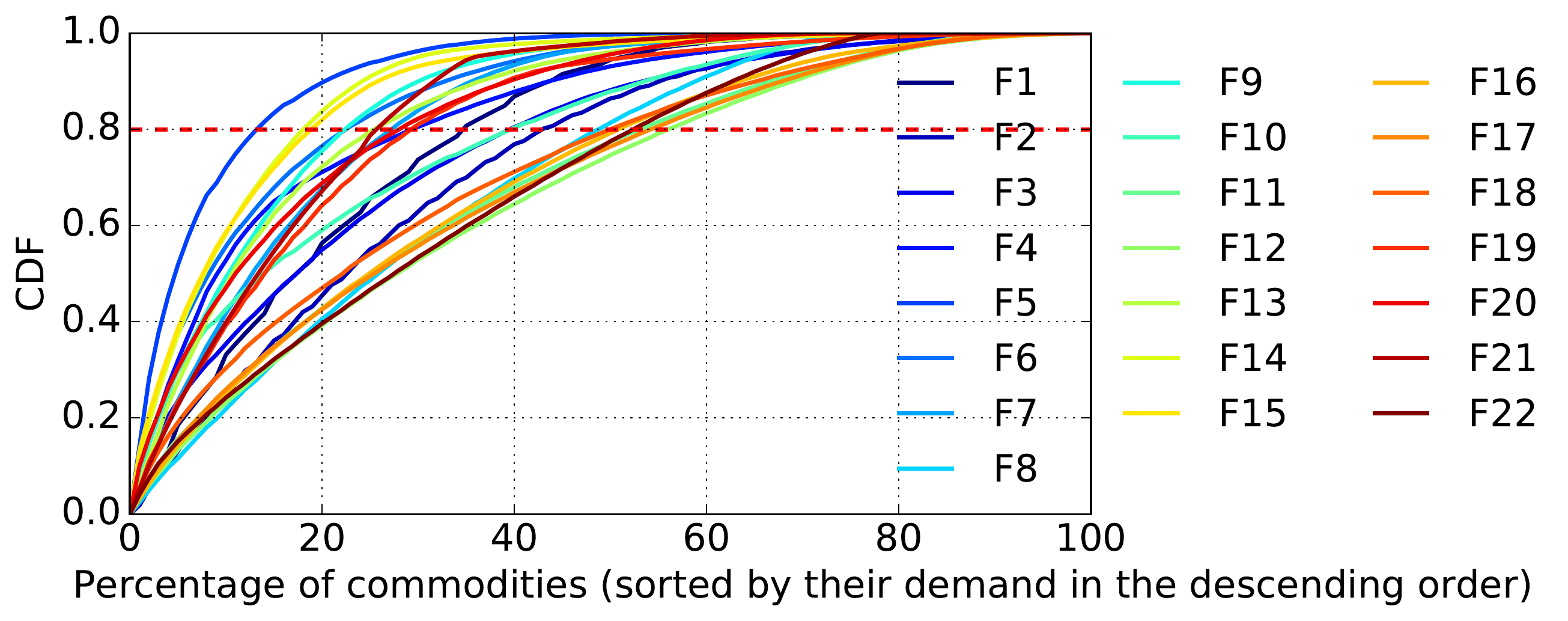}
  \setlength{\abovecaptionskip}{2pt}
 \setlength{\belowcaptionskip}{-10pt}
  \caption{\small Average pod-level traffic skew over one month}
  \label{fig:skewness}
\end{figure}

\parab{Skewed inter-pod traffic}
For some fabrics, inter-pod traffic can be significantly \textit{skewed}: a few pod-pairs account for a significant fraction of inter-pod traffic (\cite{projecttor}
reported similar skew in ToR-level traffic). \figref{fig:skewness} shows that, for
11 of our 22 fabrics, 30\% of pod-pairs (``commodities'') account for 80\% of the traffic. (In other fabrics, the distribution is more uniform.) 

Traffic skew motivates two design decisions (\secref{sec:design:joint_solver}).
First, \sysname \textit{reconfigures DCNI topology}
to match the skewed demand.
This \emph{Topology Engineering} (ToE) gives higher DCNI capacity to
pod-pairs that carry higher traffic volumes; relative to a uniform
DCNI, ToE can reduce network congestion and more efficiently use costly resources.
Prior ToE work has either augmented the DCNI with OCS-based
reconfigurable paths for specific flows~\cite{helios}, or has focused on ToR-level reconfiguration; see \secref{sec:related}; we believe reconfiguring to handle pod-pair
demands, using a single DCNI, is novel.

Second,
\sysname can use the spare pod-internal capacity, in the lower-utilization pods,
to \textit{transit} traffic between two other pods when insufficient direct capacity
exists.  \sysname adapts to the degree of skew; in fabrics with relatively uniform
traffic, \sysname generates a uniform DCNI and uses single-hop pod-to-pod paths. Those two design decisions are the core of our joint topology and routing solver (\secref{sec:design:joint_solver}). 

\begin{figure}[htb]
  \centering
  \vspace{-0.1in}
  \includegraphics[width=0.95\columnwidth]{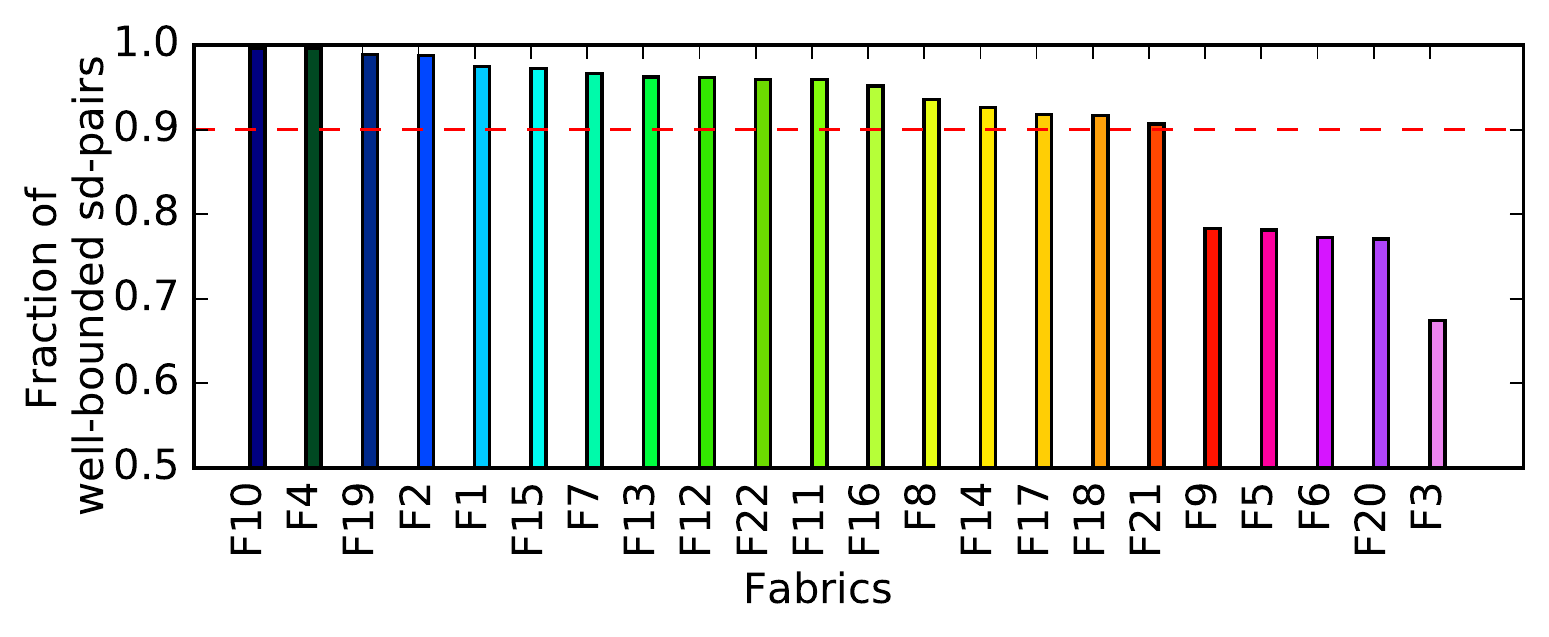}
  \setlength{\abovecaptionskip}{-4pt}
  \setlength{\belowcaptionskip}{-15pt}
  \caption{\small Fraction of well-bounded pairs; higher is better}
  \label{fig:pct_boundable_sd_pairs}
\end{figure}

\parab{Large predictable traffic component} For many fabrics, we have observed some traffic
predictability at long time-scales.  As a feasibility study,
we ``trained'' on the maximum pod-to-pod demands over a sliding aggregation window of 7 days,
then ``tested'' predictability.  We quantify this, per pod-pair, as
the ratio of demand on the \emph{next} day over
the prior-7-day maximum (demand-to-max ratio, DMR, for short). A pod-pair is \emph{well-bounded}
if its 99-th percentile DMR is below 1 -- \ie 99\% of the next day's demand
does not exceed the previous week's maximum; otherwise it is \emph{poorly-bounded}
\figref{fig:pct_boundable_sd_pairs} plots the fraction $p$ of well-bounded pod-pairs for 22 fabrics, measured over one month.
For most fabrics, most pod-pairs are well-bounded; for 17 fabrics,
$p>0.9$.  
We refer to fabrics with $p>0.9$ as \emph{mostly-bounded}.
Even for the least-predictable fabric, F3, $p=0.68$.

This suggests that it might be feasible to reconfigure the DCNI for some fabrics, no more than once
per day, based on, say, a week's worth of data, especially if we can route inter-pod demands to leverage statistical multiplexing.

\begin{figure}[htb]
    \centering
    \vspace{-0.1in}
    \includegraphics[width=\columnwidth]{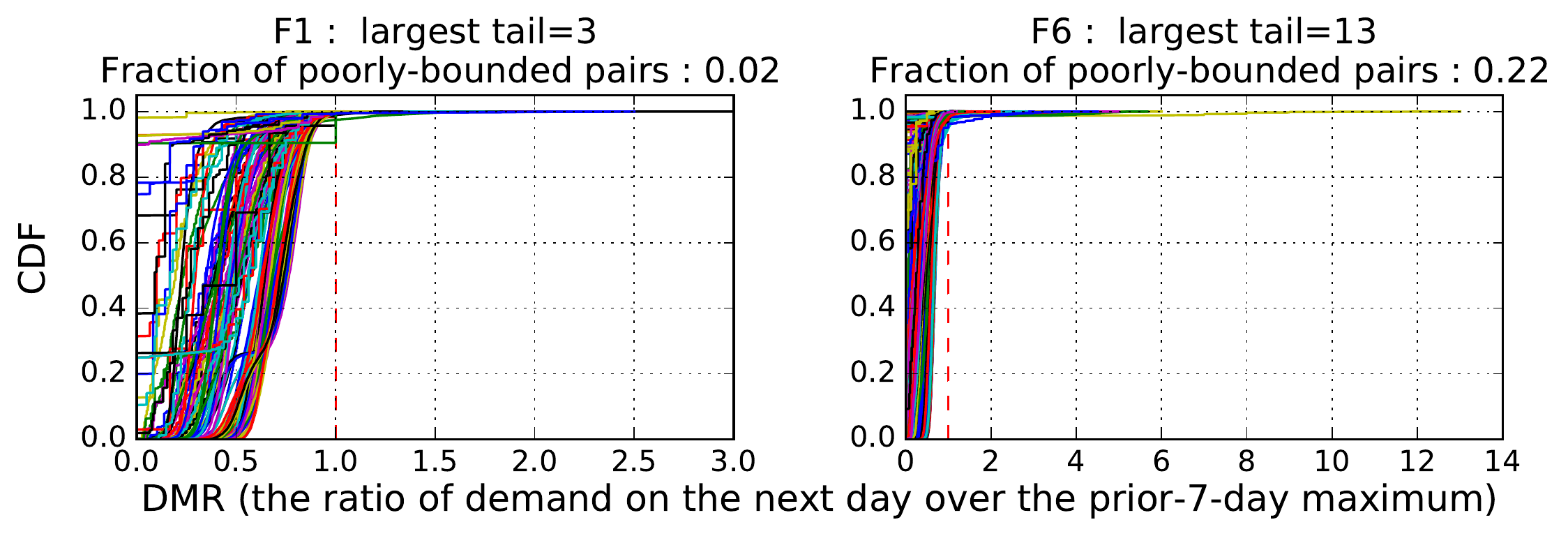}
    \setlength{\abovecaptionskip}{-8pt}
    \setlength{\belowcaptionskip}{-18pt}
    \caption{\small CDFs of demand-to-max ratio (DMR)}
    \label{fig:traffic_tail}
\end{figure}

\parab{Small unpredictable traffic component: long tail distribution for some pod-pairs}
Though most pod-pairs' demands, at most times, are well-bounded, a few are not, and these have long-tailed DMR distributions.
\figref{fig:traffic_tail} shows the pod-pair DMR distributions for two representative fabrics:
F1, %
with $p=0.98$ and F6, with $p=0.78$.
The maximum DMRs for F1 and F6 are 3 and 13.
A long tail implies sudden traffic-pattern changes.
To handle sudden changes, we could try to rapidly reconfigure the topology and routing, but that is difficult and risky; we prefer to
proactively embed sufficient inherent robustness in the topology and routing, which motivates the risk-based design in \secref{sec:design:joint_solver}.

\vspace{-0.1in}
\section{Measuring success}
\label{sec:metrics}

Networks exist to serve applications, and the gold standard for network
evaluation has been \emph{Flow Completion Time} (FCT)\cite{DukkipatiMcKeown2006}.
However, FCTs can be difficult to measure directly, and especially difficult to
simulate at datacenter scale.   It is much easier and efficient to measure network-level metrics,
such as link utilization and per-link loss rates, that are observable via
mechanisms such as SNMP, scalably and without privacy concerns.  We can
aggregate these metrics over intervals chosen as a compromise between fidelity
and feasibility.

\sysname's traffic modeling and solver use aggregated inter-pod traffic
traces as inputs.   
Our simulation-based evaluation (\secref{sec:simulation})
likewise generates link utilizations, from which we can compute MLUs and ALUs.

Intuitively, these link-utilization metrics should be correlated with FCTs,
\emph{but is that intuition correct}?
Large changes in MLU (e.g., changing MLU from 1\% to
99\%) presumably harm FCTs, but what about the smaller changes we actually see
when comparing \sysname to other designs, or when comparing parameter settings?

\begin{figure*}[htb]
\centering
\begin{minipage}{0.4\columnwidth}
  \centering
  \includegraphics[width=0.95\columnwidth]{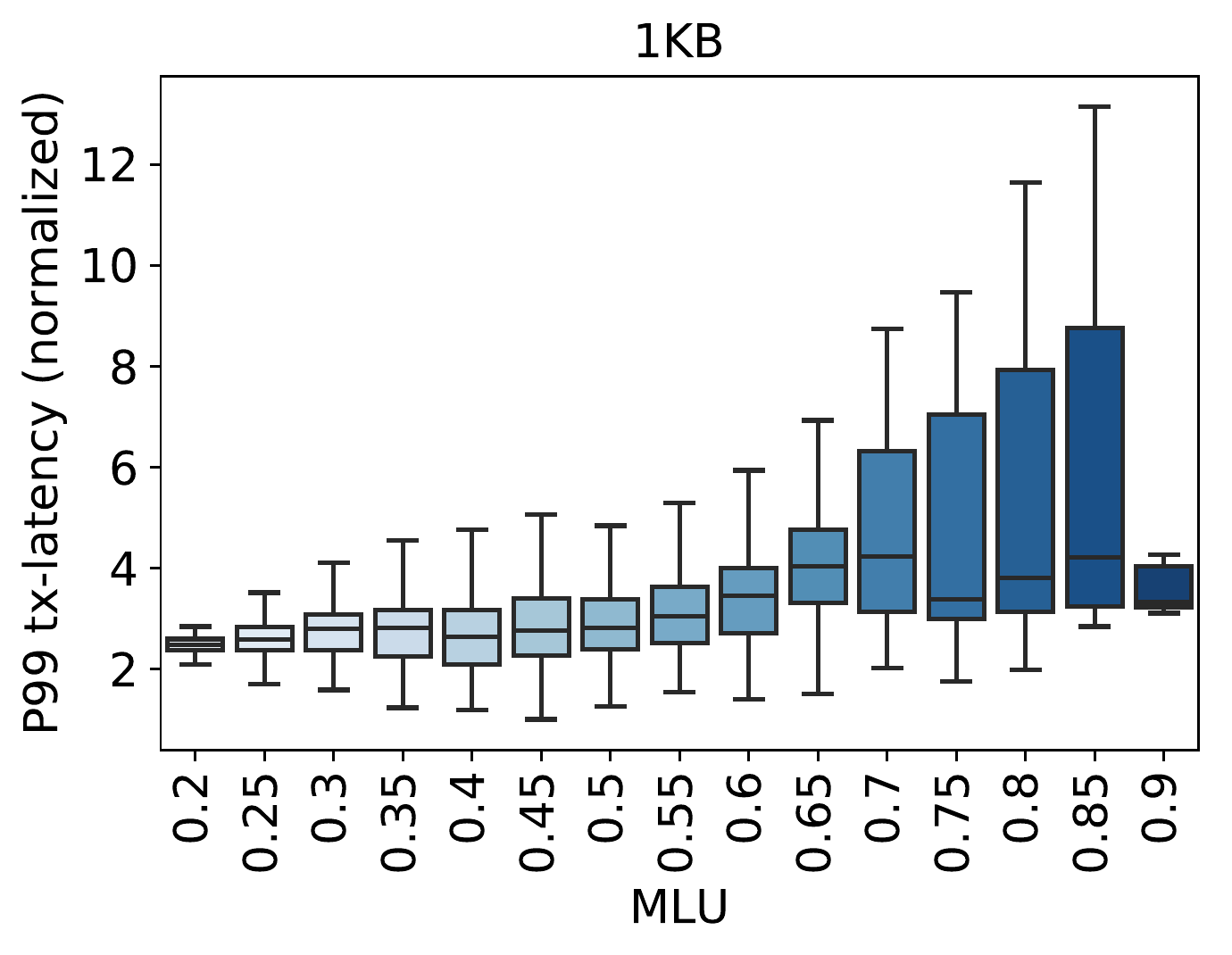}
  \label{fig:tx_latency_1kb_99p_MLU_spineful}
\end{minipage}
\begin{minipage}{0.4\columnwidth}
  \centering
  \includegraphics[width=0.95\columnwidth]{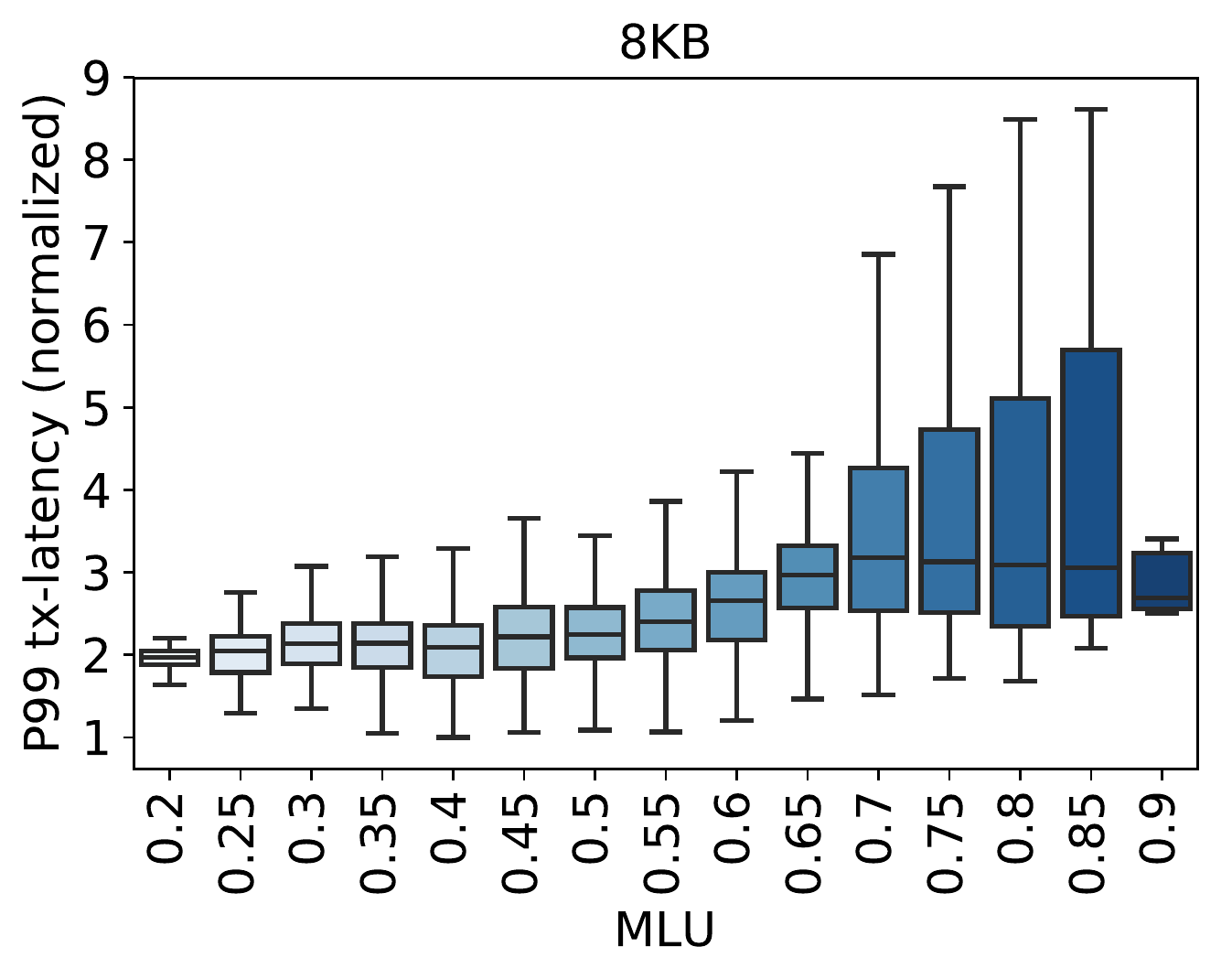}
  \label{fig:tx_latency_8kb_99p_MLU_spineful}
\end{minipage}
\begin{minipage}{0.4\columnwidth}
  \centering
  \includegraphics[width=0.95\columnwidth]{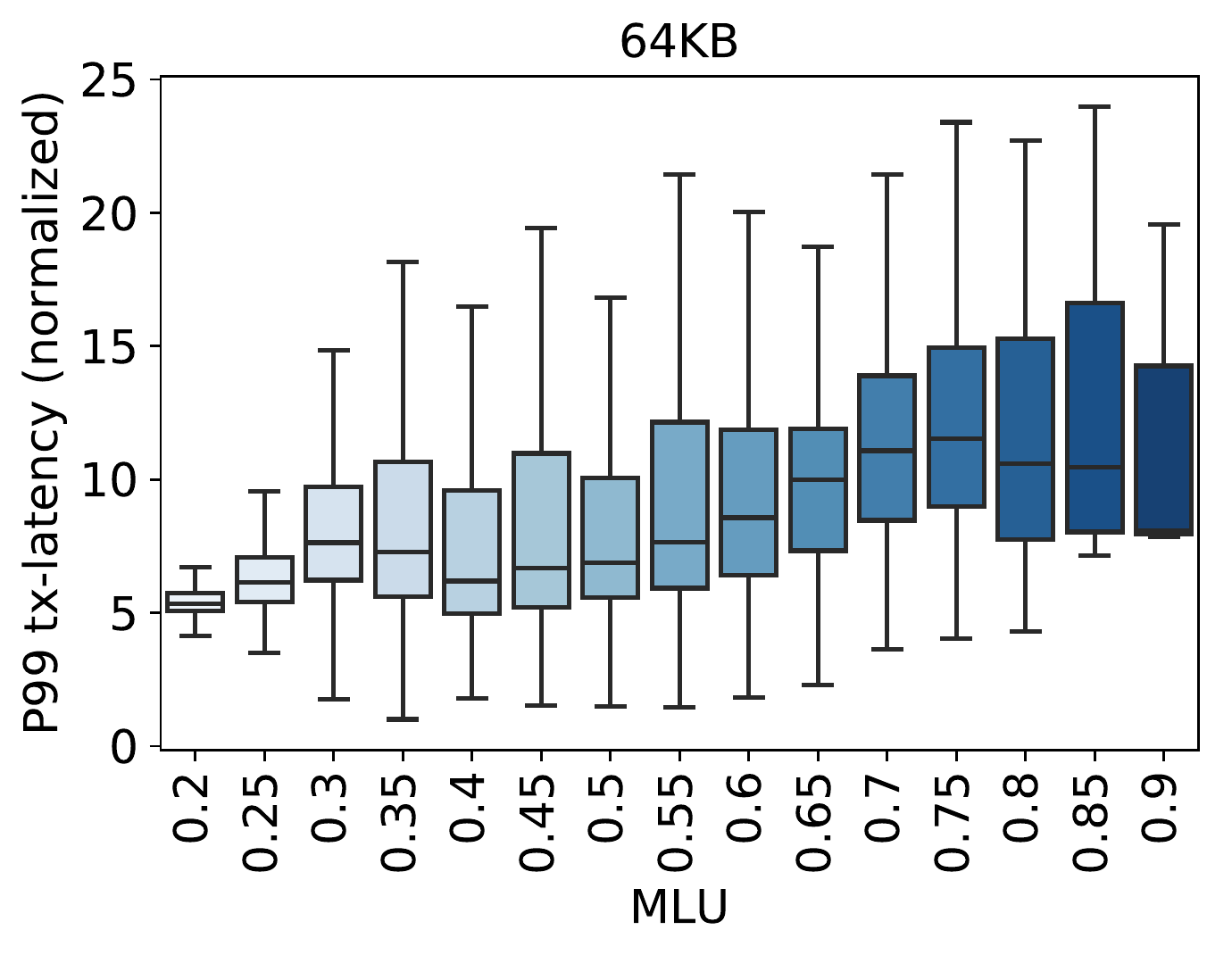}
  \label{fig:tx_latency_64kb_99p_MLU_spineful}
\end{minipage}
\begin{minipage}{0.4\columnwidth}
  \centering
  \includegraphics[width=0.95\columnwidth]{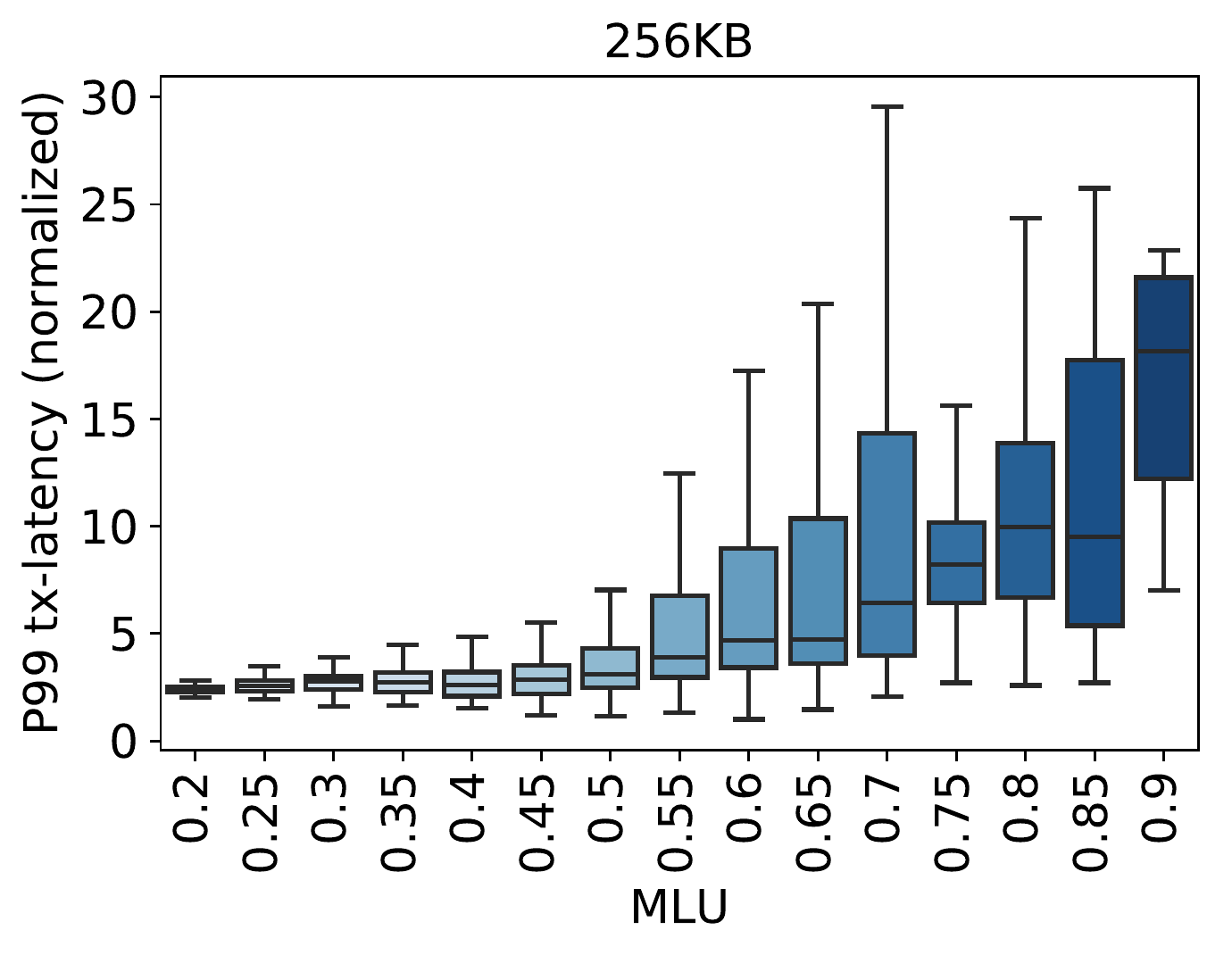}
  \label{fig:tx_latency_256kb_99p_MLU_spineful}
\end{minipage}
\begin{minipage}{0.4\columnwidth}
  \centering
  \includegraphics[width=0.95\columnwidth]{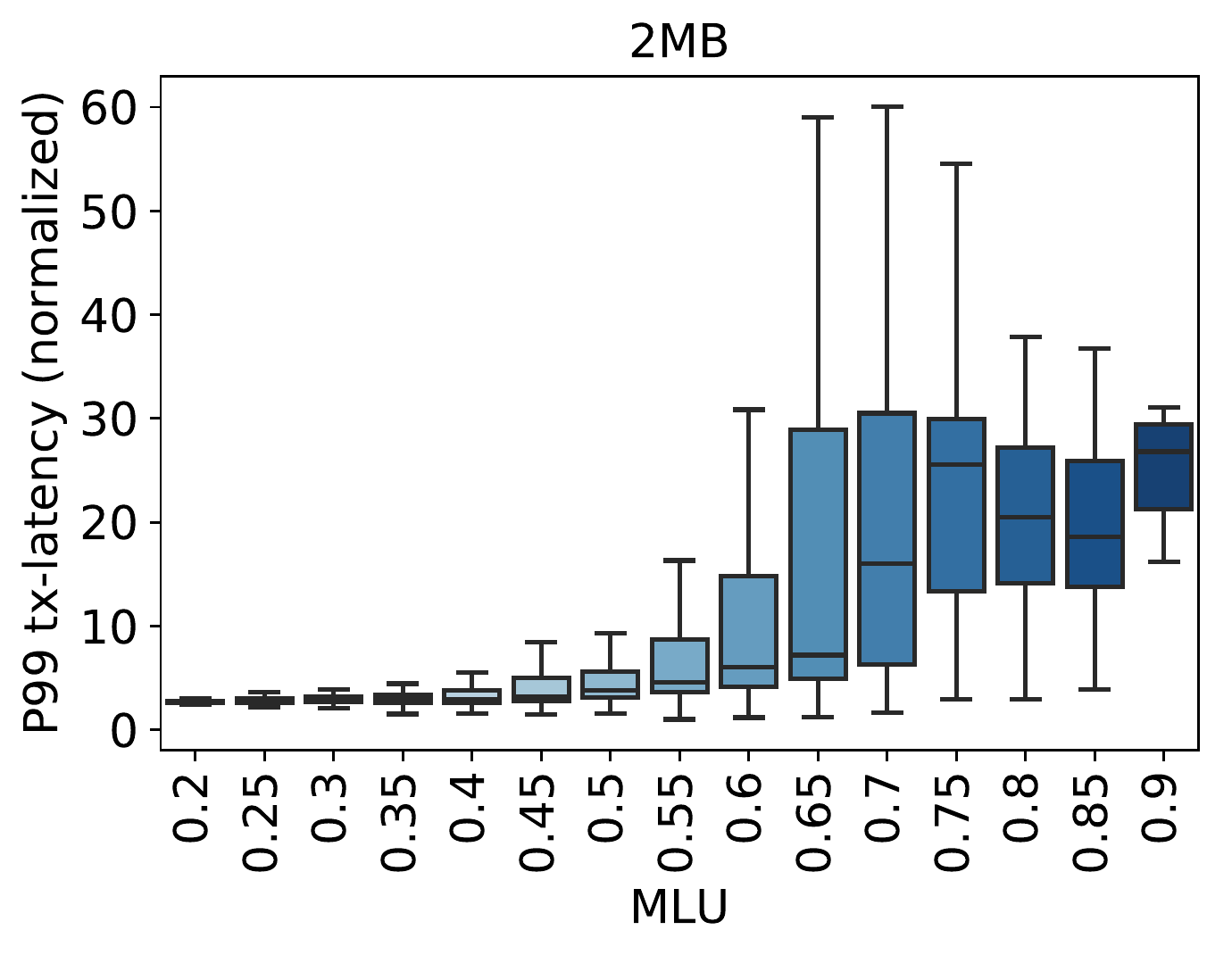}
  \label{fig:tx_latency_2mb_99p_MLU_spineful}
\end{minipage}
\setlength{\abovecaptionskip}{-2pt}
\setlength{\belowcaptionskip}{-4pt}
\caption{FCTs vs p99 MLUs on production fabrics}
\label{fig:fct_vs_mlu} 
\end{figure*}

\begin{figure*}[htb]
\centering
\begin{minipage}{0.4\columnwidth}
  \centering
  \includegraphics[width=0.95\columnwidth]{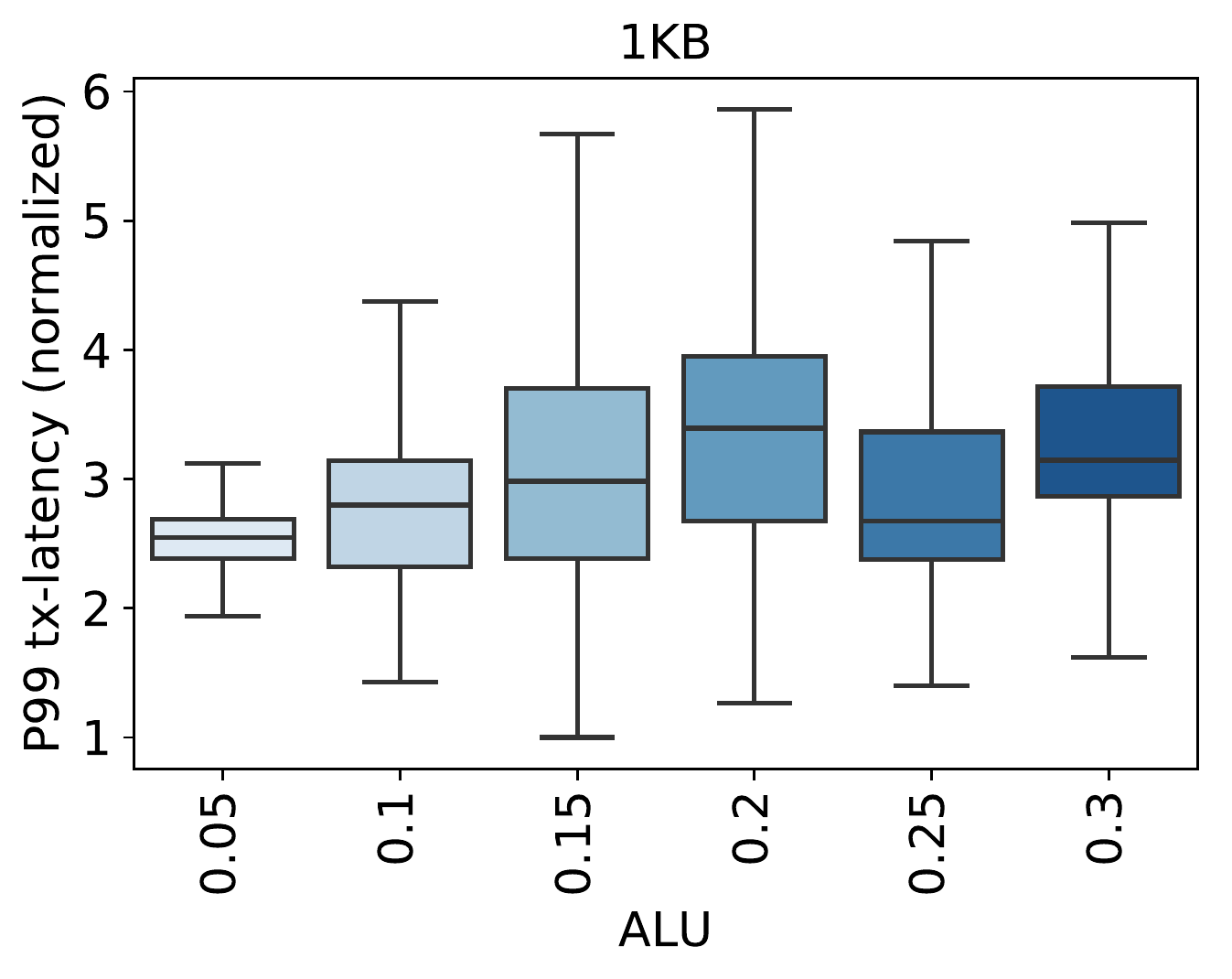}
  \label{fig:tx_latency_1kb_99p_ALU_spineful}
\end{minipage}
\begin{minipage}{0.4\columnwidth}
  \centering
  \includegraphics[width=0.95\columnwidth]{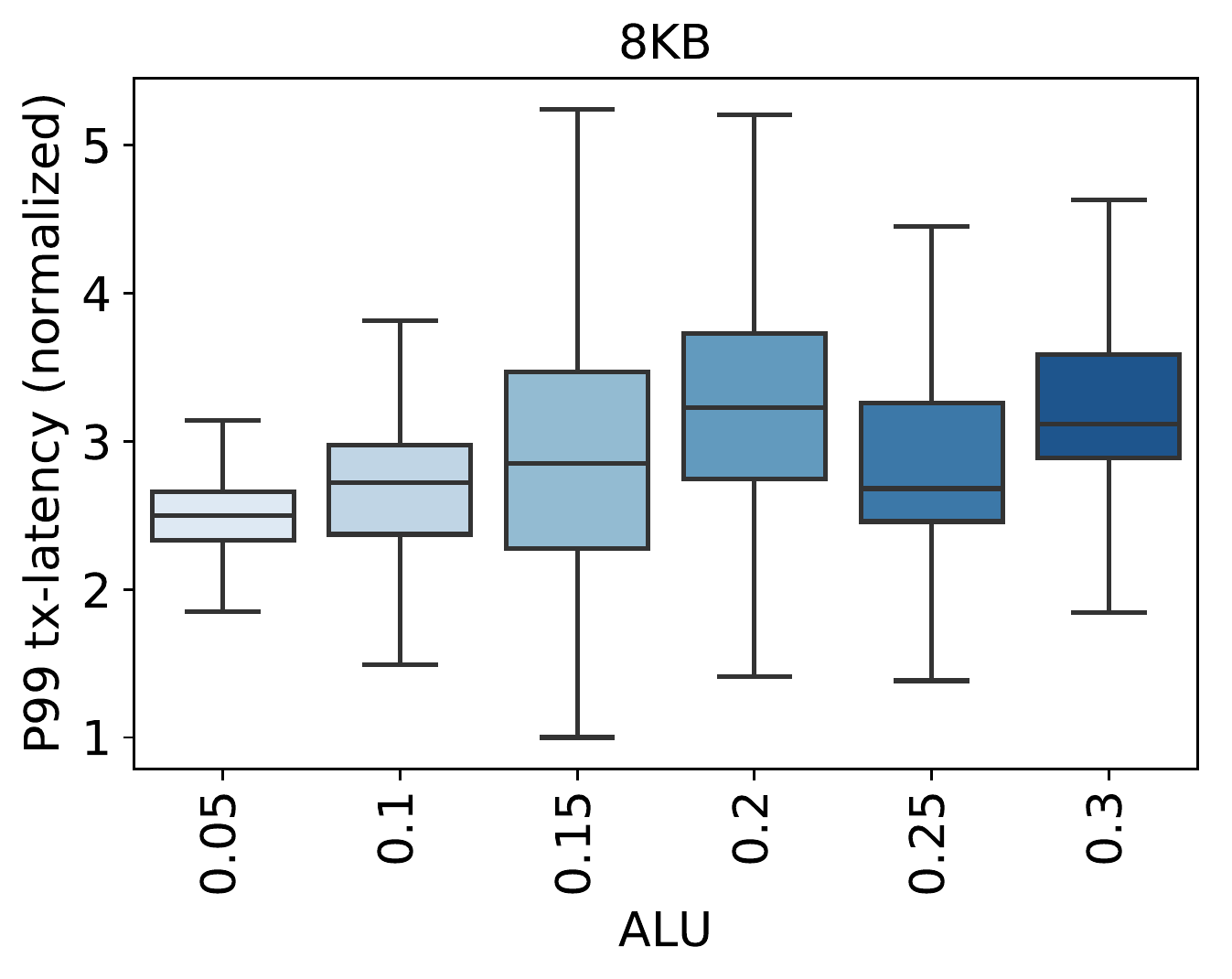}
  \label{fig:tx_latency_8kb_99p_ALU_spineful}
\end{minipage}
\begin{minipage}{0.4\columnwidth}
  \centering
  \includegraphics[width=0.95\columnwidth]{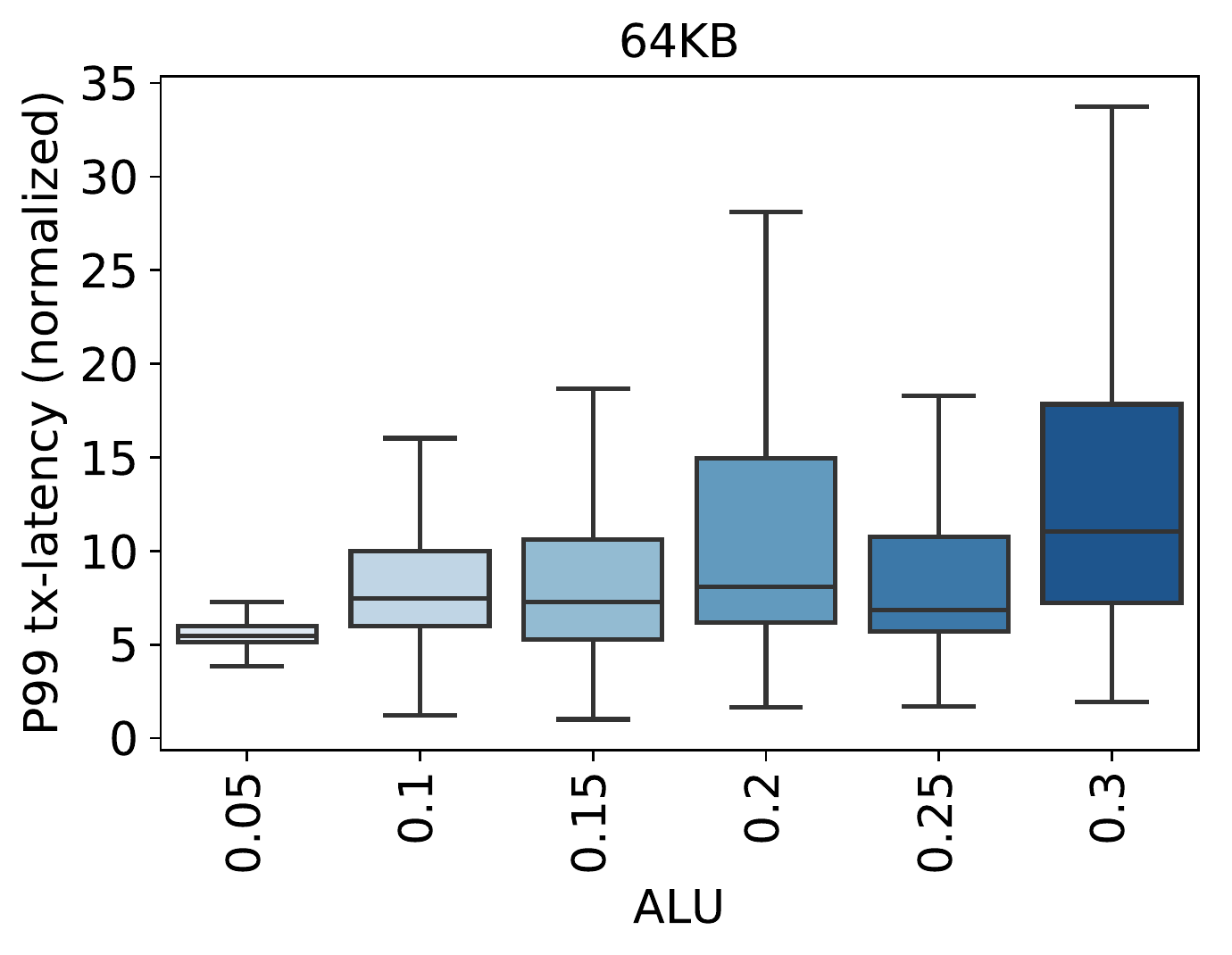}
  \label{fig:tx_latency_64kb_99p_ALU_spineful}
\end{minipage}
\begin{minipage}{0.4\columnwidth}
  \centering
  \includegraphics[width=0.95\columnwidth]{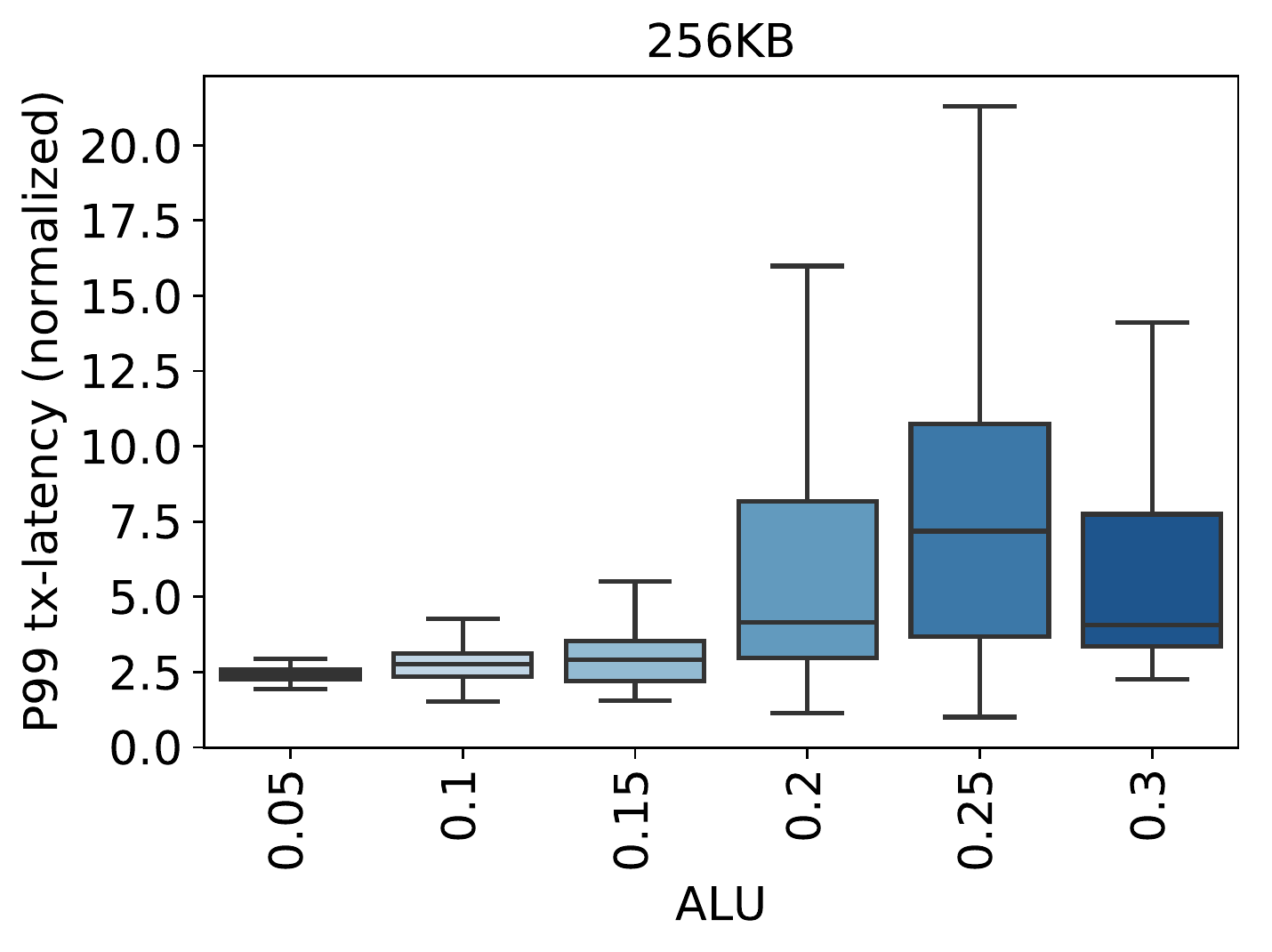}
  \label{fig:tx_latency_256kb_99p_ALU_spineful}
\end{minipage}
\begin{minipage}{0.4\columnwidth}
  \centering
  \includegraphics[width=0.95\columnwidth]{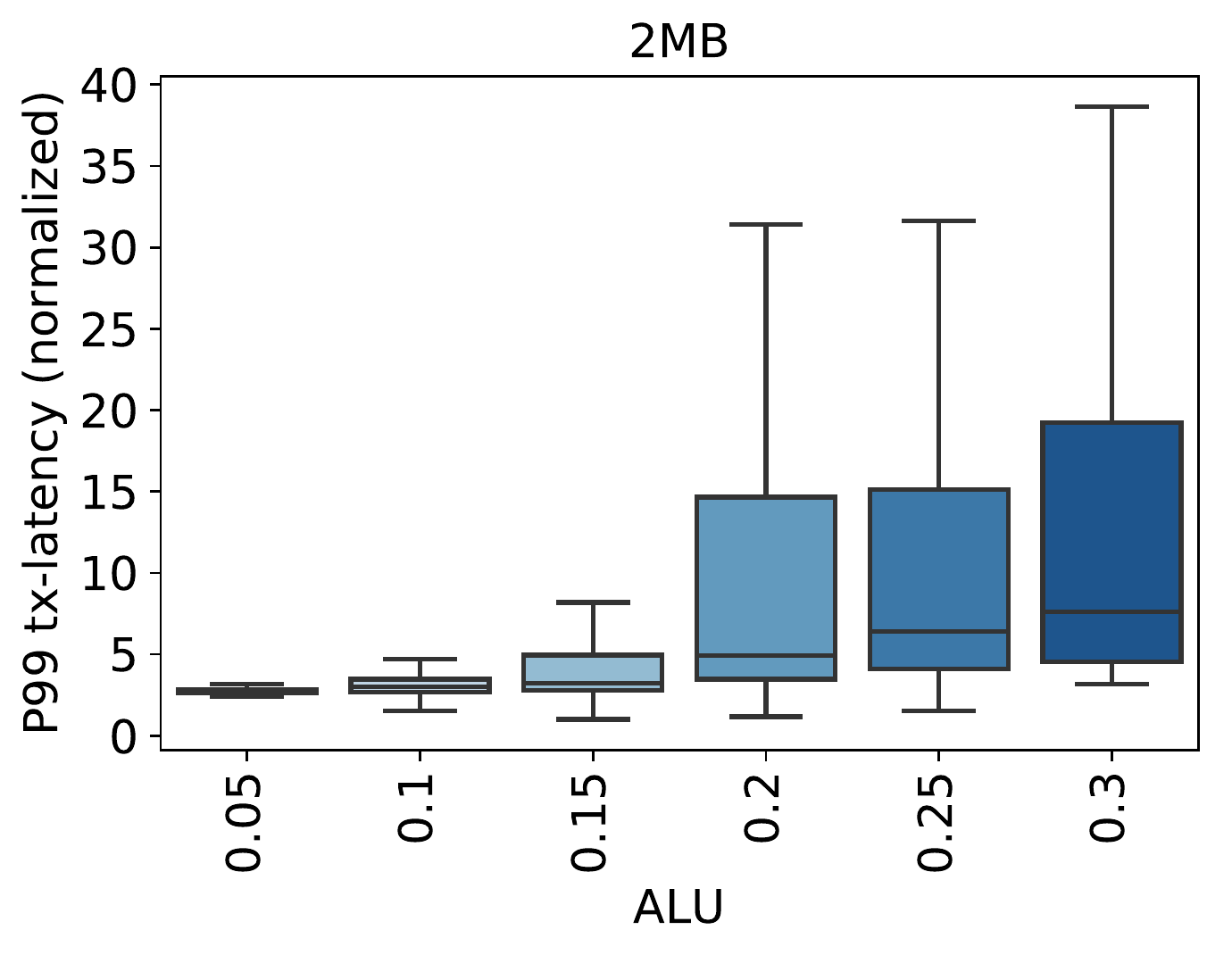}
  \label{fig:tx_latency_2mb_99p_ALU_spineful}
\end{minipage}
\setlength{\abovecaptionskip}{-2pt}
\setlength{\belowcaptionskip}{-4pt}
\caption{FCTs vs p99 ALUs on production fabrics}
\label{fig:fct_vs_alu}
\end{figure*}
\vspace{-0.1in}
\subsection{Correlation between FCT and MLU}
\label{sec:fctcorrelations}

Because we do have FCT data for some (not all) applications on our network, we
could study the correlation between several FCT-based metrics and network
metrics (ALU and MLU) in production.   
\textbf{This is not an exhaustive or rigorous study,}
which remains a good topic for future work.

Using data aggregated from 32 Fat-Tree datacenter fabrics over 7 days, we
collected FCT metrics (transmission latencies for various message sizes) for flows
between servers on different pods of the same fabrics; this focuses on the effects
of DCNI utilizations, which we collected simultaneously.

\figref{fig:fct_vs_mlu} plots FCTs for several message-size buckets\footnote{The message size shown at the top of each graph, in \figref{fig:fct_vs_mlu} and \figref{fig:fct_vs_alu}, is the upper bound for the message-size bucket represented by that graph.}, vs. DCNI MLU (in 5\% buckets),
normalized to the best sample for each size.
The results suggest that p99 FCTs increase with MLU.
This is consistent with prior work showing that, at high link utilizations, packet loss rates increase~\cite{swift},
which would be likely to increase FCTs.
However, our experiments cannot conclusively establish the relationship between FCT and MLU, since there may be confounding factors (\eg offered load); clarifying this relationship is future work.

\figref{fig:fct_vs_alu} suggests a weaker effect of ALU on FCTs, until the ALU
exceeds about 20\%, where ALUs clearly appear to affect long-message FCTs.

In \secref{sec:all_link_correlations} we plot FCTs \vs link utilization metrics for all links
in each network, not just the DCNI links.  However, those results are less indicative
of whether the DCNI-only simulated utilizations in \secref{sec:evaluation} would be predictive of FCT benefits.

\subsection{FCT vs. frequency of overloaded links}
\label{sec:fctvsbalance}

MLU and ALU do not tell the whole story.  Two \sysname solutions could have identical
MLU and ALU, while one had relatively few overloaded links, while another
had many overloaded links.
One might expect the second to exhibit a much higher
total loss rate, and therefore would be likely to have much worse FCTs~\cite{PadhyeEtAl1998, Luan2019}.

From measurements of loss rates \vs link utilization by others~\cite{swift}, we believe that links loaded
above 0.8 (in a measurement interval, e.g. 5 minutes) should be treated as ``overloaded.''
Therefore, one metric we will apply in our evaluations is the fraction of overloaded links (``Overloaded Link Ratio,'' OLR).
We attempted to find a correlation between FCTs and DCNI OLRs in our own fabrics, but these were generally too lightly loaded to provide enough data.
(\figref{fig:fct_vs_olr_all_links} in \secref{sec:all_link_correlations} does suggest a correlation for
all-links OLRs.)

\section{\sysname Design}
\label{sec:design}

\gemini predicts for each fabric, using historical traffic data, a reconfiguration strategy that optimizes fabric metrics. This strategy determines both the DCNI topology as well as routing paths for inter-pod traffic. 

\subsection{\sysname Overview} 
\label{sec:design:overview}

\parab{Approach} \sysname addresses the challenges described in \secref{sec:motivation} as follows. (i) It optimizes the topology based on link utilization metrics; these have been shown to be correlated with loss rates~\cite{swift}. (ii) It models traffic demand using a \textit{collection} of historical traffic matrices, and uses this model to derive topology and routing \textit{configurations}. (iii) It jointly optimizes topology and routing configuration, which enables it to identify opportunities to optimize link utilization aggressively. Its optimization formulation accounts for pod heterogeneity. (iv) It hedges against mispredictions in fabrics with short-term variability by spreading traffic across multiple paths (the shortest path and 2-hop\footnote{For now, to minimize latency impact, we only consider 2-hop paths. We have left it to future work to explore longer paths.} paths). (v) Because fabrics vary in skew and predictability, and because hedging can increase path stretch, \sysname selects determines the best reconfiguration strategy (whether to use ToE or not, whether to use hedging or not) for each fabric by simulating, on historical traffic, the impact of these choices on link utilization. (vi) It incorporates several techniques to scale the search for the best configuration to large fabrics.

\begin{figure}[t]
{		 
\centering		
\includegraphics[width=1\columnwidth]{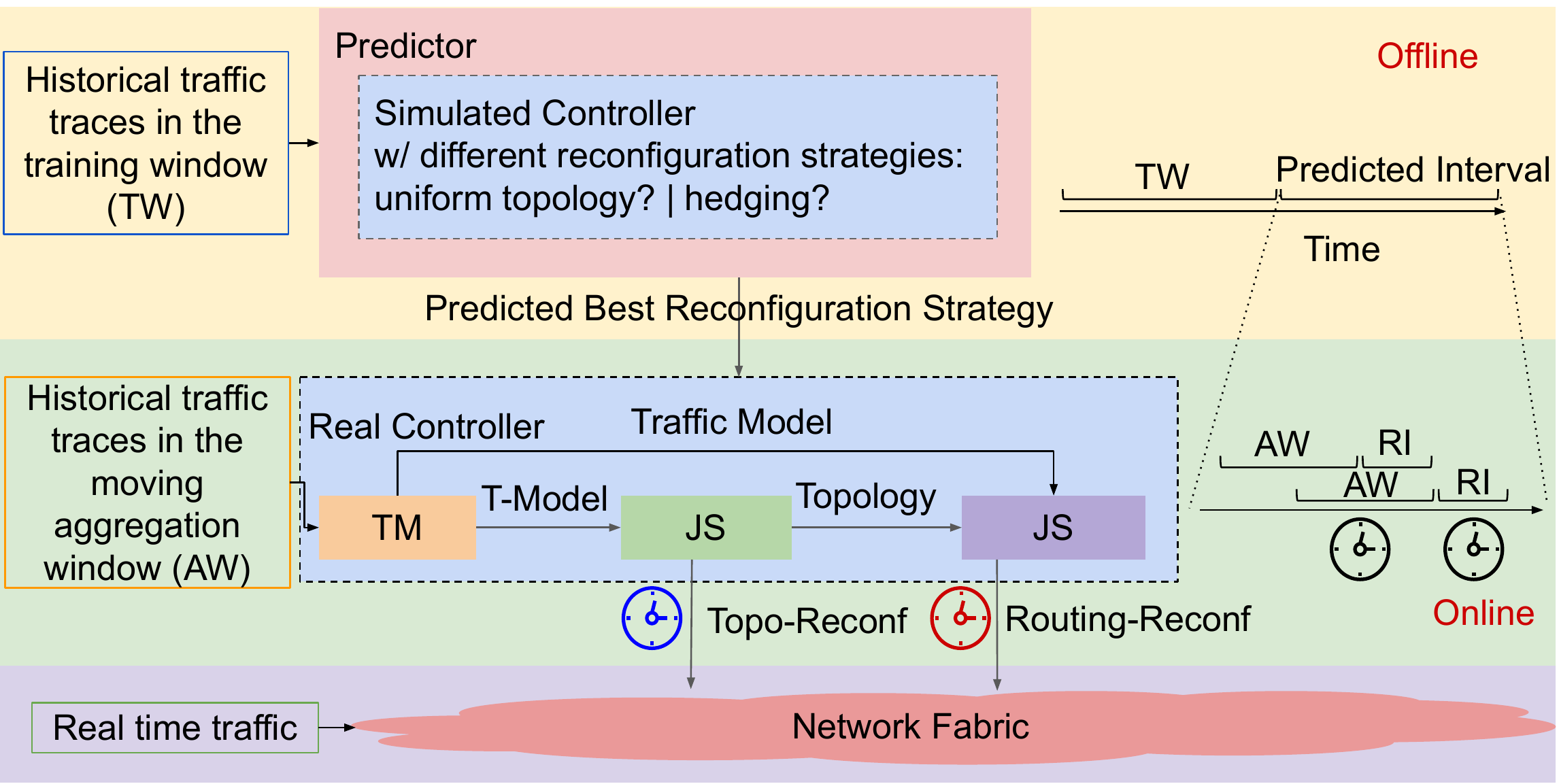}
\caption{\small \sysname architecture. TM: traffic modeler; JS: joint topology and routing solver; RI: reconfiguration interval.}
\label{fig:gemini_arch}	 
}
\end{figure}

\parab{Gemini Architecture} 
\gemini consists of two components, a \textit{Predictor} and an extended SDN \textit{Controller} (\figref{fig:gemini_arch}).
The Predictor determines, given a historical trace of traffic matrices over a \textit{training window} (e.g., one month) for a fabric, the best \textit{reconfiguration strategy} to use over the next \textit{predicted interval}, which we set to one month, leveraging the long-term predictability described in \secref{sec:motivation}. A reconfiguration strategy consists of two decisions: (a) whether to reconfigure the topology (b) whether to hedge routing.

Using the predicted strategy, to adapt to significant short-term variability (\secref{sec:motivation}), the Controller adapts routing configurations at finer timescales, based on a moving \textit{aggregation window}'s worth of traffic (e.g., one week; see \secref{sec:eval:sensitivity} for a sensitivity analysis).
The Controller always updates routing configuration using a \textit{routing reconfiguration interval} (15 minutes, in our simulations); hedging, if included in the strategy, is done via routing adjustments. Reconfiguring routing can take several seconds (\secref{sec:intro}), so a 15-minute interval is feasible.

If the strategy also includes topology reconfiguration, the Controller changes the DCNI topology at a fixed \textit{topology reconfiguration interval }. \secref{sec:eval:sensitivity} discusses how results vary with intervals ranging from one day to multiple weeks, and shows that once per month is sufficient.  This is feasible using patch panels; if faster reconfiguration is required, commercial OCS switches could be employed. (\secref{sec:eval:sensitivity} shows that routing needs to be reconfigured much more often than topology.)

Note that the Predictor and Controller both use the same components (\figref{fig:gemini_arch}): a \textit{traffic modeler} that abstracts traces into a compact traffic model, and a \textit{joint topology and routing solver} that produces solutions necessary for prediction and configuration. We first describe these components, then describe how the Predictor and Controller use these. Before doing so, we discuss what metric(s) \sysname seeks to optimize.

\subsection{Minimizing Link Utilization}
\label{sec:design:metrics}

Recent work has focused on minimizing flow completion time (FCT) as the objective of reconfiguration. In large datacenters with centralized (SDN) control~\cite{jupiter, orion}, reconfiguring the DCNI to optimize individual flows does not scale, since several million flows can be active at any instant. Also, flow size information might not be available for scheduling~\cite{flowsize}. Instead, \sysname makes reconfiguration decisions based on inter-pod traffic demand. This prevents us from using FCT as an optimization goal. 

\begin{figure}[t]	
{		 
\centering		
\includegraphics[width=0.75\columnwidth]{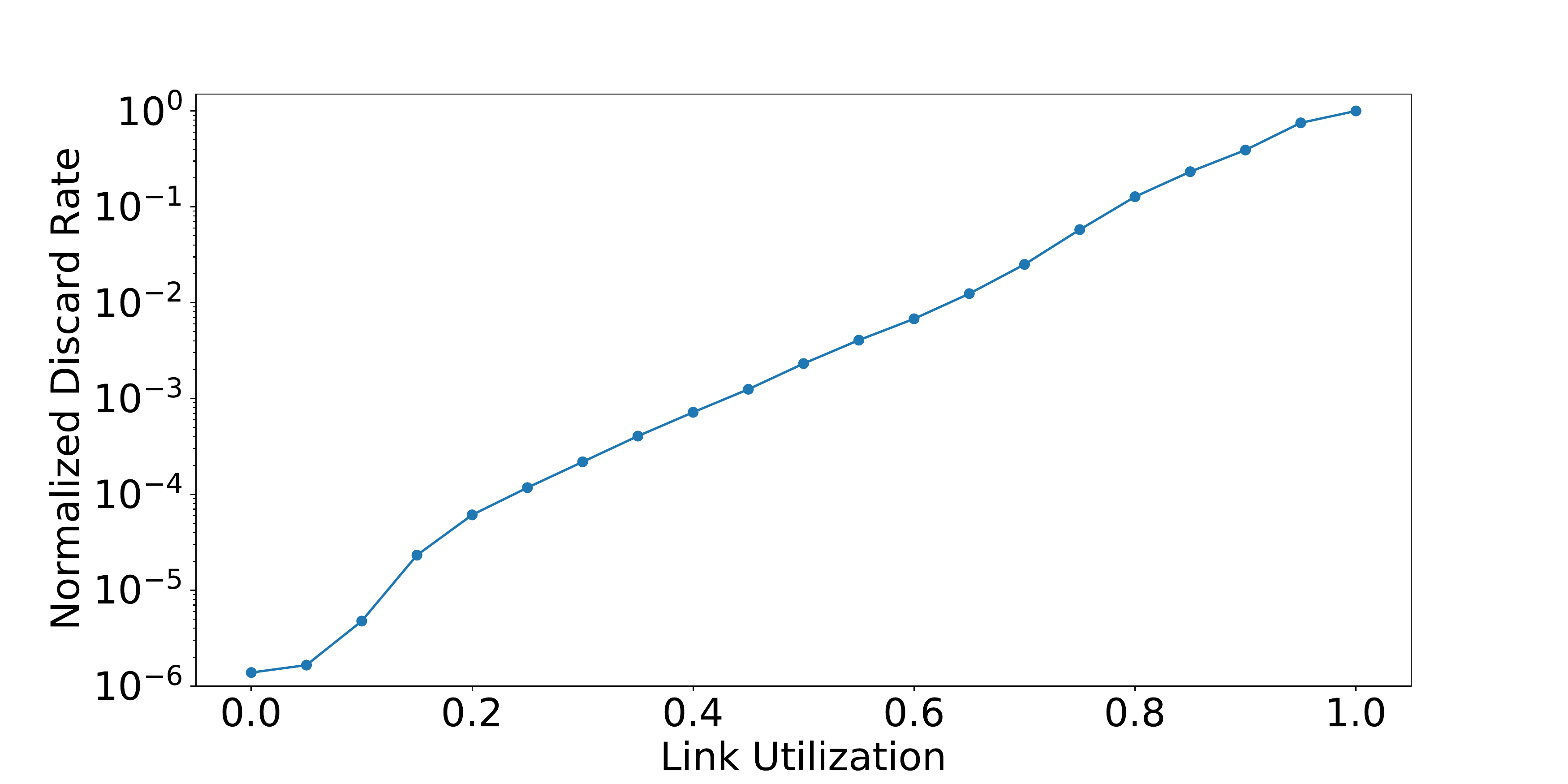}
\caption{\small Per-link relationship: utilization vs. discard rate.} %
\label{fig:util-loss}	 
}
\end{figure}
Instead, \sysname optimizes link utilization, which is (non-linearly) correlated with packet discard rate (as shown in~\cite{swift} for datacenters, and~\cite{ospf-weights} for WANs; we have also independently verified this using measured utilization and discard rate of DCNI links in all of the fabrics in our dataset, \figref{fig:util-loss}). This implies that at higher utilizations there is less headroom to tolerate bursts, which can cause high discard rates. A topology where every link operates at low to moderate utilization is likely to provide better application-perceived performance than one in which a few links operate at high utilization.

More specifically, \sysname minimizes the maximum link utilization (MLU) across the fabric, following prior work~\cite{criticalTM,cope,Applegate2003,Applegate2004} which uses the same objective for WAN traffic engineering. In addition, \sysname also aims to  minimize the network path stretch, equivalent to minimizing the average link utilization (ALU). This ensures, when possible, that traffic is routed along low latency paths. This objective is secondary, since the latency impact of an extra hop within the datacenter is lower than that of packet discards.

\subsection{Traffic Modeling}
\label{sec:design:traffic_model}

\gemini exploits historical traffic matrices available from our networks to develop a model of traffic demand in a fabric. We capture traffic matrices as average pod-to-pod bandwidth utilization every five minutes for each fabric. Even for a single fabric, an aggregation window's worth of traffic matrices can be significant, corresponding to over 2000 traffic matrices. To scale better, \gemini abstracts this traffic matrix history into a compact \textit{traffic model}, and makes configuration and prediction decisions based on this traffic model.

\gemini could have used the elementwise-maximal inter-pod traffic matrix (Maximal-TM), where each element is the maximum demand during the aggregation window for that pod-pair. This is a pessimistic choice, and can result in an inefficient use of network resources, since the maximum demand for pod-pairs might not all occur at the same time.

\begin{figure}[t]	
{		 
\centering		
\includegraphics[width=\columnwidth]{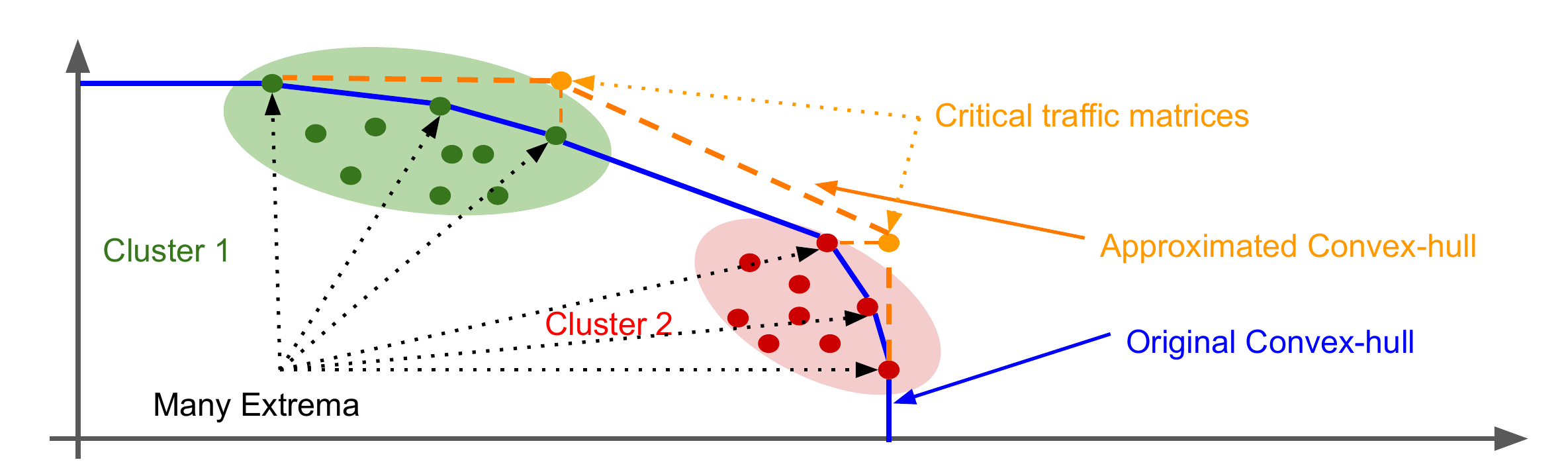}
\caption{\small Convex-hull-based Prediction.}	
\label{fig:polytope}	 
}
\end{figure}

Instead, it uses an
\textit{approximate convex hull of all traffic matrices within the aggregation window}, building upon prior work on robust routing~\cite{Applegate2004,Applegate2003, cope}. That work, given an arbitrary topology, seeks a routing assignment for a collection of traffic matrices $\mathcal{T}$ such that the maximum MLU is minimized. Those papers show that it suffices to consider the traffic matrices at the extrema of the convex hull of $\mathcal{T}$. \sysname uses this property, but departing from this prior work, seeks to \textit{jointly optimize both topology and routing} (\secref{sec:design:joint_solver}). While this property reduces the number of traffic matrices to consider in the joint optimization, the number of extrema on the convex hull can be very large for large aggregation windows necessary to ensure long timescale reconfiguration (\figref{fig:polytope}).

To further reduce computational complexity, \sysname leverages the technique used in~\cite{criticalTM}. Specifically, it groups traffic matrices into $k$ clusters. Then, from all the traffic matrices in a cluster, we generate a \textit{critical traffic matrix} whose elements are the element-wise maxima of the cluster's traffic matrices. Those critical TMs are extrema of an approximated convex hull which is strictly larger than the original one as shown in \figref{fig:polytope}. Note that the Maximal-TM is a special case in our approach where $k=1$. Optimizing over these critical TMs can minimize MLU when the future traffic falls within the approximated convex hull.

The traffic model uses two parameters, the traffic aggregation window and the number of critical TMs. We evaluate their impact in~\secref{sec:eval:sensitivity}.

\subsection{Hedging}
\label{sec:design:hedging}

To mitigate the impact of short-term unpredictability, \sysname incorporates a technique we call \textit{hedging}. This section describes the intuition underlying hedging; the next section formalizes this intuition.

\begin{figure}
{
\centering		
\includegraphics[width=1\columnwidth]{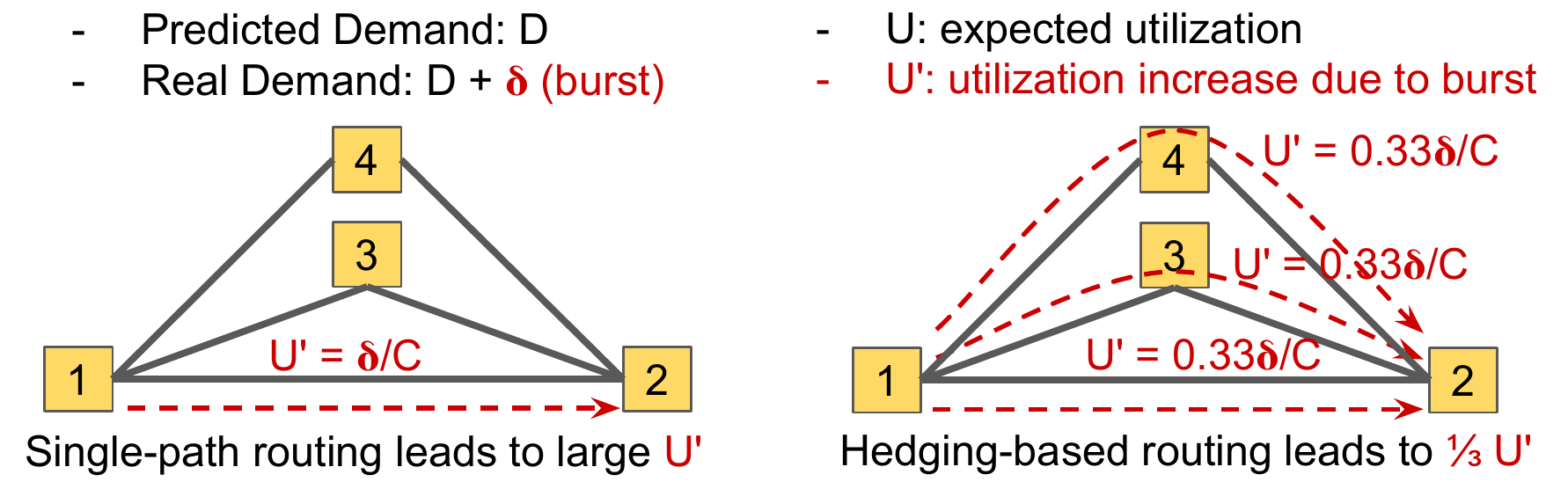}
\caption{\small Hedging-based routing and topology. The example shows that hedging-based routing and topology engineering should be used for handling traffic bursts.}
\label{fig:gemini_motivation_examples}	 
}
\end{figure}

\figref{fig:gemini_motivation_examples} illustrates the idea of using hedging to reduce utilization surge due to a burst $\delta$. The topology has only four pods. For simplicity, we only focus on a particular pod-pair, 1-2. Those three figures show the utilization increase due to a traffic burst $\delta$. The leftmost figure shows that when a single shortest-path routing is used in a uniform topology, the burst can lead to $\delta/C$ utilization increase on a single trunk. However, if we split the traffic on three paths (one one-hop path and two two-hop paths) as shown in the middle, the burst is spread across these paths leading to only $0.33\delta/C$ utilization increase. Alternatively, if we assign more capacity on trunk 1-2 as shown in the rightmost figure, we achieve the same utilization increase as in the middle figure. We call both strategies \textit{hedging}: the former effects hedging through traffic engineering, the latter through topology engineering.

To use hedging in practice, \sysname must decide: 1)  the traffic split ratio of a demand (defined as $f$) on each path and 2) the right capacity allocated on each trunk. Motivated by the example of \figref{fig:gemini_motivation_examples}, we quantify the risk $r$ of a burst $\delta$ over a trunk with capacity $C$ as $f\delta/C$, which is exactly the utilization increase due to $\delta$ on the trunk. Minimizing $r$ forces bursts to be spread out over more paths, which reduces (“hedges”) the risk that any single path will be overloaded or may cause more capacity to be allocated on some paths. One can estimate $\delta$ for each pod-pair based on past traffic data. However, in our experience, assigning the same $\delta$ for all pod-pairs reduces MLU effectively. Therefore, we leave more accurate estimate of per pod-pair $\delta$s to future work.

Hedging has the undesirable side-effect of using two-hop paths, increasing path stretch. \sysname disables hedging for well-bounded fabrics, as discussed in \secref{sec:design:predictor}.

\subsection{Joint Solver Design}
\label{sec:design:joint_solver}

The core of \sysname is a solver that searches for the optimal topology and routing configuration, given the traffic model described in \secref{sec:design:traffic_model}, and is designed to scale to large fabrics. %

\begin{figure*}[htb]
{		 
\centering		
\includegraphics[width=2\columnwidth]{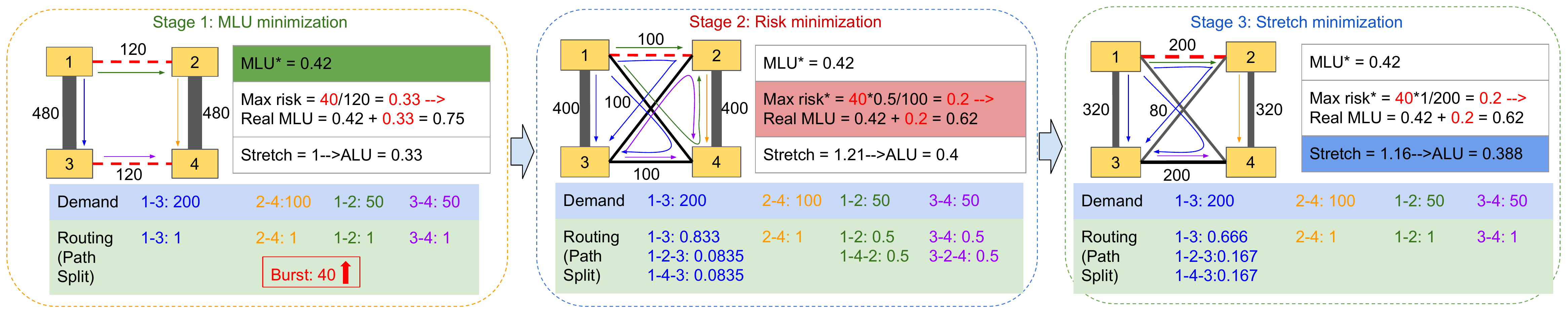}
\caption{\small Multi-stage optimization example. Red dashed edges: trunks with maximum risk.}
\label{fig:multistage_optimization}
}
\end{figure*}

\parab{Notation} We model the pod-level network as a directed graph $G(\set{V},\set{E})$, where the vertex set $\set{V}$ represents pods and edge set $\set{E}$ represents directed logical trunks between pods. 

The traffic model $\mathcal{T}$ contains $m$ critical
traffic matrices. In the $t$-th matrix, the $i,j$-th entry, denoted by $d_{i,j,t}$ denotes the demand between pods $i$ and $j$. Also, we use $f_{i,j,p}$ to denote the path split ratio of $d_{i,j, t}$ on path $p\in \mathcal{P}_{i,j}$.

\parab{Modeling pod heterogeneity} Pods can be heterogeneous, as a result of incremental expansion~\cite{googleexpansion}. We model this as follows: the $i$-th pod has a fixed radix $x_i$ (number of ports\footnote{In this paper, we only consider the DCNI-connected ``uplink'' ports.}) and all ports of a pod have the same uplink rate $s_i$. 
(Uplink and downlink rates are the same.)
Pod $i$'s ports are partitioned into different subsets, such that each subset of ports is connected via fiber links to a distinct pod.
The set of $n_e$ links, connecting two pods constitutes a \textit{trunk} between the pods. 
The capacity of the trunk $C_{e}$, is determined by $n_e\cdot s_e$, where $s_e = min(s_i, s_j)$.

\begin{figure}
{
\centering		
\includegraphics[width=1\columnwidth]{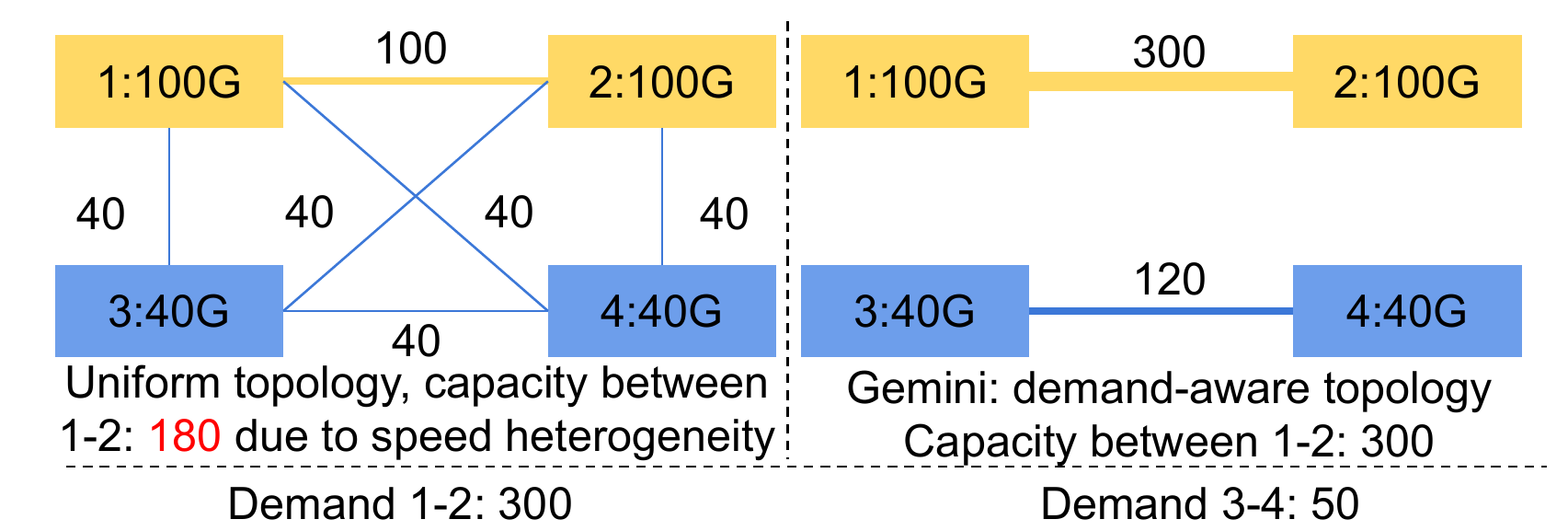}
\caption{\small Due to pod speed heterogeneity (40G/100G), the uniform topology on the left can not support the demand between pod 1 and pod 2. Since \sysname is demand-aware, it can find the feasible topology as shown in the right figure.}
\label{fig:mix_line_rate}	 
}
\end{figure}
Modeling this heterogeneity explicitly allows \gemini to avoid connecting ports with different speeds (\eg a 100G port to a 40G port) which wastes capacity. This, together with ToE, allows \gemini to satisfy demands that may be impossible to satisfy with a uniform topology, defined as a topology with same number of links between each pod-pair, but not necessarily the same link speeds. Consider the example of~\figref{fig:mix_line_rate}, in which there is a demand of 300 between pods 1 and 2 (each with 100G ports), and 50 between pods 3 and 4 (each with 40G ports). This demand cannot be satisfied with the uniform topology on the left, but can, with the topology on the right.

\parab{Three-stage optimization} The solver has three optimization stages, each of which optimizes for a particular objective. We use \figref{fig:multistage_optimization} to illustrate these stages.

\parab{Stage 1: Minimize MLU} The first stage generates a (potentially non-uniform) topology and routing that minimizes the MLU across all extreme traffic matrices in the traffic model. Denote by $u^*$ the resulting MLU. 
\eqnref{eq:opt_u_main_texts} ensures that all link utilizations are smaller than $u^*$ across all traffic matrices in $\mathcal{T}$. \eqnref{eq:port_speed} models the consequence of the pod speed heterogeneity: the speed of a link connecting two speed-heterogeneous pods equals to the smaller speed of two end pods. \eqnref{eq:deg_constraint} makes sure that the total number of links originating from pod $i$ is smaller than the pod radix, $R_i$. 

\begin{flalign}
 \text{min} & \quad u  \nonumber \\
 \text{s.t.} & \sum_{i,j \in \set{V}}\sum_{\{p | p\in \set{P}_{i,j}, e\in p\}} f_{i,j,p} \cdot d_{i,j,t} \leq u \cdot C_{e}, \forall e \in \mathcal{E}, \forall t\in \mathcal{T} \label{eq:opt_u_main_texts} \\
 &C_e = n_e \cdot s_e, \forall e=(i,j) \in \mathcal{E}, s_e = min(s_i, s_j) \label{eq:port_speed}\\
 & \sum_{\{e|e \text{ originates from pod } i\}} n_{e} \leq R_i, \hspace{10mm}\forall i \ \in \set{V},  \label{eq:deg_constraint}\\
 &\text{Flow constraints for $f_{i,j,p}$ and } n_e>=0 \label{eq:other}
\end{flalign}

In our example, as shown in the leftmost figure in~\figref{fig:multistage_optimization}, the first stage produces a highly skewed topology based on the skewed TM and shortest-path routing. However, this solution can create high risk on thin trunks (\eg 1-2 and 3-4); given a traffic burst of 40 units, the utilization on those trunks could increase up to 33\%.

\parab{Stage 2: Enable hedging} This stage addresses risk hedging, formalizing the intuition described in \secref{sec:design:hedging}. It achieves hedging by minimizing the maximum risk (\secref{sec:design:hedging}) over all trunks. \eqnref{eq:opt_u} ensures that all trunk utilization is smaller than $u^*$ obtained in stage 1 and \eqnref{eq:risk_all} ensures that the maximum risk for all pod-pairs is less than $r$. 
\begin{flalign}
 \text{min} & \quad r  \nonumber \\
 \text{s.t.} & \sum_{i,j \in \set{V}}\sum_{\{p | p\in \set{P}_{i,j}, e\in p\}} f_{i,j,p} \cdot d_{i,j,t} \leq u^* \cdot  C_{e}, \forall e \in \mathcal{E}, \forall t\in \mathcal{T} \label{eq:opt_u} \\
 &f_{i,j,p} \cdot \delta \leq r \cdot C_{e}, \forall i,j \in \set{V}, \forall p \in \set{P}_{i,j}, \forall e\in p \label{eq:risk_all}\\
 &\text{Equations (\ref{eq:port_speed})-(\ref{eq:other})} \nonumber
\end{flalign}
In our example (middle figure in~\figref{fig:multistage_optimization}), hedging reduces the MLU on trunk 1-2 and 3-4 from 0.75 to 0.62 by jointly adjusting the topology and routing. However, because it uses many two-hop paths to hedge, it results in a larger path stretch and ALU compared to the first stage solution.

\parab{Stage 3: Minimize path stretch} 
This stage re-arranges the computed routing and topology solution to minimize \textit{path stretch} while maintaining $u^*$ and $r^*$ from previous stages. The path stretch is defined as the total network load over all links divided by the total demand, where the total load equals to the summation of loads on all links. Since the total demand is a constant, we only keep the total load as the objective in the formulation. \eqnref{eq:opt_u} and \eqnref{eq:risk_all} ensure that MLU and the maximum risk for all pod-pairs are less than $u^*$ and $r^*$ respectively. Note that since $u^*$ and $r^*$ are constants, the formulation is a linear program.
\begin{flalign}
 \text{min} & \quad  \sum_{t\in \mathcal{T}} \sum_{e \in \mathcal{E}} \sum_{i,j \in \set{V} } \sum_{\{p | p\in \set{P}_{i,j}, e\in p\}} f_{i,j,p} \cdot d_{i,j,t} \nonumber \\
 \text{s.t.} & \sum_{i,j \in \set{V}}\sum_{\{p | p\in \set{P}_{i,j}, e\in p\}} f_{i,j,p} \cdot d_{i,j,t} \leq u^* \cdot C_{e}, \forall e \in \mathcal{E}, \forall t\in \mathcal{T} \label{eq:opt_u} \\
 &f_{i,j,p} \cdot \delta \leq r^* \cdot C_{e}, \forall i,j \in \set{V}, \forall p \in \set{P}_{i,j}, \forall e\in p \label{eq:risk_all}\\
 &\text{Equations (\ref{eq:port_speed})-(\ref{eq:other})} \nonumber
\end{flalign}

The rightmost figure in~\figref{fig:multistage_optimization} illustrates the topology and routing solution after stage 3, which enables more traffic to go through shortest paths without increasing $u^*$ and $r*$.

\parab{Scaling the solver}
The formulations in stage 1 and stage 2 are both non-linear: MLU and risk are multiplied by the trunk capacity (\eg \eqnref{eq:opt_u_main_texts} and \eqnref{eq:risk_all}). To make the solver tractable, we conduct a binary search over a range of values for MLU and risk. For a fixed value of MLU/risk, the stages result in linear programs that can be solved efficiently. Each step bounds the objective function above or below; we stop when the gap between the two bounds is below a threshold. 

We reduce the search space for the binary searches via several bounds. For MLU, the lower bound is the maximum ratio of aggregate demands on a pod over the pod's block capacity; the upper bound can be approximated by a simple algorithm, such as VLB over a uniform topology. The lower bound for risk is achieved when $\delta$ is equally split on all possible paths which have the same capacity.

\subsection{Predictor and Controller operation}
\label{sec:design:predictor}

Hedging can reduce MLU and allow \gemini to be more robust to misprediction,
which is beneficial for fabrics with unpredictable traffic (\ie those with few well-bounded pairs).
However, it can force traffic over longer paths, increasing ALU as in~\figref{fig:multistage_optimization}.
\gemini uses the predictor to derive configurations with and without hedging, then picks the better configuration (one with lower MLU, or, if the MLUs are comparable, the one with lower ALU). Thus, for a fabric with largely predictable traffic, \gemini avoids using hedging because it increases ALU. However, for one with unpredictable traffic, it selects a hedging-based configuration because that has lower MLU.

To select a strategy, the Predictor needs a goal that %
depends on the network operator's objectives.  Our operators prioritize
the strategy that reduces the p99.9 MLU to within 5\% of the best strategy, and breaks ties
via the p99.9 ALU.\footnote{In this paper, we define the p99.9 MLU with respect to a period, such
as 1 month, by selecting the most-utilized link in each five-minute measurement interval, and
reporting the 99.9th percentile of these per-interval MLUs.  The p99.9 ALU is the 99.9th percentile
over the ALUs for each interval.}

As discussed earlier, both Predictor and Controller use the traffic model and the solver, but in slightly different ways.
The Predictor takes a training window's worth of data (1 month), and runs the solver \textit{offline} for four strategies combining two binary choices:
\emph{Uniform} vs. \emph{Non-uniform} pod-to-pod topology, and 
\emph{Hedging} vs. \emph{no-hedging} for risk minimization.
(The Predictor can disable topology reconfiguration by fixing
$n_e$ in the joint formulation to the number of trunks in the uniform topology, rather than setting it as an optimization variable.) The Predictor simulates each strategy, over the training window, to estimate link utilizations.

The Controller reconfigures the fabric \textit{online} using the predicted strategy.  If the strategy includes topology reconfiguration,
then every topology reconfiguration interval, it computes a new topology
that network operators can use to implement a reconfiguration.
Note that
the Controller \emph{always} uses periodic routing reconfiguration;
at each routing reconfiguration interval, it computes a routing solution and invokes the fabric SDN controller to update the switches.

\subsection{Practical Considerations}
\label{sec:design:others}

\parab{Physical realization with Patch Panels}
To realize a physical topology with patch panels as shown in~\figref{fig:physical-gemini}, we need to address two issues: (a) \gemini's solver might output fractional trunks; \sysname must round  fractional trunks to integers, while maintaining the same pod radixes
and (b) any suggested topology reconfiguration should \textit{not} require moving fibers between patch panels. In other words, any logical topology should be realized by optical paths through fixed-radix patch panels. To round the fractional solution, we find a rounding algorithm \algoref{al:rounding} (see \secref{sec:appendix:physical-gemini}) which has the following property:
\begin{theorem}\label{th:rounding}
Given a graph $G(V,E)$ that has even node degrees, arbitrary edge weights $n_e$, $\forall e \in E$, and no self-loops, \algoref{al:rounding} can construct a graph $G'(V,E)$ with no self-loops in $O(|V|^2)$ time, with the same node degrees of $G$ and with $n'_e \in \{\lfloor n_e \rfloor, \lfloor n_e \rfloor + 1\}$.
\end{theorem}
To avoid moving fibers between patch panels, we use this theorem to distribute links between pods and patch panels:
\begin{theorem}\label{th:implement}
If the radix of every pod is $2^k$, any topology that has integer numbers of inter-pod trunks can be constructed using $2^p$ patch panels ($p < k$), by connecting $2^{k - p}$ ports of every pod to every patch panel.
\end{theorem}
\secref{sec:rounding} discusses \algoref{al:rounding} and proofs of both theorems.

\parab{Handling major workload changes}
Predictions based on historical traffic might not handle large demand increases (\eg adding a new large-scale service to a datacenter).  We address this primarily via an approval process for admitting large-scale workloads, together with a continual process for adding new capacity (pods and DCNI resources) on a live network~\cite{condor,rewiring,fatclique}. \sysname's topology-reconfiguration strategy can integrate exogenous predictions of future demand to inform our regular capacity-augmentation processes.

\parab{Wiring complexity} Modern datacenters use fiber bundling to reduce wiring complexity~\cite{jupiter,fatclique}, using a layer of patch panels between the pod and spine layers~\cite{rewiring}. A spine-free topology's restriping layer permits fiber bundling in much the same way, even in the presence of pod heterogeneity (the proof of~\thmref{sec:appendix:physical-gemini} describes one approach to this).

\parab{Topology restriping/expansion} Restriping or expanding non-blocking Clos topologies requires rewiring fibers. Algorithms for these operations~\cite{rewiring} generate multi-step rewiring plans, to ensure that at each step fabric capacity does not drop below an operator-specified threshold. Spine-free topologies can use these algorithms,
with minimal modifications.

\vspace{-0.1in}
\section{Evaluation}
\label{sec:evaluation}

In this section, we 
validate \gemini on a testbed (\secref{sec:testbed}), and more
extensively simulate \gemini using traces from 22 production fabrics (\secref{sec:simulation}).
\textbf{In simulations, \sysname consistently improves link-utilization metrics (\secref{sec:metrics}) over various baselines.}

\subsection{Testbed Evaluation}\label{sec:testbed}
We used a testbed, exposed to a production workload (including
storage, search, computation, and video serving), to validate our simulator,
and to test that \sysname works reliably.   This gave us some ability to
compare \sysname to a baseline, but operational constraints limited what experiments we could run, and these were
not randomized controlled trials.

The testbed has 12 pods, each with 256 100G ports, using a non-blocking Clos design similar to Jupiter~\cite{jupiter}. Pods are connected by a DCNI with multiple patch panels as in~\figref{fig:physical-gemini}. 

\parab{Reconfiguration} The \gemini system
interfaces with the fabric's centralized control plane, which performs routing state (re)-configuration and collects traffic matrices. The fabric uses WCMP~\cite{wcmp} to route traffic within pods. In this implementation, \gemini reconfigures routing once every 8 hours, and reconfigures topology in multiple steps to reduce the capacity degradation at each step, using an algorithm similar to~\cite{rewiring} to generate the rewiring plan at each step.

\parab{Configurations}
We initially configured the testbed with a uniform topology, with hedging enabled, and
collected traffic data for two weeks.  Based on that 
traffic, 
\gemini's Predictor recommended a (non-uniform, no-hedge) configuration. We then reconfigured the topology accordingly, and collected data for another two weeks.
We refer to the initial configuration as \emph{baseline} and the latter as \emph{predicted-best}.
Both configurations use demand-aware routing.

\begin{figure}[htb]
{
\centering		
\vspace{-2ex}
\includegraphics[width=0.8\columnwidth]{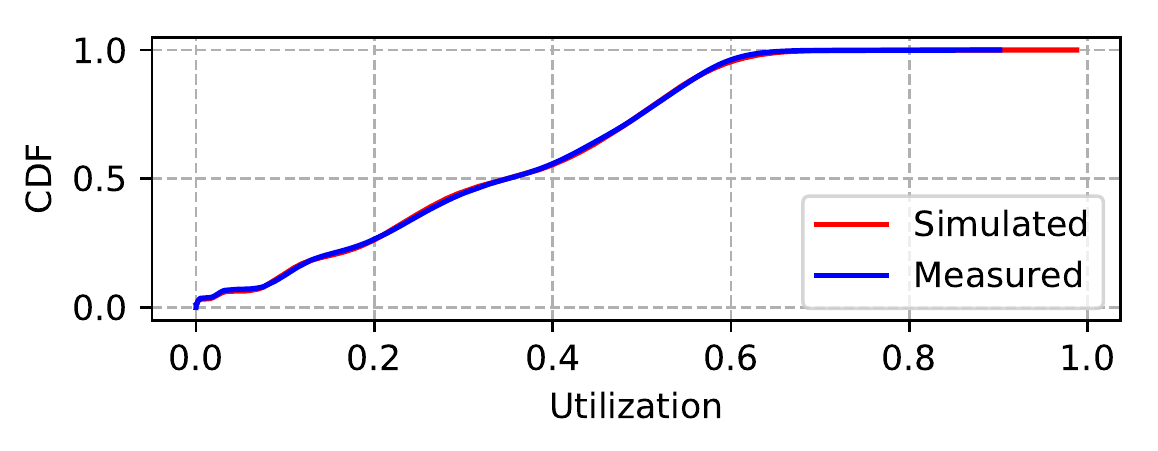}
\begin{spacing}{0.1}
{\footnotesize p99.9 utilization: simulated $=$ 0.716 vs. measured $=$ 0.701 ($2\%$ error);\\
p99.9 MLU: simulated $=$ 0.836 vs. measured $=$ 0.843 ($<1\%$ error)
}
\end{spacing}
\setlength{\belowcaptionskip}{-8pt}
\caption{\small Simulated \vs measured testbed utilization}
\label{fig:sim_accuracy} 	
}
\end{figure}

\parab{Simulator validation}
Using the testbed 
%of \secref{sec:testbed} 
we compared our simulator's predicted link utilizations against
measurements, for a (different) 27-day period, with the testbed
restored to the ``baseline'' configuration: (Uniform, hedging).
\figref{fig:sim_accuracy} shows that, at least in this configuration, the simulator agrees closely with
ground truth.
We have no reason to expect the simulator to be less accurate in other configurations.
(Note also that the Predictor depends on the accuracy of the simulator, as shown in \figref{fig:gemini_arch}.)

\begin{figure}[htb]
{
\centering	
\vspace{-2ex}
\includegraphics[width=1\columnwidth]{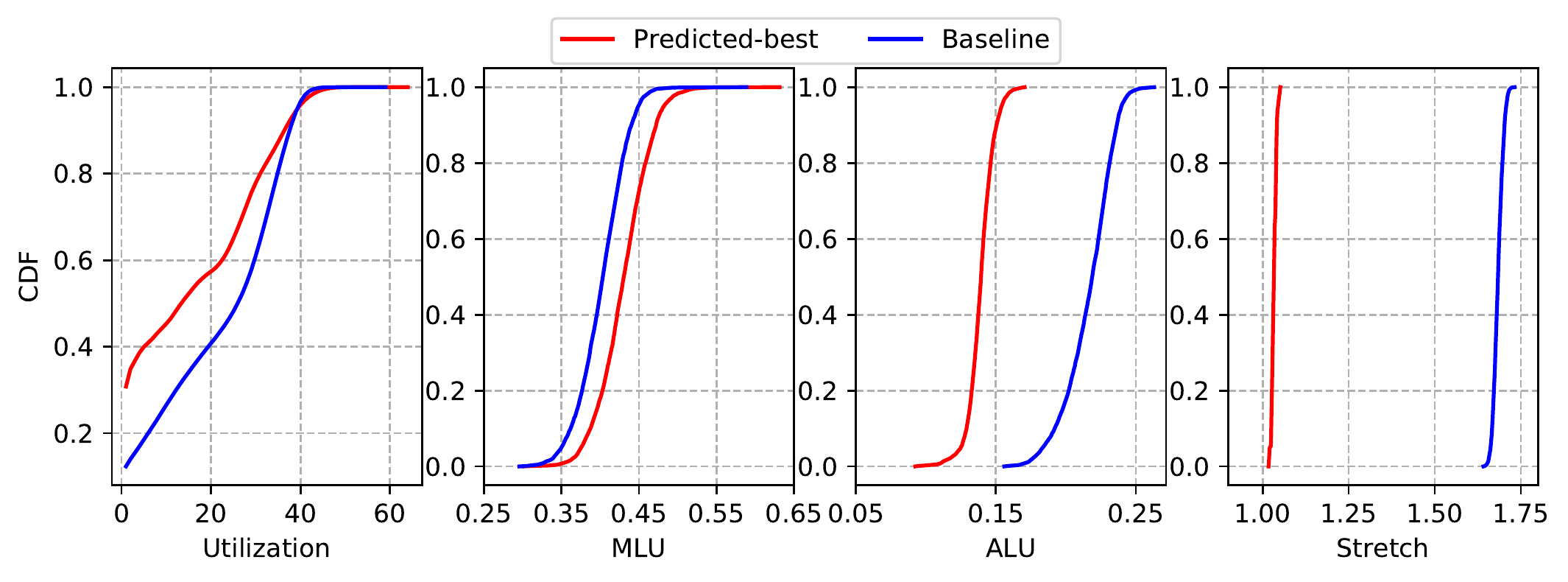}
\setlength{\abovecaptionskip}{-10pt}
\setlength{\belowcaptionskip}{-8pt}
\caption{\small Comparing \emph{baseline} vs. \emph{predicted-best} configs}
\label{fig:canary_before_after}
}
\end{figure}

\parab{Utilization metrics}
Due to a data-retention mistake, we have no direct measurements of link utilizations during these testbed
trials.   We did retain directly-measured traffic matrix traces, so we reconstructed link utilizations from these traces using the simulator.

\parab{Results} 
\emph{Because we used production traffic, the two configurations were under somewhat different
workloads (predicted-best had 17\% higher traffic, and a DMR of 5.49 \vs 1.67), 
so these results are ``suggestive'' of improvements, but not proofs.}

\figref{fig:canary_before_after} shows CDFs of (simulated) link metrics comparing
the \emph{baseline} and \emph{predicted-best} configurations, showing apparent reductions in 
utilization, ALU, and stretch, when using \emph{predicted-best}, but increases in MLU.
Some of the shift might be caused by the use of different workloads, but we lack data to
separate that from the topology effect.   However, the large improvement
in stretch is consistent with the use of a non-uniform topology that avoids most transit routing.

\secref{sec:testbed_fcts} provides some additional results from the testbed.

\vspace{-0.1in}
\subsection{Large-scale Simulation}
\label{sec:simulation}

To better understand the performance of \gemini at scale, we present results from a trace-driven evaluation, using data from production fabrics. To do this, we feed historical traffic matrices from these fabrics to a simulator
using a Map-Reduce implementation of \gemini's algorithms (\secref{sec:design}).

\parab{Datasets} 
We use six months' worth of 5-minute traffic matrices from 22 fabrics in our whole fleet.\footnote{For brevity, we often use the term ``fabric'' to refer to the combination of a workload trace and the specific fabric where it was obtained.}
These fabrics span a range of topology sizes, utilization levels and traffic characteristics; some of the fabrics mix several line rates and/or pod radices in their DCNI.\footnote{Absolute numbers for fabric size, throughput, latency, or overall loss rates are proprietary, so we cannot publish these.} The choice of a 5-minute window for traffic matrices can obscure traffic dynamics within the window, but this window size has long been an industry-standard collection interval~\cite{RFCSNMP, vl2, fbtraffic,microburst}.

\vspace{-0.05in}
\parab{\gemini Prediction Methodology}
For a given fabric, \gemini's Predictor chooses the best of  four possible demand-aware strategies listed in \secref{sec:design:predictor}. For each fabric, we used a 1-month training window.\footnote{We chose a 1-month training window partly for convenience; a sensitivity study for this parameter is future work.}

\vspace{-0.05in}
\parab{Baselines}
We compare \gemini against three demand-oblivious baselines.
The first, \textit{(Uniform, VLB)} connects pods directly using a uniform topology with Valiant-loading-balancing based (VLB) routing. 
VLB splits traffic equally across all N one-hop and two-hop paths between ingress and egress pods~\cite{vlb_heterogeneous}. The second, \textit{Same-cost Clos}, is a
2:1 oversubscribed-Clos DCNI running ECMP. 
\textit{(Uniform, VLB)} and \textit{Same-cost Clos} have the same total DCNI cost (including switches, cables, and optical transceivers) as the one that \gemini uses. 
The third, \textit{Full Clos}, is a Clos topology with a spine layer, with twice \gemini's total DCNI cost. 

\vspace{-0.05in}
\parab{Parameters} Our baseline simulations use traffic models built with 12 critical traffic matrices, an aggregation window of a week, a topology reconfiguration rate of 1/day, and a routing reconfiguration period of 15 min.  In \secref{sec:eval:sensitivity}, 
we evaluate \gemini's sensitivity to these parameters.

\vspace{-0.05in}
\parab{Success metrics}
We compare configurations and strategies on several metrics: MLU, ALU, and Overloaded Link
Ratio (OLR) as discussed in \secref{sec:metrics}.  An ideal strategy would improve all three of these vs. the other options.

\subsubsection{\textbf{Benefits of Demand-Awareness.}} \label{sec:simulation:aware_oblivious}

\begin{figure}[htb]
    \vspace{-0.1in}
    \centering
    \includegraphics[width=1\columnwidth]{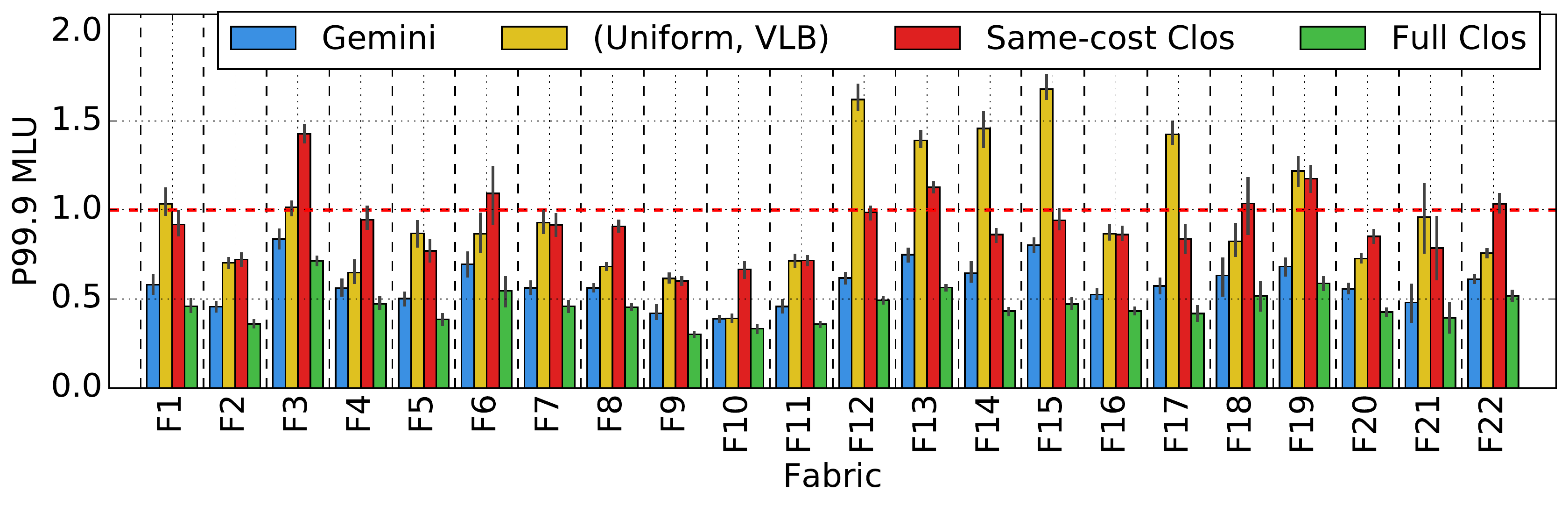}
    \begin{spacing}{0.2}
    {\footnotesize Bars above MLU$=$1.0 represent demands that cannot be feasibly routed}	
    \end{spacing}
\setlength{\belowcaptionskip}{-12pt}
    \caption{P99.9 MLU impact of demand awareness}
    \label{fig:MLU-aware}
\end{figure}
\begin{figure}[htb]
    \centering
    \includegraphics[width=1\columnwidth]{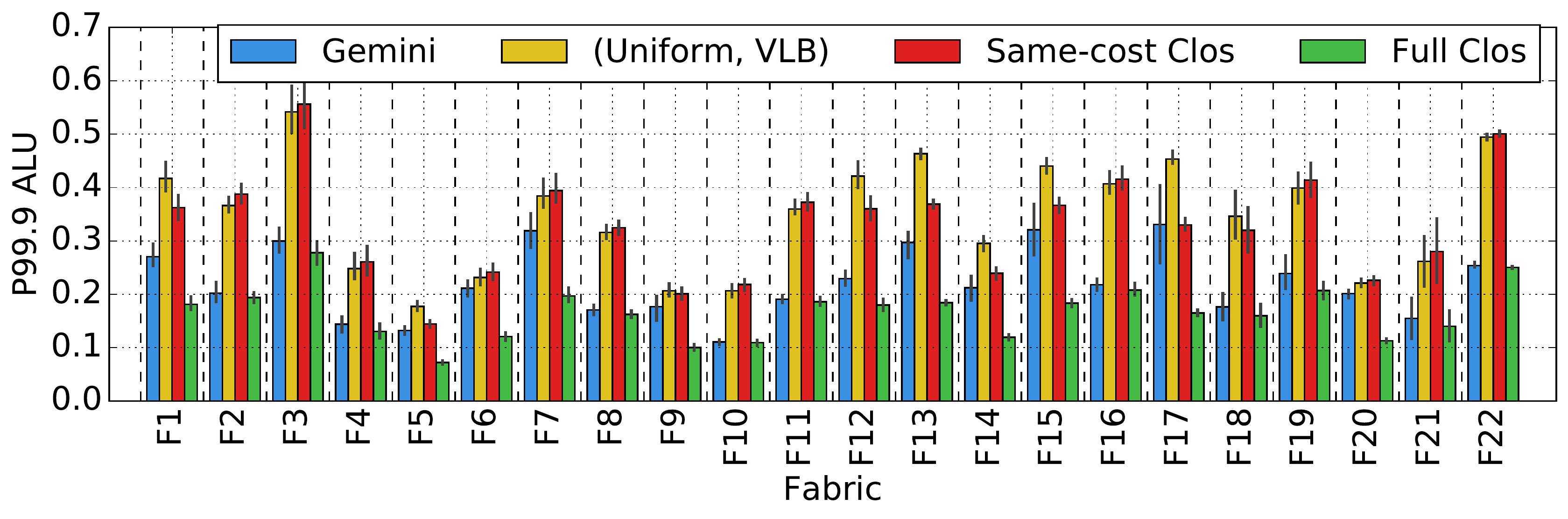}
\setlength{\abovecaptionskip}{-10pt}
\setlength{\belowcaptionskip}{-12pt}
    \caption{P99.9 ALU impact of demand awareness}
    \label{fig:ALU-aware}
\end{figure}
\begin{figure}[htb]
    \centering
    \includegraphics[width=1\columnwidth]{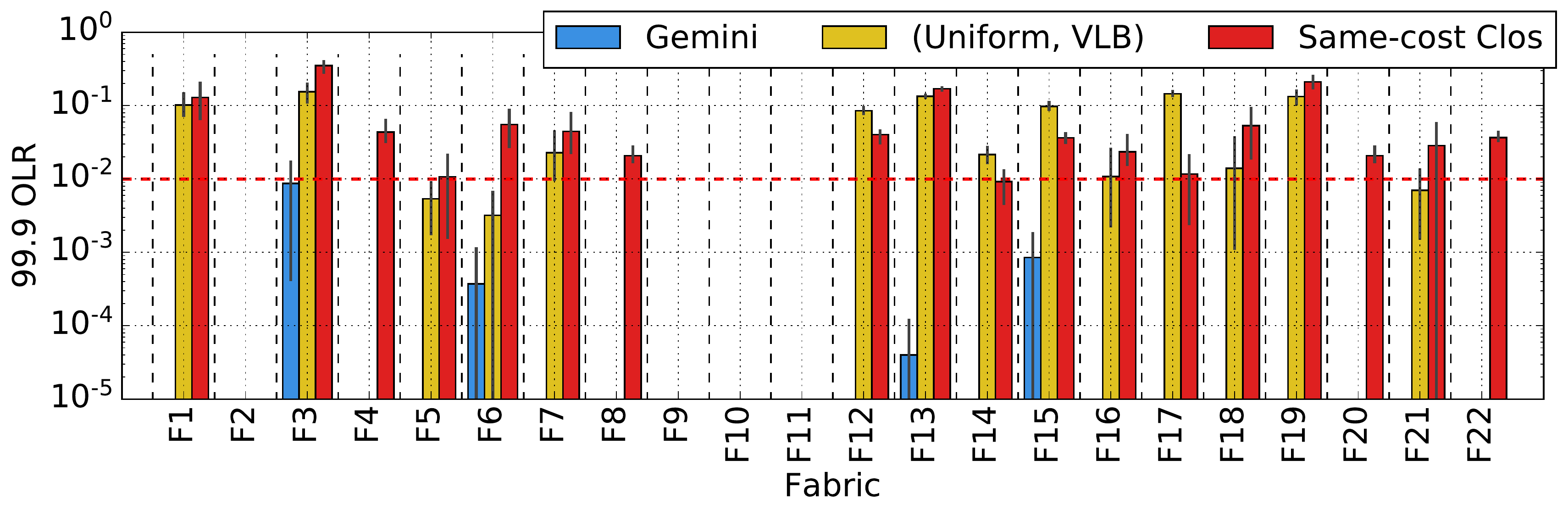}
    \setlength{\abovecaptionskip}{-10pt}
    \setlength{\belowcaptionskip}{-8pt}
    \caption{P99.9 OLR impact of demand awareness}
    \label{fig:OLR-aware}
\end{figure}

Our simulations demonstrate that \gemini's demand-aware approach is superior to demand-oblivious approaches
on MLU, ALU, and OLR. 
Our evaluation here compares \sysname to other designs using p99.9 values for MLU (\figref{fig:MLU-aware}),
ALU (\figref{fig:ALU-aware}), and OLR (\figref{fig:OLR-aware}).
These figures aggregate results (with error bars) over all 5 months.

\vspace{-0.05in}
\parae{\sysname has lower p99.9 MLU than cost-competitive approaches}
Across all fabrics, \gemini's p99.9 MLU is always comparable to, or lower than, (Uniform, VLB) and Same-cost Clos.\footnote{Our use of fabrics with mixed line rates and radices explains why
same-cost Clos sometimes outperforms VLB, which can suffer from hot spots in such cases.}
The average difference between these two alternatives and \gemini is 42\% and 30\%, respectively. For about half of the fabrics, the demand-oblivious approaches do not result in feasible routing: the total demand on at least one link exceeds capacity (this is shown as an MLU greater than 1). \gemini, on the other hand, is able to accommodate traffic demands on every fabric.

Same-cost Clos performs worse because the oversubscription is uniform at all pods, so ``hot'' pods which generate high demand can see high utilization. For such pods, \gemini would provision additional trunk capacity. (Uniform, VLB) has high MLU because all pods need to act as transit for all other pods; this can increase MLU for hot pods significantly.

Compared to \gemini, Full Clos has smaller p99.9 MLU across all the fabrics. This is because Full Clos has twice the number of switches and links, can perfectly load balance traffic at each pod, and no pods carry transit traffic for other pods. Interestingly, for 17 of 22 fabrics, \gemini's p99.9 MLU is at most 30\% higher than Full Clos. 
For these fabrics, traffic is predictable, so \sysname's configuration is able to achieve low p99.9 MLU because it is well matched to the traffic demand. 

\vspace{-0.05in}
\parae{\gemini's p99.9 ALU is comparable to, or lower than, competing approaches} Across all fabrics, \gemini's p99.9 ALU is always comparable to, or lower than, (Uniform, VLB) and Same-cost Clos for the same reasons discussed before. For highly-variable
fabrics where \gemini decides to use hedging, which spreads traffic across longer transit paths,
the p99.9 ALU is twice that of Full Clos.
For the other fabrics, \gemini's p99.9 ALU is similar to Full Clos.
\emph{Thus, \gemini appears to provide a better tradeoff between DCNI cost and ALU than any
of the demand-oblivious options.}

\vspace{-0.05in}
\parae{\gemini's OLR is better than any same-cost option}
\figref{fig:OLR-aware} shows that \sysname's OLR is always significantly better than
(Uniform, VLB) and Same-cost Clos.  The figure uses a log scale, and so omits bars where
OLR$=$0 (no links exceeded 0.8 utilization), which is always the case for Full Clos and
for \sysname on many fabrics.  \sysname's OLR is always below 1\%, suggesting that FCTs
would remain low.

\begin{figure}[htb]
    \centering
    \vspace{-2ex}
    \includegraphics[width=1\columnwidth]{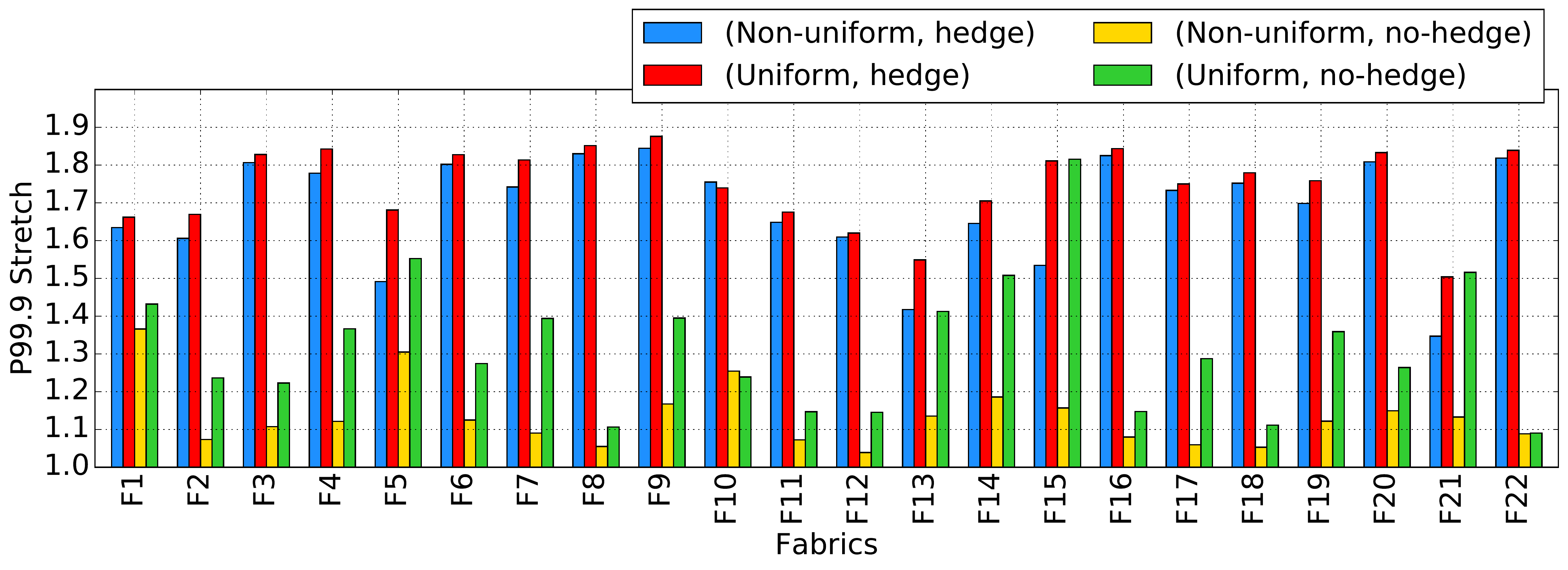}
    \setlength{\abovecaptionskip}{-10pt}
    \setlength{\belowcaptionskip}{-14pt}
    \caption{P99.9 stretch
    for one arbitrarily-chosen month (“M3”) in our study; other months are qualitatively similar.}
    \label{fig:Stretch-toe}
\end{figure}

\parae{\sysname's stretch is typically low}
Clos networks (with spines) and VLB networks always have an extra hop, compared to the
best-case \sysname configurations.  While \sysname sometimes does add a second hop, for
transit routing, \figref{fig:Stretch-toe} shows that the p99.9 stretch remains below 2, and when hedging
is not necessary, the p99.9 stretch is usually close to 1.0 (ideal) because \sysname can
exploit non-uniform topology.

\vspace{-0.05in}
\subsubsection{\textbf{Prediction quality}}
\begin{figure}[htb]
    \centering
    \vspace{-3ex}
    \includegraphics[width=1\columnwidth]{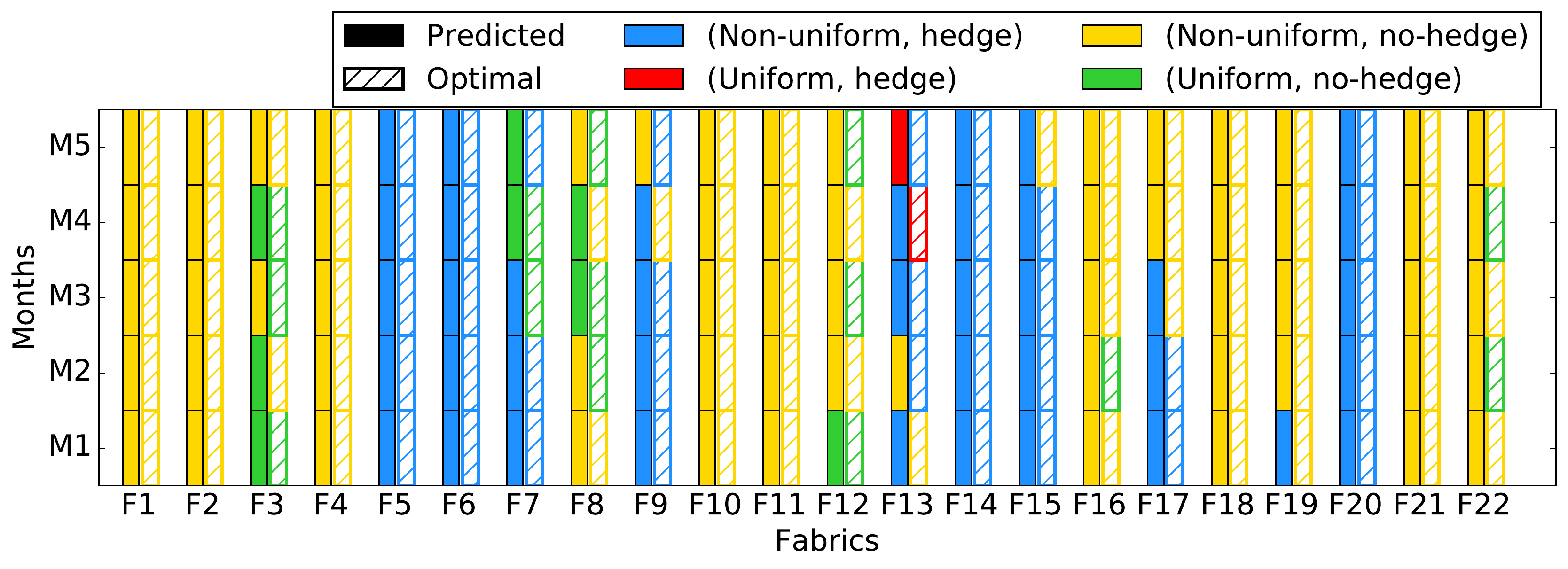}
    \setlength{\abovecaptionskip}{-10pt}
    \setlength{\belowcaptionskip}{-8pt}
    \caption{\small Gemini's predicted strategy \vs optimal strategy}
    \label{fig:algo-comp}
\end{figure}
Our simulations demonstrate that \sysname usually makes correct predictions, that when these predictions are correct
they are beneficial, and when they are wrong, the cost is relatively low.  (Here, we use data from all five
months of predictions.)

\vspace{-0.05in}
\parae{\gemini usually recommends optimal strategies}
In \figref{fig:algo-comp}, solid bars represent \gemini's recommendations and hashed bars
show the optimal (in hindsight) strategy, for each fabric and each month; matching colors
mean a correct prediction.  \sysname overall correctly predicts 81\% of the choices here,
and (for these 5 months) predicts perfectly for 11 fabrics, and near-perfectly for 4 others.
\secref{subsec:prediction} discuss the benefits of good predictions and the costs of bad ones.

\vspace{-0.05in}
\parae{Correct predictions are beneficial}
\label{subsec:prediction}
Earlier we showed that \sysname typical predicts the optimal strategy; here we show
the benefits of correct predictions.

\begin{figure}[htb]
    \centering
    \vspace{-2ex}
    \includegraphics[width=1\columnwidth]{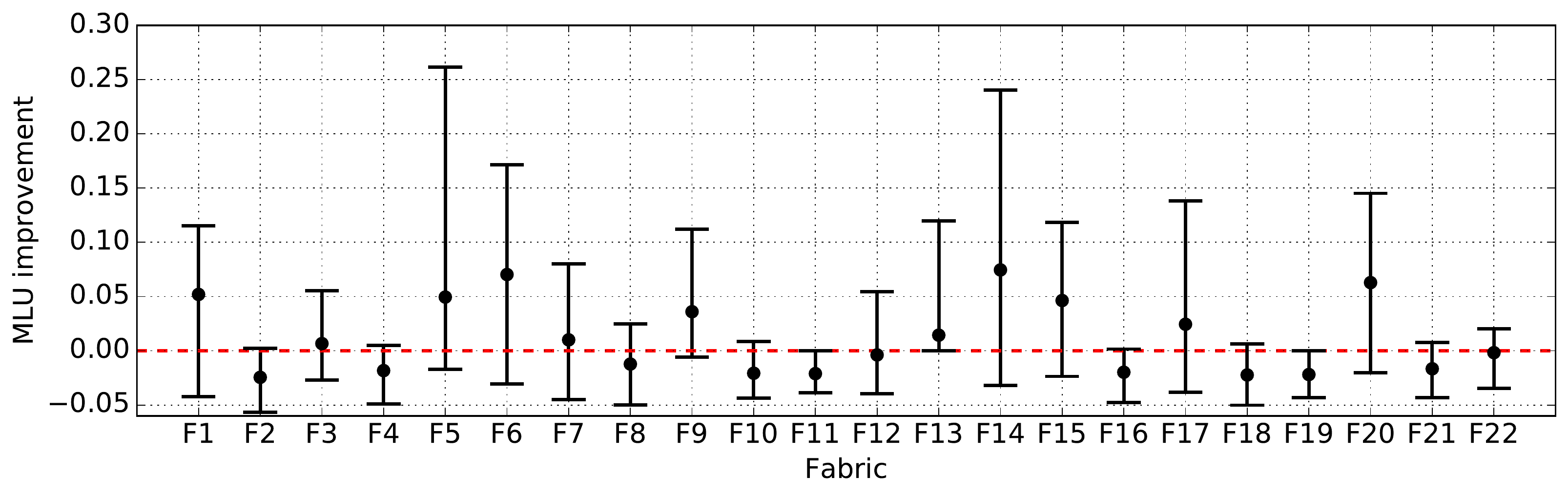}
    \begin{spacing}{0.4}
    {\footnotesize MLU benefit from correct predictions (higher is better)}	
    \end{spacing}
    \vspace{1.5ex}
    \includegraphics[width=1\columnwidth]{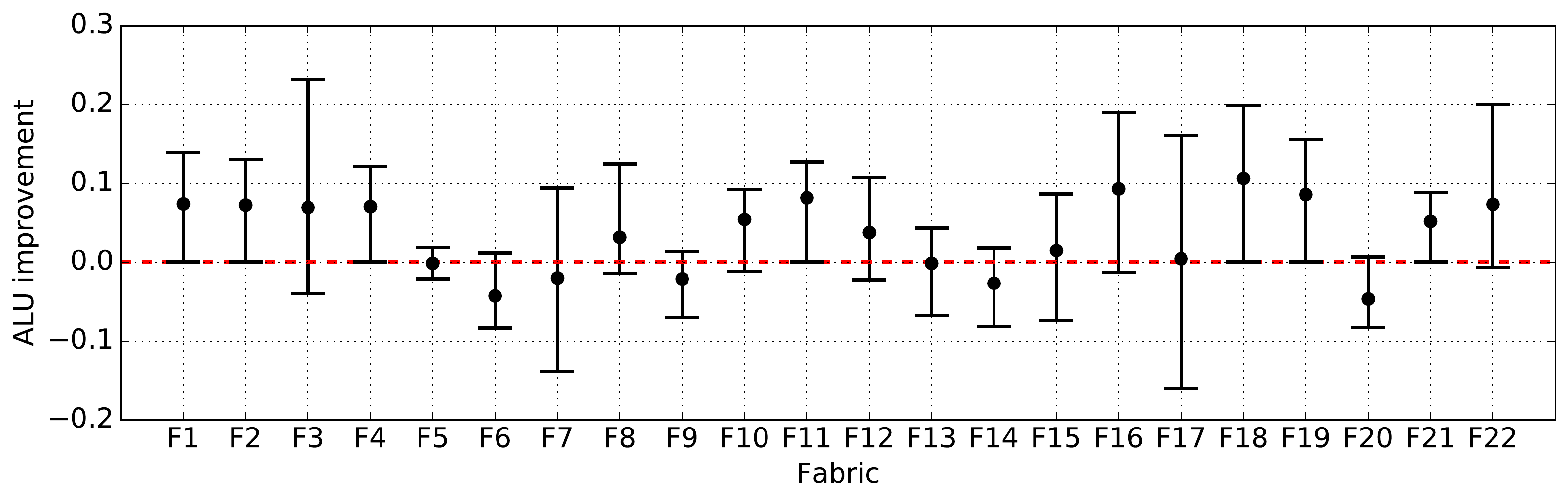}
    \begin{spacing}{0.4}
    {\footnotesize ALU benefit from correct predictions (higher is better)}	
    \end{spacing}
    \setlength{\belowcaptionskip}{-8pt}
    \caption{Benefits from correct predictions}
    \label{fig:correct-benefits}
\end{figure}

\figref{fig:correct-benefits} 
shows the MLU and ALU improvements from correct predictions. Each vertical bar plots, for one fabric, the range,
across 5 months, of improvements for MLU (respectively ALU) compared to every other strategy.
In most cases, both MLU and ALU benefit from correct predictions; when MLU is worsened, it is still within
the 5\% cushion applied by the Predictor (\secref{sec:design:predictor}), when it trades off MLU for improved
ALU (e.g., fabrics F5, F6, F7, F9, F13, F14, F15, F17, F20).

\begin{figure}[htb]
    \centering
    \vspace{-2ex}
    \includegraphics[width=1\columnwidth]{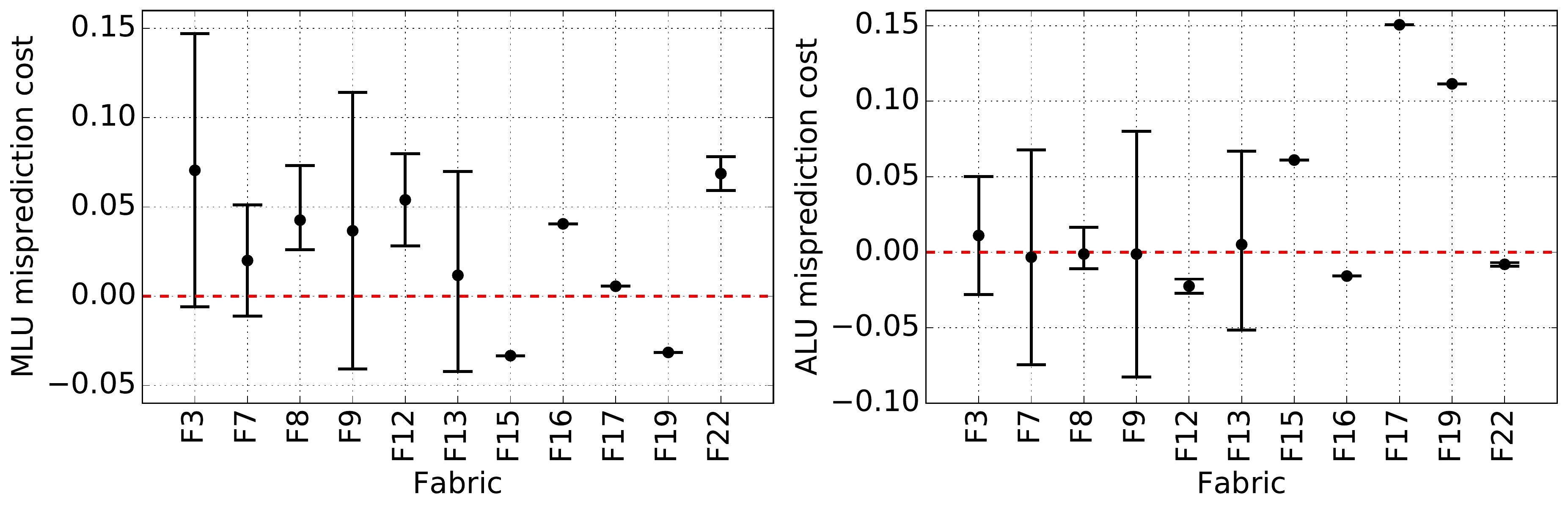}
    \begin{spacing}{0.4}
    {\footnotesize p99.9 values; higher is worse}	
    \end{spacing}
    \setlength{\belowcaptionskip}{-8pt}
    \caption{\small Misprediction cost, MLU (left) and ALU (right)}
    \label{fig:mlu-cost}
\end{figure}

\vspace{-0.05in}
\parae{Misprediction costs are small}
For fabrics where \sysname mispredicted at least once,
\figref{fig:mlu-cost} shows the difference from optimal in p99.9 MLU (left) and MLU (right); the whiskers show
the range across the mispredicted months.
The worst-case increases are 15\% for both MLU and ALU, but most MLU increases are within the Predictor's
5\% cushion; most ALU increases are also below 5\%. In many cases, a misprediction still improves
either MLU or ALU over the optimal strategy.

Taken together, these results suggest that, while there is room for improvement in \gemini's prediction accuracy,
even its mispredictions are mostly harmless.

\subsection{Sensitivity Analyses}
\label{sec:eval:sensitivity}

\begin{figure}[t]
    \centering
    \includegraphics[width=1\columnwidth]{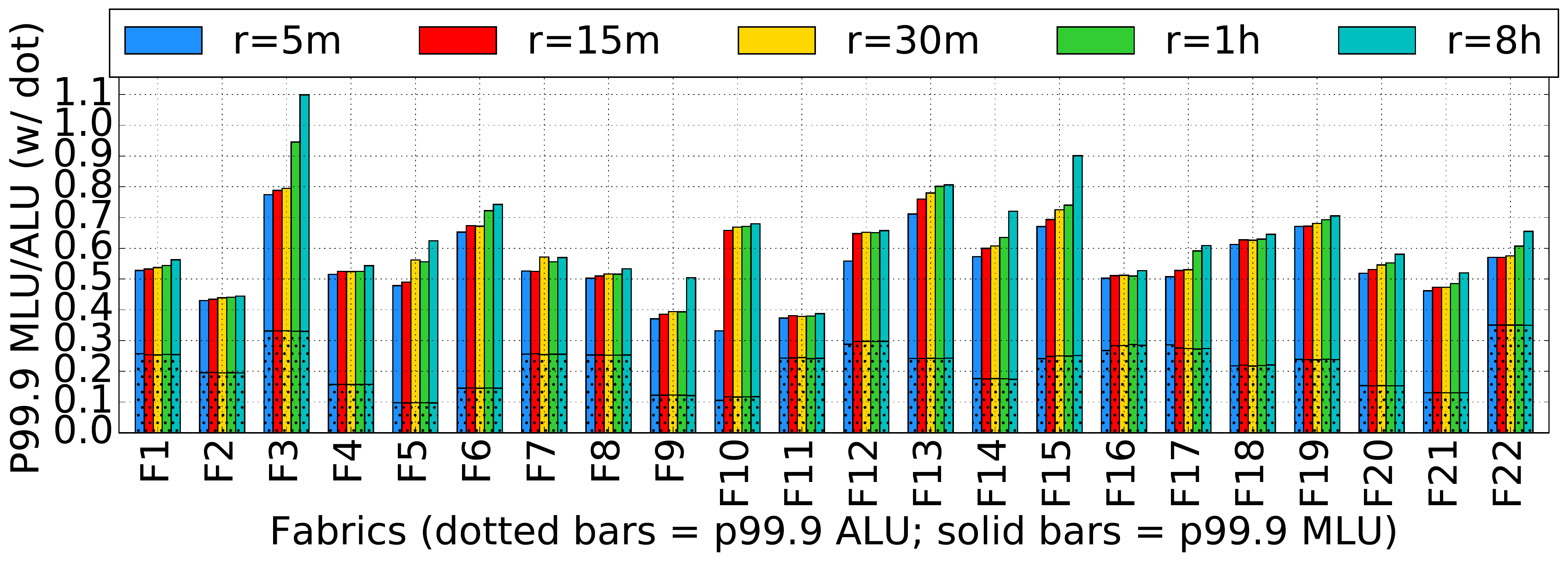}
    \caption{\small MLU/ALU vs. routing reconfiguration interval (r)}
    \label{fig:rreconf}
\end{figure}

\parab{Routing Reconfiguration Frequency}
\figref{fig:rreconf} shows the impact of the routing reconfiguration interval ($r$) on MLU and ALU, for 
$r$ between 5 minutes and 8 hours.  (For this experiment, we fix the topology reconfiguration interval to 1 day. We only show the (Non-uniform, hedge) strategy for month M3; other strategies in all months are qualitatively similar.)
For half of the fabrics, p99.9 MLU decreases as $r$ decreases; the rest are insensitive to $r$. The p99.9 ALU is insensitive to $r$.  $r=$15min is typically sufficient.

\begin{figure}[t]
    \centering
    \vspace{-2ex}
    \includegraphics[width=0.9\columnwidth]{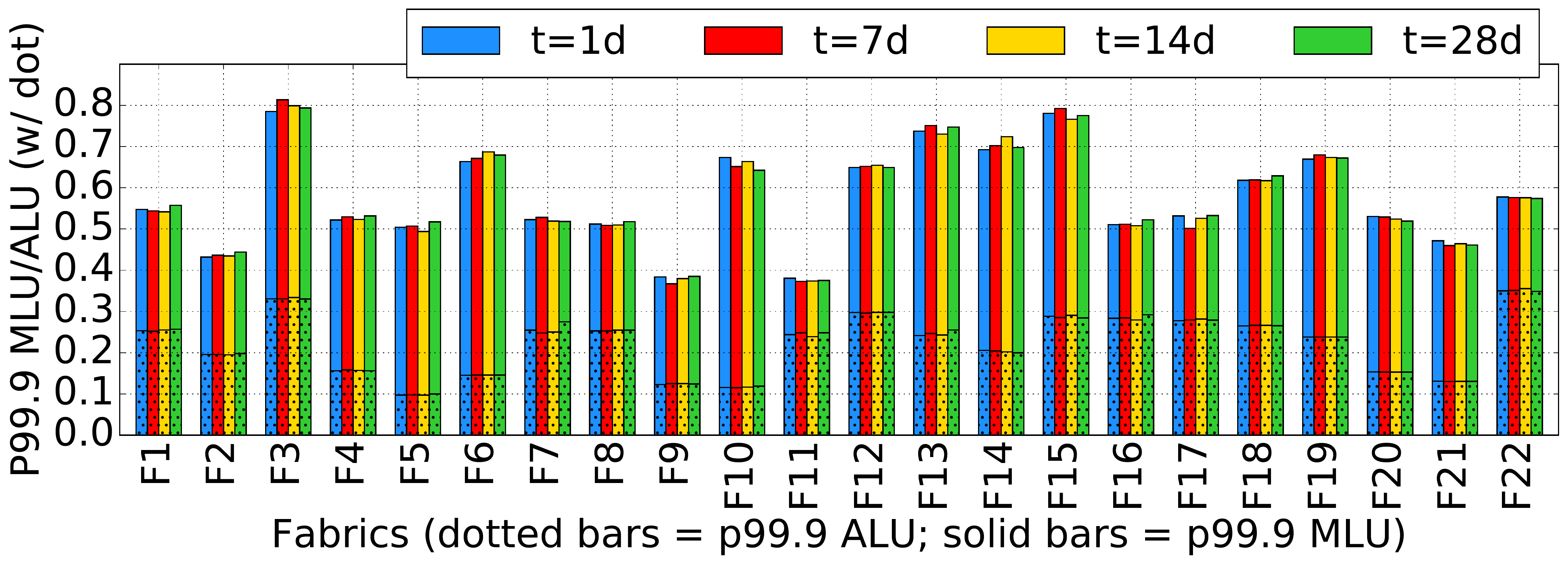}
    \caption{\small MLU/ALU vs. topology reconfig. interval (t)}
    \label{fig:treconf}
\end{figure}

\parab{Topology reconfiguration frequency}
We studied the effects of several topology reconfiguration intervals ($t$) between 1 day and 4 weeks, with $r$ fixed at 15 minutes. Our experimental results show that for all fabrics, both MLU and ALU are independent of $t$, implying that topology reconfiguration can be infrequent (for our fabrics/workloads), and can be implemented with inexpensive patch panels.

\begin{figure}[t]
    \centering
    \includegraphics[width=0.9\columnwidth]{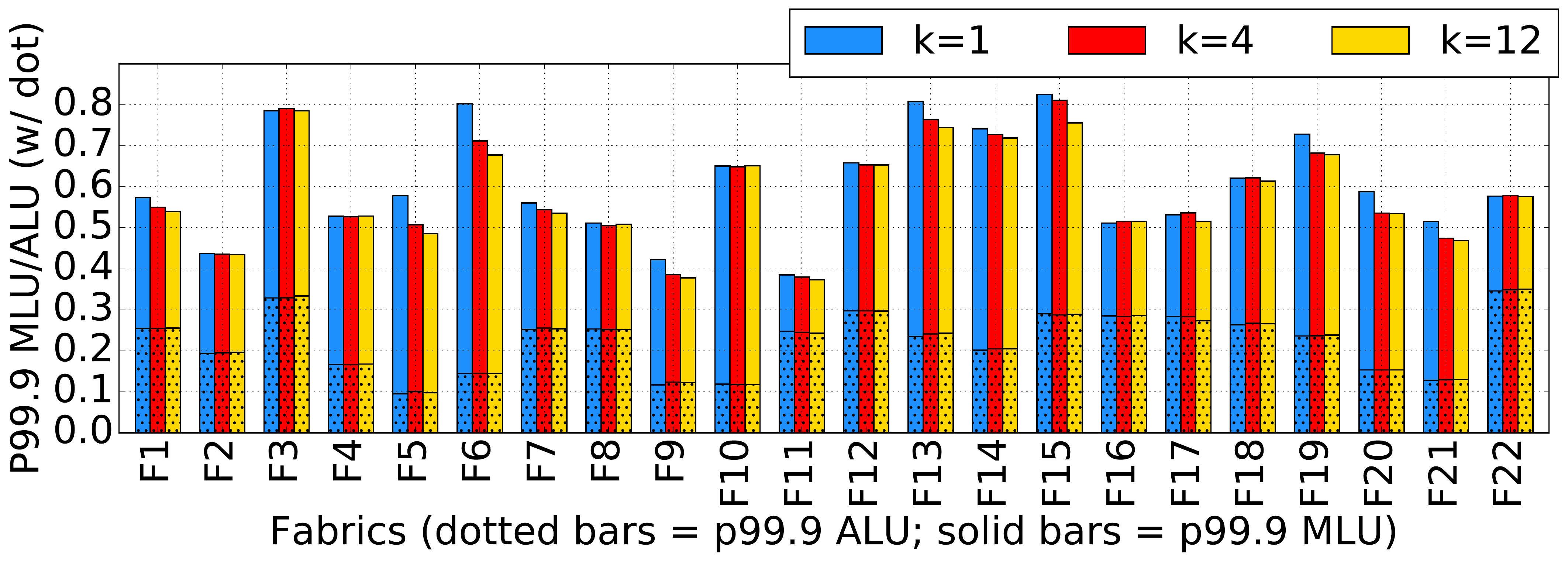}
    \caption{\small MLU/ALU vs. number of matrices (k) in M3}
    \label{fig:MLU-ALU-TM}
\end{figure}

\parab{Impact of multiple critical traffic matrices}
In its traffic model, \gemini clusters traffic matrices (TM), then selects critical TMs. We simulated MLU and ALU against the number of clusters $k$ (or critical TMs), for $k=$ (1, 4 and 12). We observed that as $k$ increases, p99.9 MLU decreases by 5\% in average in half of fabrics without increasing p99.9 ALU, but we observe diminishing returns:
$k=12$ is a sweet spot.

\begin{figure}[t]
    \centering
    \includegraphics[width=0.9\columnwidth]{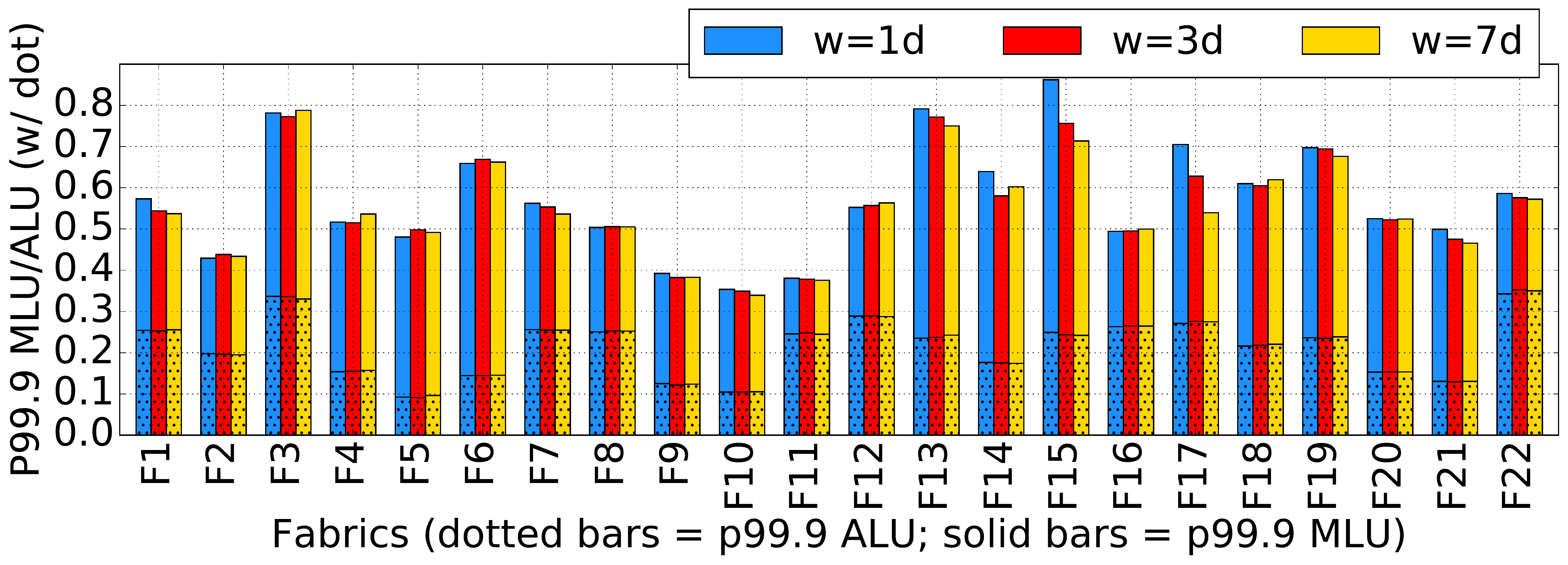}
    \caption{\small MLU/ALU vs. aggregation window (w) in M3}
    \label{fig:MLU-ALU-Window}
\end{figure}

\parab{Impact of traffic aggregation window}
In its traffic model, \gemini clusters traffic matrices within a traffic aggregation window $w$. We simulated MLU and ALU for $w=$ 1 day, 3 days, and 7 days;
we fixed $r=$15min and $t=$1 day.
Increasing $w$ seems to improve MLU in 20\% of the fabrics; it has no apparent affect on ALU.

\vspace{-0.1in}
\section{Related Work}
\label{sec:related}
\textbf{Non-blocking topologies:} Previous datacenter work~\cite{fattree, jupiter, jellyfish, xpander, fatclique} focused on non-blocking topologies. For Clos-based designs like Jupiter~\cite{jupiter}, simple demand-oblivious routing schemes exist. 
Practical routing for other proposed topologies is an open question~\cite{beyondfattree}.

\textbf{Reconfigurable designs with commercial OCS:} Closest to our work is Helios~\cite{helios}, which augmented the spine layer with a reconfigurable OCS, and segregated long flows to be routed via the OCS. Unlike Helios, \sysname leverages %
models from real-traffic observations and makes reconfiguration decisions on inter-pod demand, not individual flows.

\textbf{$\mu$Sec-level reconfigurability:} Much work has focused on fine-timescale reconfiguration~\cite{helios, projecttor, firefly, solstice, mordia, rotornet, opera, sirius}.  In contrast, \sysname relies only on commodity hardware proven to work at large scales.  It also attempts to minimize link utilizations, \textit{vs.}
directly minimizing FCTs, as collecting flow-size information in real time can be hard~\cite{flowsize,opera, elephant-flow-detection}.

\textbf{Robust routing design:} Prior work has focused on robustness to failure~\cite{smore, te-failure-recovery}. More relevant to \sysname is research focusing on robustness to traffic variations;
some~\cite{Applegate2003, Applegate2004} are demand-oblivious and perform poorly relative to demand-aware approaches in practice~\cite{cope}. 
Demand-aware approaches (e.g., \cite{cope, criticalTM}) reconfigure based on multiple traffic matrices, as \sysname does, but focus on routing for fixed wide-area networks. \sysname jointly optimizes topology and routing, and scales to large datacenters with highly variable traffic.

\section{Summary}
\label{sec:conclusion}

Our measurements show that real datacenter traffic is variable, skewed, and diverse across fabrics; some are more bursty than others, but many are boundable, allowing a prediction-based approach.
This led us to support demand-aware routing, to address workload variability; reconfigurable, non-uniform topology, to address workload skew; and hedging to accommodate unpredicted bursts.
\sysname, which jointly optimizes topology and routing based on traffic models,
improves various metrics over several demand-oblivious baselines, and can support realistic workloads using a spine-free topology that is half the cost of traditional FatTree.

\clearpage
\bibliographystyle{ACM-Reference-Format}
\bibliography{topo.bib}

\clearpage
\newpage
\appendix
\section*{Appendices}
\label{sec:appendix}

\section{Physical realization using patch panels}
\label{sec:rounding}

The topology computed by the joint solver may have a fractional number of inter-pod links. To physically realize \sysname (\figref{fig:physical-gemini}), we develop an algorithm to round the fractional links to integers, and to decompose the graph into subgraphs, each of which is constructed by inter-connections of ports within a patch panel.

The following theorem explains how to round the fractional trunks to integers, while maintaining the same node degrees.
\begin{theorem}\label{th:rounding}
Given a graph $G(V,E)$ that has even node degrees, arbitrary edge weights $n_e$, $\forall e \in E$, and no self-loops, a graph can be constructed in $O(|V|^2)$ time, that has the same node degrees, integer edge weights, either $\lfloor n_e \rfloor$ or $\lfloor n_e \rfloor + 1$, $\forall e \in E$, and no self-loops.
\end{theorem}

We also show that by connecting an equal number of links between each pod to each patch panel, any logical topology of inter-pod connections can be realized by only reconfiguring links within patch panels (\ie without moving fibers between patch panels).
\begin{theorem}\label{th:implement}
If the radix of every pod is $2^k$, any topology that has integer numbers of inter-pod trunks can be constructed using $2^p$ patch panels ($p < k$), by connecting $2^{k - p}$ ports of every pod to every patch panel, and pairing ports in the patch panels to construct inter-connections between pods.
\end{theorem}

The proofs of these theorems can be found in Appendix~\secref{sec:appendix:physical-gemini}. The constructive proofs provide a polynomial-time algorithm to construct a DCNI using commercially available patch panels with fixed port-count. Suppose that there are $n$ pods; we can support any inter-pod connections with patch panels whose port count is at least $2n$. 
The result can be generalized to the case where the radix of pods are different powers of two, by considering a pod with radix $2^{k_1}$ to be equivalent to $2^{k_1 - k_2}$ smaller pods with radix $2^{k_2}$, ($k_1 > k_2$).

\subsection{Proofs of theorems}
\label{sec:appendix:physical-gemini}
\emph{Proof of Theorem \ref{th:rounding}: }
Given a graph $G(V,E)$ that has even node degrees $x_v$, $\forall v \in V$, arbitrary edge weights $n_e$, $\forall e \in E$, and no self-loops, Algorithm \ref{al:rounding} computes a graph that has integer edge weights close to $n_e$ (either $\lfloor n_e \rfloor$ or $\lfloor n_e \rfloor + 1$) $\forall e \in E$, while maintaining the same node degrees.

\begin{algorithm}[h] 

\caption{Rounding fractional edges to integers while maintaining node degrees.}
\label{al:rounding}

\begin{enumerate}

\item Round down the value of each edge $n_e$ to the largest integer not exceeding $n_e$, i.e., $\lfloor n_e \rfloor$. Denote the graph by $G(V, E^0)$.

\item Compute the node degrees $y_v$ in $G(V, E^0)$. Let $z_v = x_v - y_v$ be the residue degree of node $v$. Sort nodes in the descending order of residue degrees $v_i (z_i)$, $\forall i \in \{1,2,\dots,|V|\}$. 

\item Connect one edge between $v_1$ and each of the next $z_1$ nodes that have the largest residue degrees, i.e. $v_2, v_3, \dots, v_{z_1+1}$. Let the resulting graph be $G(V, E^1)$.

\item Repeat Steps 2 and 3 until all residue degrees are zero.

\end{enumerate}

\end{algorithm}
Since there are at most $|V|$ iterations of Steps 2 and 3, and it takes $O(|V|)$ time for integer sorting and connecting at most $|V|$ edges in each iteration, \algoref{al:rounding} runs in $O(|V|^2)$ time. We next prove that the algorithm outputs a graph that satisfies the degree and edge constraints. %

By Erdos-Gallai Theorem \cite{choudum1986}, given node degree sequence $z_1 \geq z_2 \geq z_n$ whose sum is even, the following inequality is sufficient for the existence of a simple graph without parallel edges or self-loops that satisfies the degree sequence.
\begin{equation} \label{eq:degree}
 \sum_{i=1}^k z_i \leq k(k-1) + \sum_{i=k+1}^n \min(z_i, k), ~~ 1 \leq k \leq n. 
\end{equation} 

Moreover, if the inequality holds, by Theorem 5 of \cite{hakimi1962}, the algorithm of iteratively connecting the node with the largest degree $z_1$ and the next $z_1$ nodes of the largest degrees constructs a simple graph $G(V,E')$. 
Since there is at most one edge between any pair of nodes in a simple graph, $G(V,E^0 \cup E')$ a graph with degrees $x_v$, 
$\forall v \in V$, and integer edges $\lfloor n_e \rfloor$ or $\lfloor n_e \rfloor + 1$, $\forall e \in E$. 
Moreover, $G(V,E^0 \cup E')$ does not have self-loops since there is no self-loop in $E^0$ or $E'$.   

It remains to prove that the sum of $z_1 \geq z_2 \geq z_n$ is even and that inequality (\ref{eq:degree}) holds. The total degrees of nodes in $G(V,E^0)$ is even, because every edge contributes an additional degree to two nodes. Moreover, $x_v$ is even, $\forall v \in V$. Therefore, $\sum_{v \in V} z_v = \sum_{v \in V} x_v - \sum_{v \in V} y_v$ is even.

The residual degrees $z_i$ are non-zero only if the fractional edges adjacent to $i$ are rounded down by a value smaller than 1. Thus, there exists fractional edges $0 \leq w_{ij} < 1$ that satisfy $\sum_{i=1}^n w_{ij} = z_j$ and $\sum_{j=1}^n w_{ij} = z_i$.

\begin{eqnarray}
    \sum_{i=1}^k z_i &=& \sum_{i=1}^k \sum_{j=1}^n w_{ij} = \sum_{i=1}^k \sum_{j=1}^k w_{ij} + \sum_{i=1}^k \sum_{j=k+1}^n w_{ij} \nonumber \\
    &\leq& k(k-1) + \sum_{i=1}^k \sum_{j=k+1}^n w_{ij} \nonumber \\
    &\leq& k(k-1) + \sum_{j=k+1}^n \min(z_j, k). \nonumber
\end{eqnarray}

The first inequality holds because $w_{ij} < 1$ and $w_{ii} = 0$. The second inequality holds because $\sum_{i=1}^k w_{ij} \leq z_j$ and $\sum_{i=1}^k w_{ij} \leq k$.

\emph{Proof of Theorem \ref{th:implement}: }
If every node in $G$ has degree $2r$, then the graph can be decomposed into $r$ edge-disjoint 2-factors in polynomial time (i.e., a graph where every node $v \in V$ has degree 2) \cite{petersen1891}. Therefore, a graph with uniform node degree $2^k$ can be decomposed into $2^{k-1}$ 2-factors, which can be partitioned into $2^p$ groups of size $2^{k - 1 - p}$. Since there are two edges adjacent to each node in a 2-factor, there are a total of $2^{k - p}$ edges adjacent to each node in a group of 2-factors. Edges in each group of 2-factors can be constructed by one patch panel, because every pod has $2^{k - p}$ ports connected to a patch panel and links can be arbitrarily connected between ports in a patch panel.

\section{Correlation between FCT and MLU, over All Links}
\label{sec:all_link_correlations}

\begin{figure*}[htb]
\centering
\begin{minipage}{0.4\columnwidth}
  \centering
  \includegraphics[width=0.95\columnwidth]{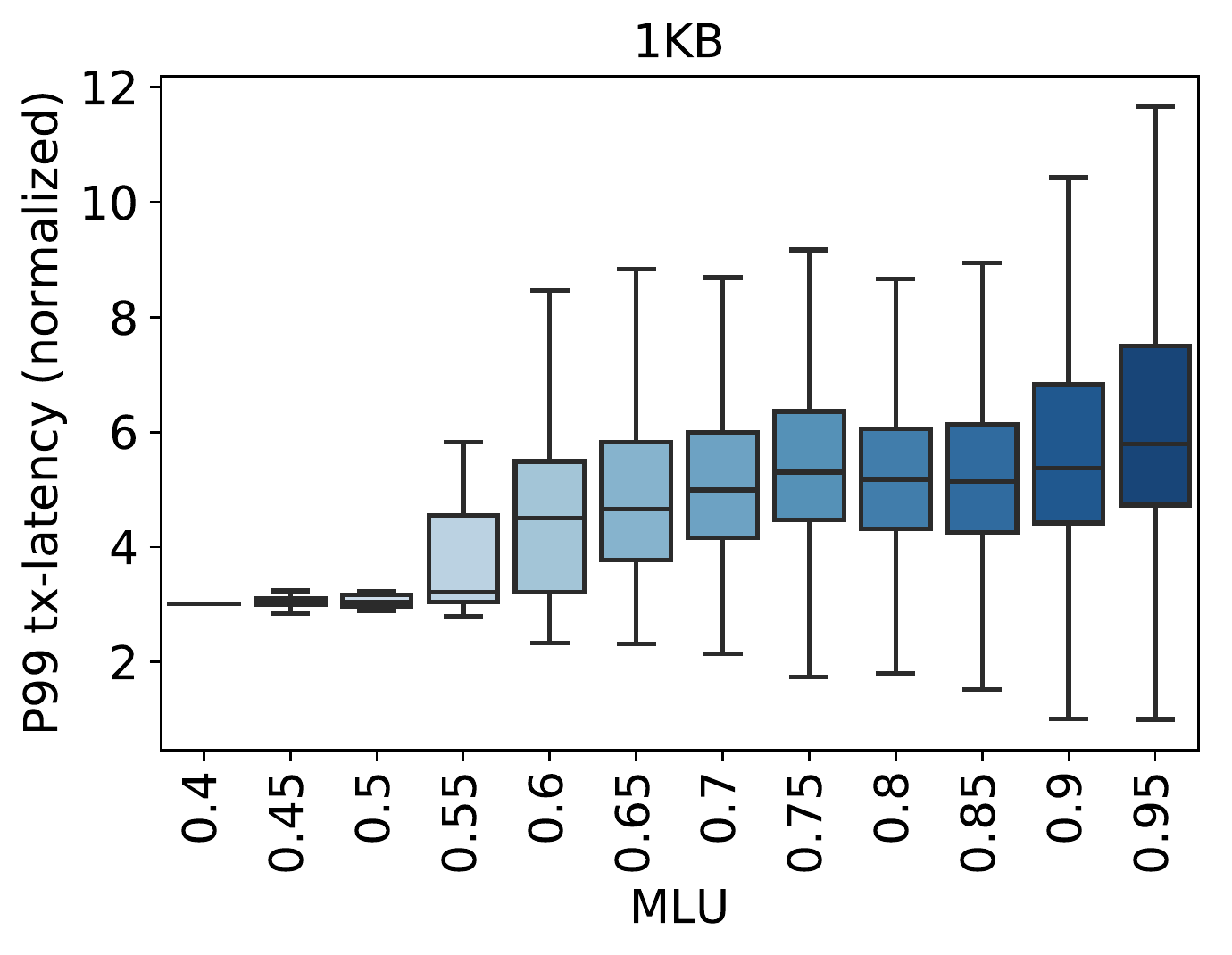}
\end{minipage}
\begin{minipage}{0.4\columnwidth}
  \centering
  \includegraphics[width=0.95\columnwidth]{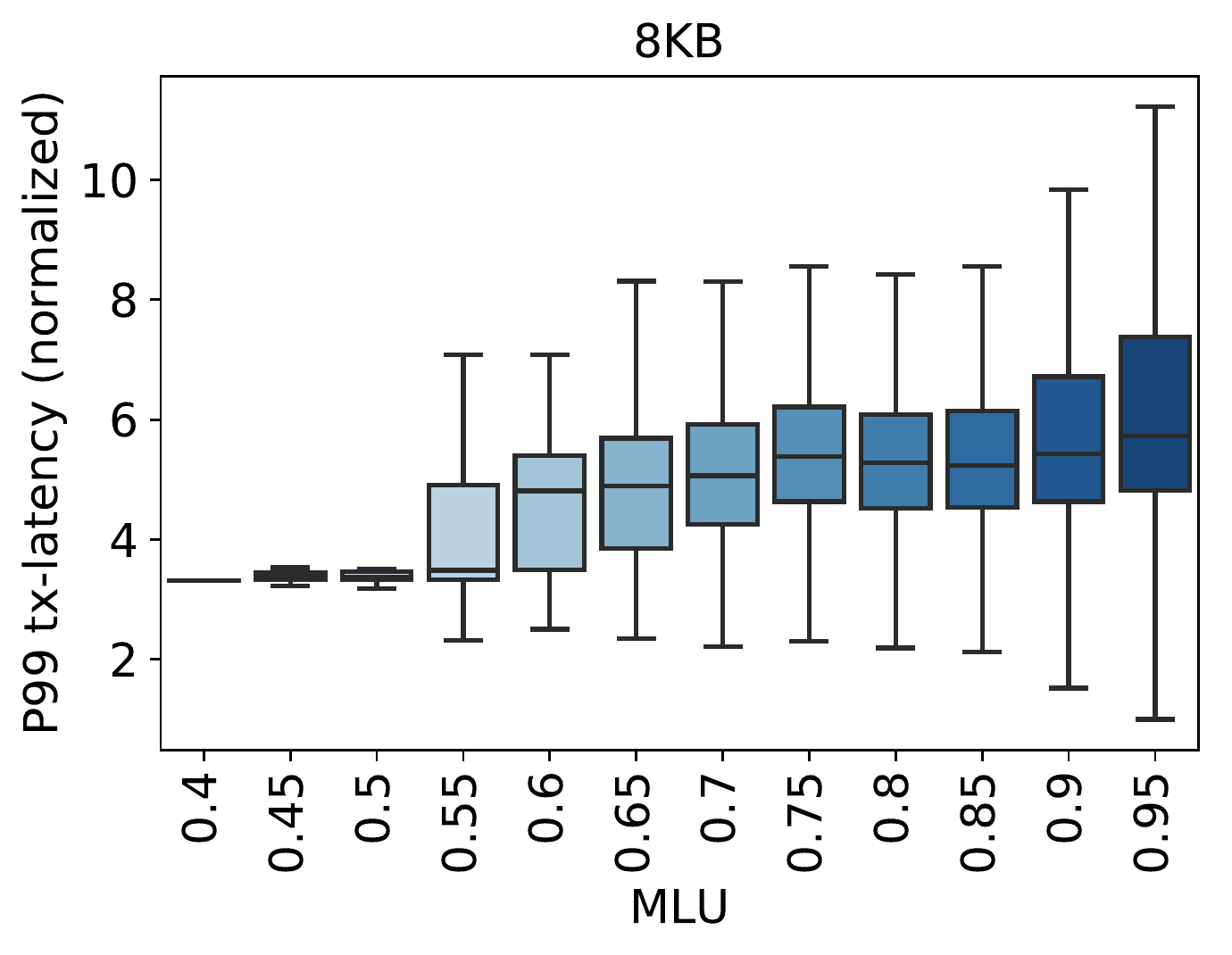}
\end{minipage}
\begin{minipage}{0.4\columnwidth}
  \centering
  \includegraphics[width=0.95\columnwidth]{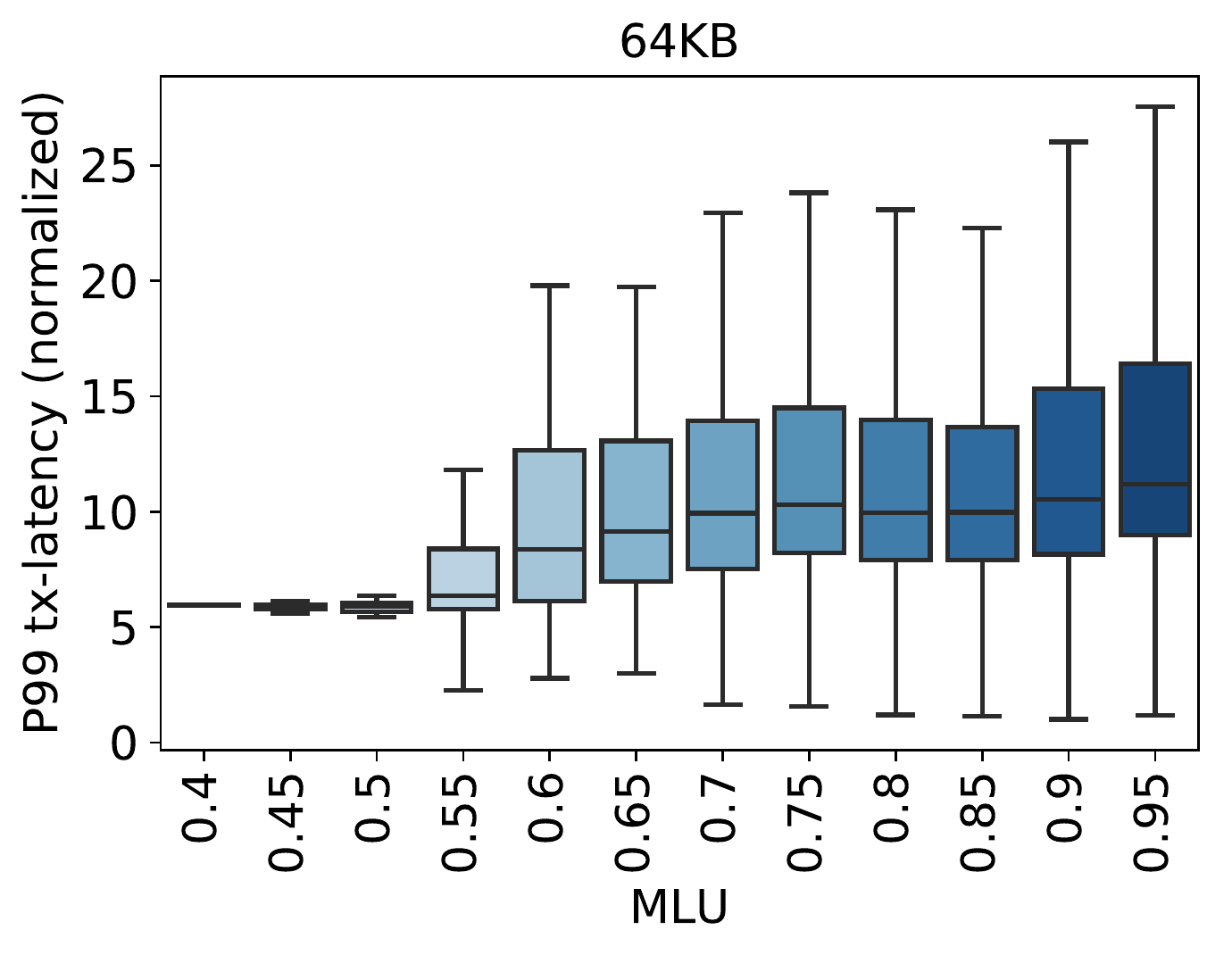}
\end{minipage}
\begin{minipage}{0.4\columnwidth}
  \centering
  \includegraphics[width=0.95\columnwidth]{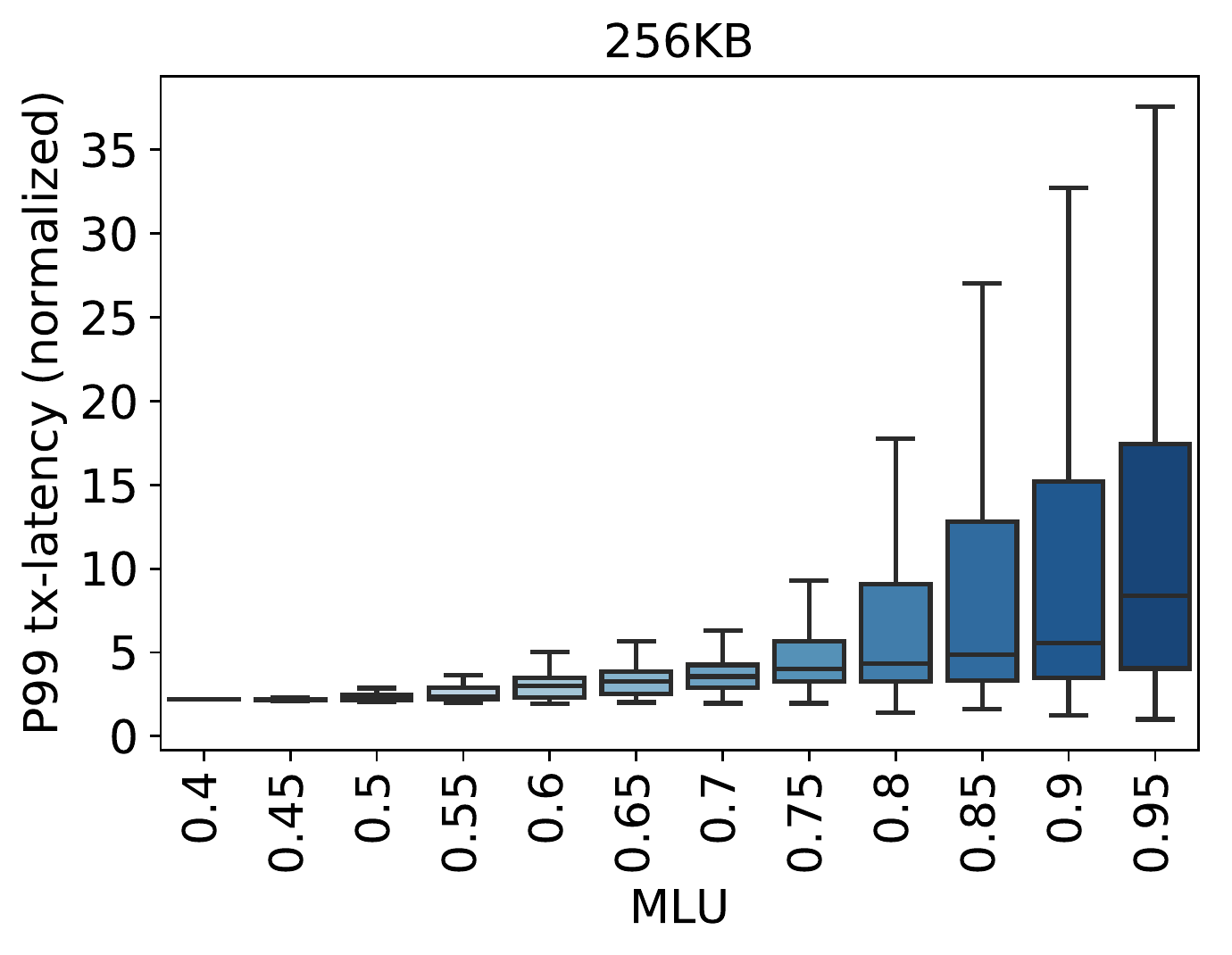}
\end{minipage}
\begin{minipage}{0.4\columnwidth}
  \centering
  \includegraphics[width=0.95\columnwidth]{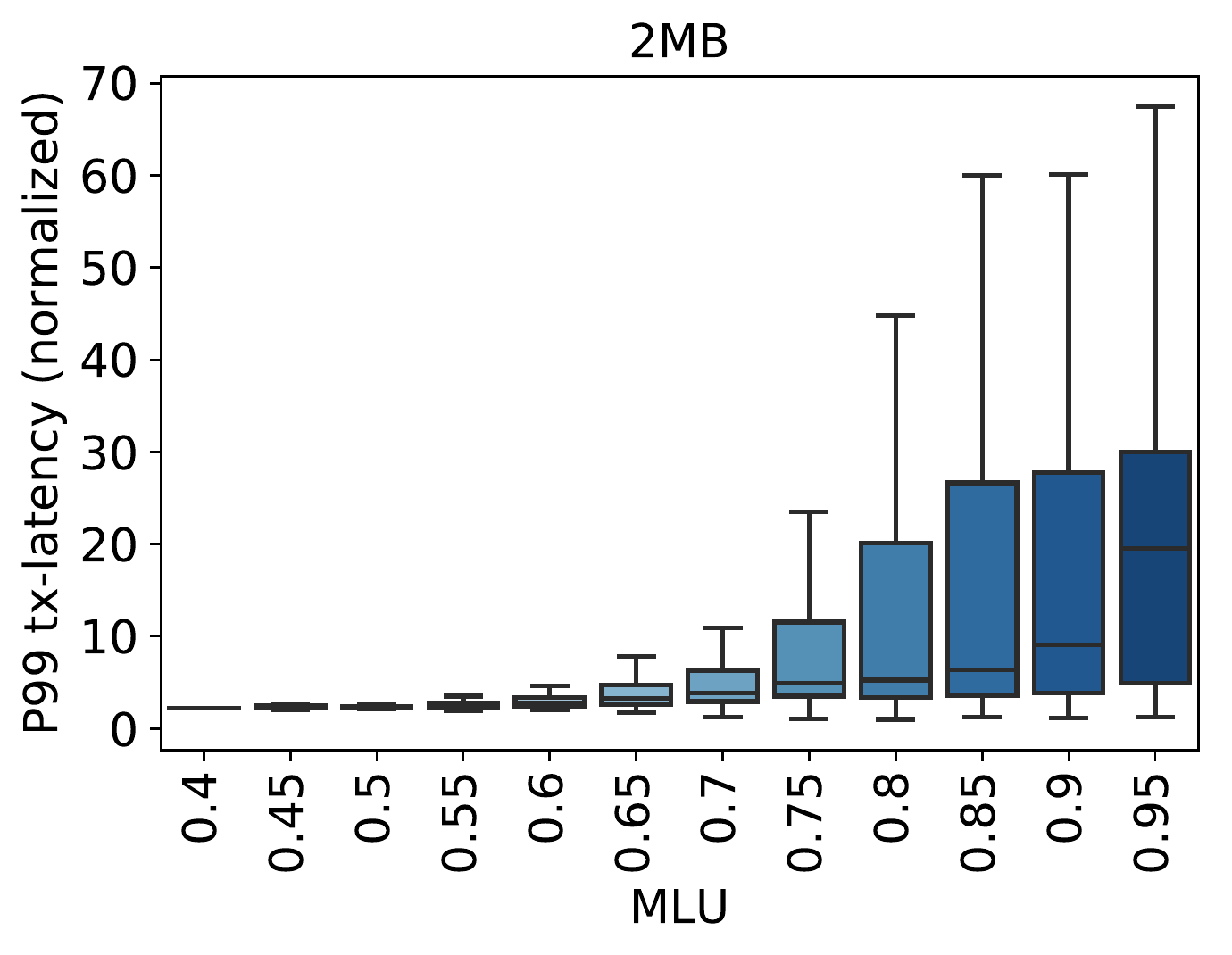}
\end{minipage}
\setlength{\abovecaptionskip}{-2pt}
\setlength{\belowcaptionskip}{-4pt}
\caption{FCTs (inter-pod flows) vs p99 MLUs (all links) on production fabrics}
\label{fig:fct_vs_mlu_all_links} 
\end{figure*}

\begin{figure*}[htb]
\centering
\begin{minipage}{0.4\columnwidth}
  \centering
  \includegraphics[width=0.95\columnwidth]{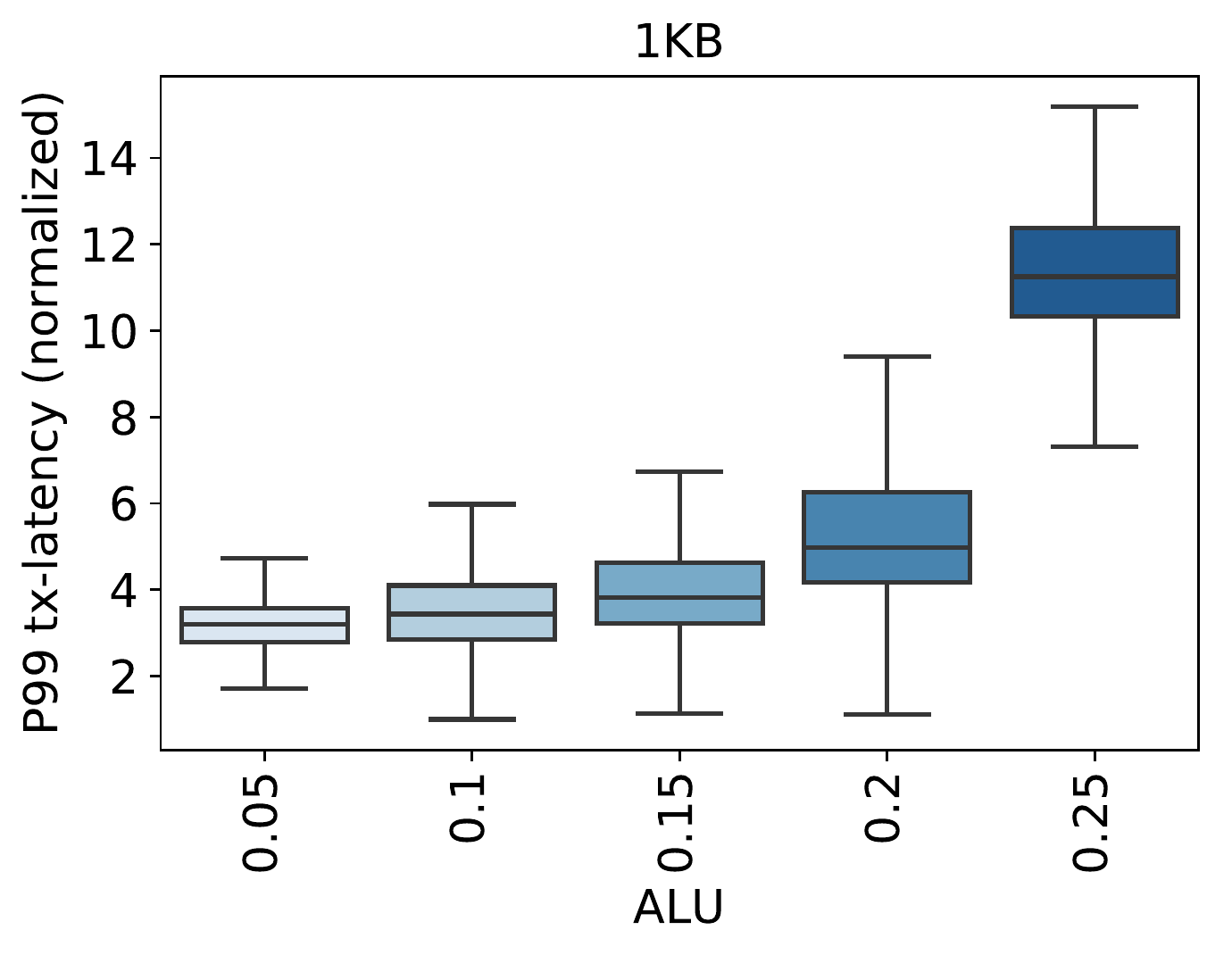}
\end{minipage}
\begin{minipage}{0.4\columnwidth}
  \centering
  \includegraphics[width=0.95\columnwidth]{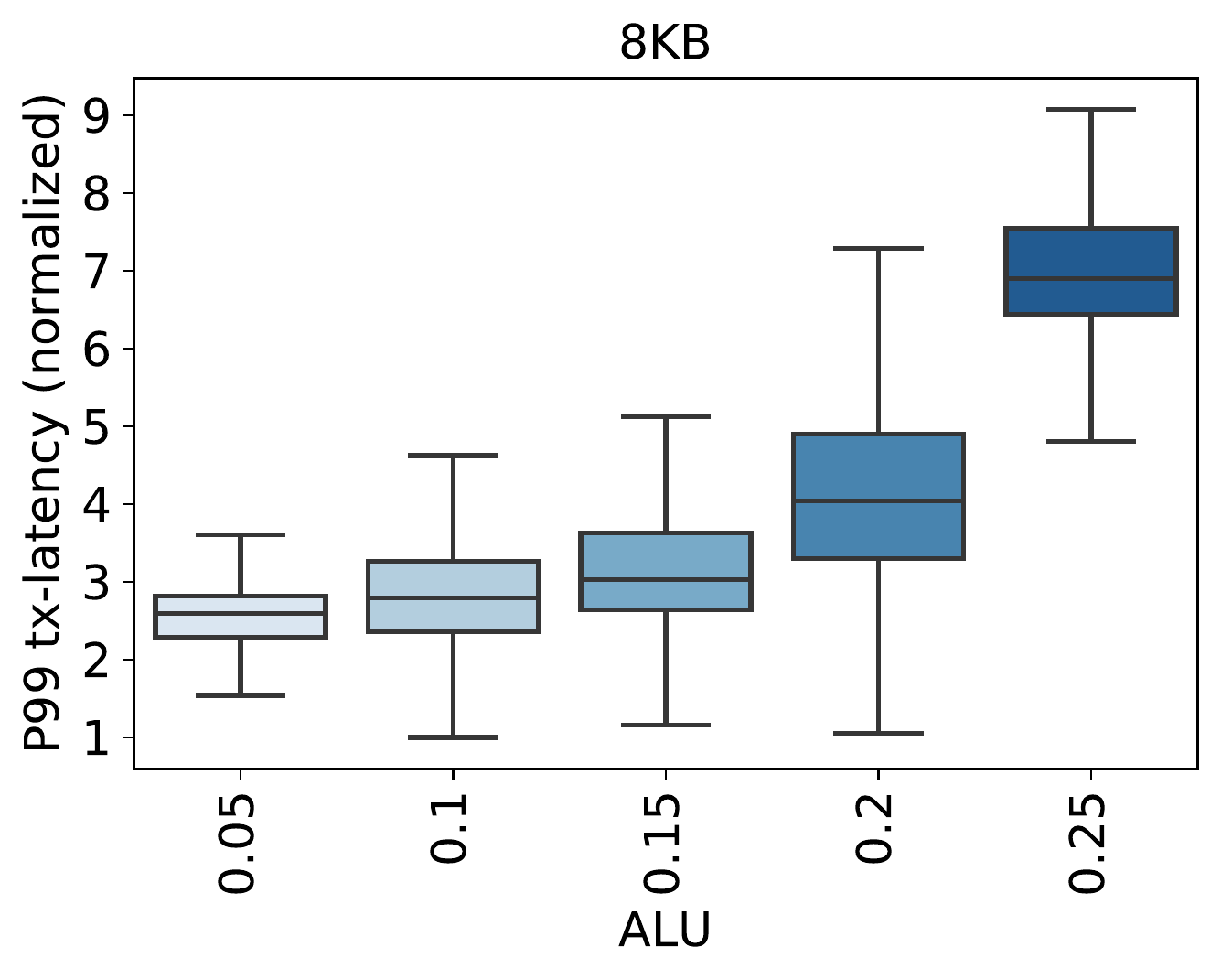}
\end{minipage}
\begin{minipage}{0.4\columnwidth}
  \centering
  \includegraphics[width=0.95\columnwidth]{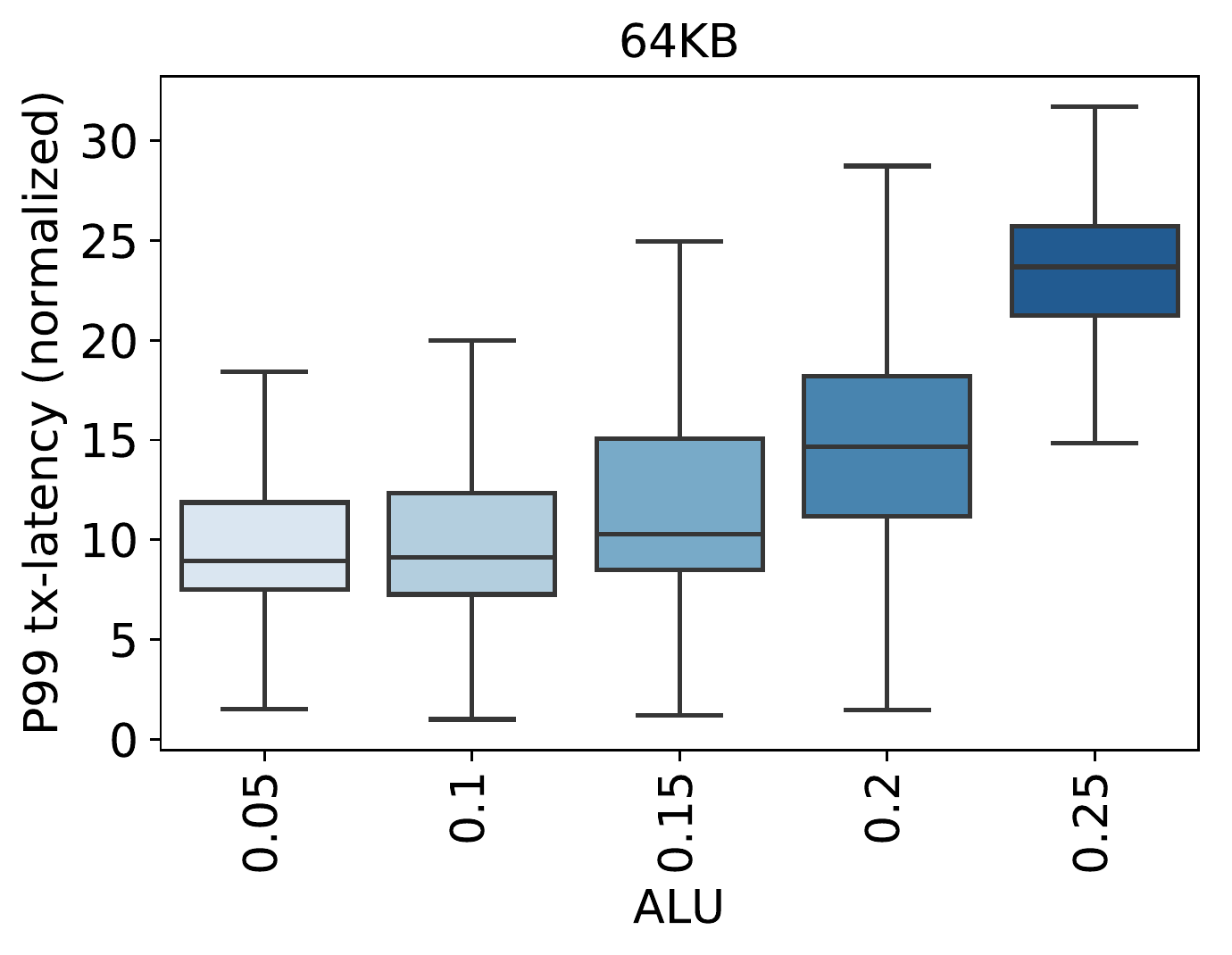}
\end{minipage}
\begin{minipage}{0.4\columnwidth}
  \centering
  \includegraphics[width=0.95\columnwidth]{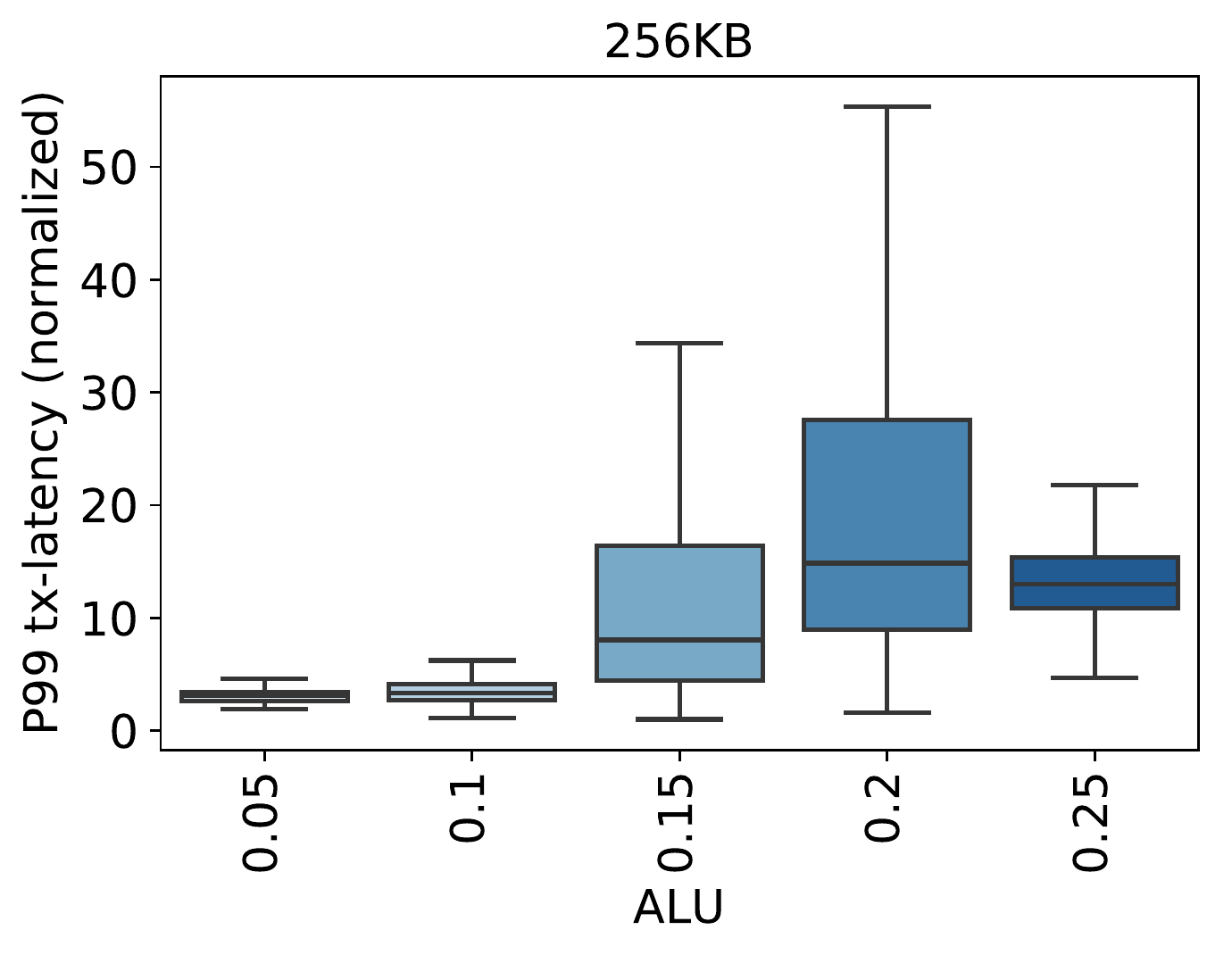}
\end{minipage}
\begin{minipage}{0.4\columnwidth}
  \centering
  \includegraphics[width=0.95\columnwidth]{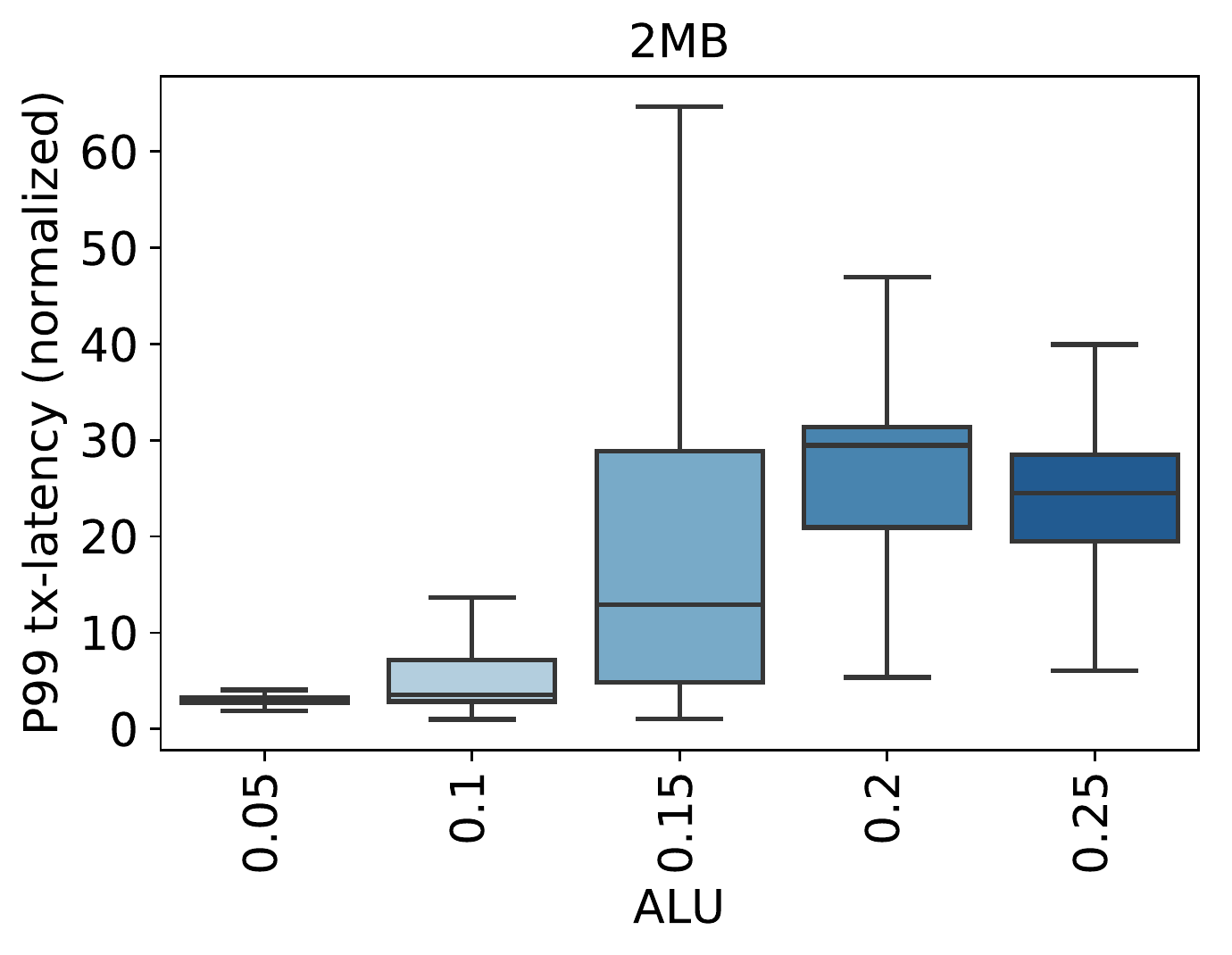}
\end{minipage}
\setlength{\abovecaptionskip}{-2pt}
\setlength{\belowcaptionskip}{-4pt}
\caption{FCTs (inter-pod flows) vs p99 ALUs (all links) on production fabrics}
\label{fig:fct_vs_alu_all_links}
\end{figure*}

\begin{figure*}[htb]
\centering
\begin{minipage}{0.4\columnwidth}
  \centering
  \includegraphics[width=0.95\columnwidth]{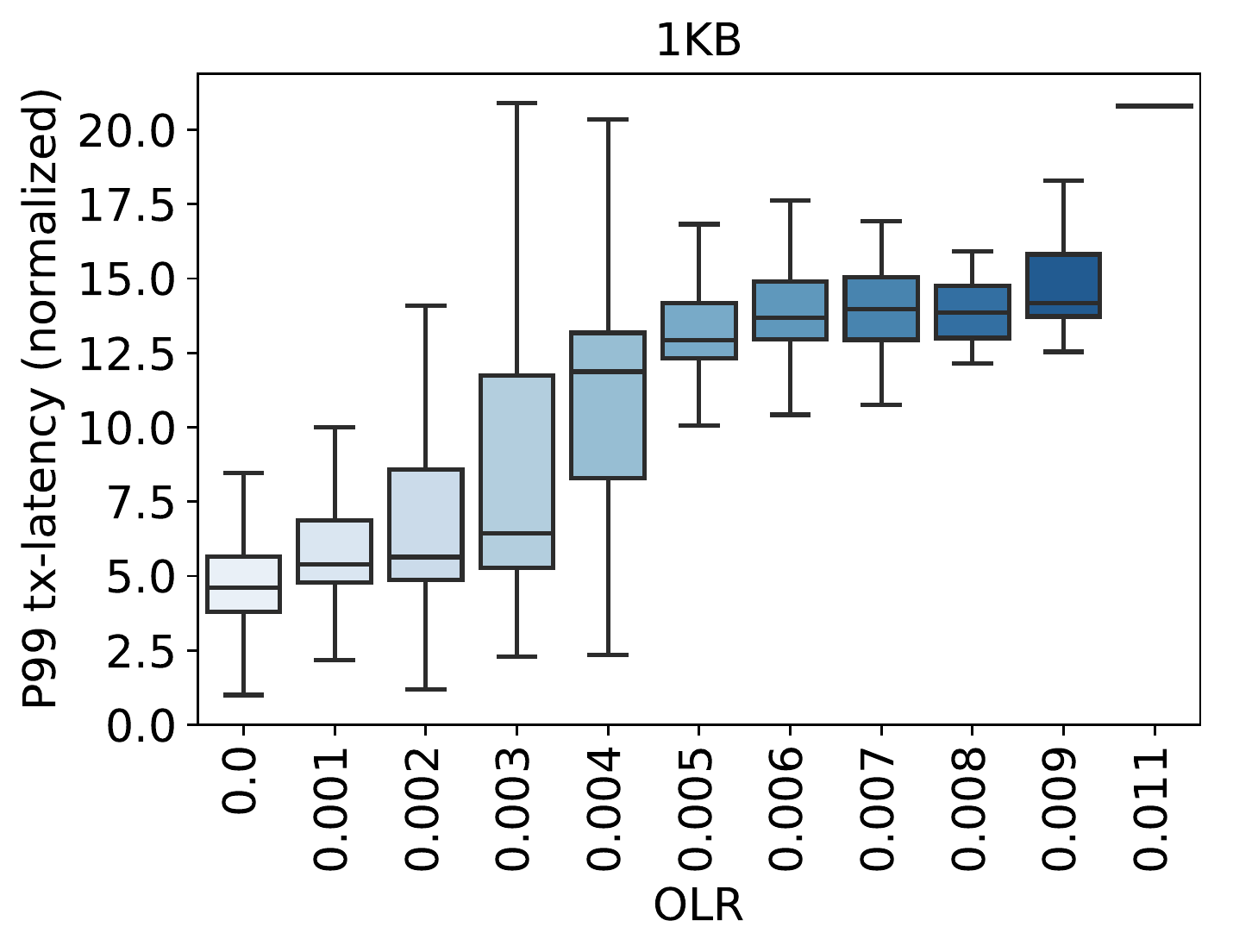}
\end{minipage}
\begin{minipage}{0.4\columnwidth}
  \centering
  \includegraphics[width=0.95\columnwidth]{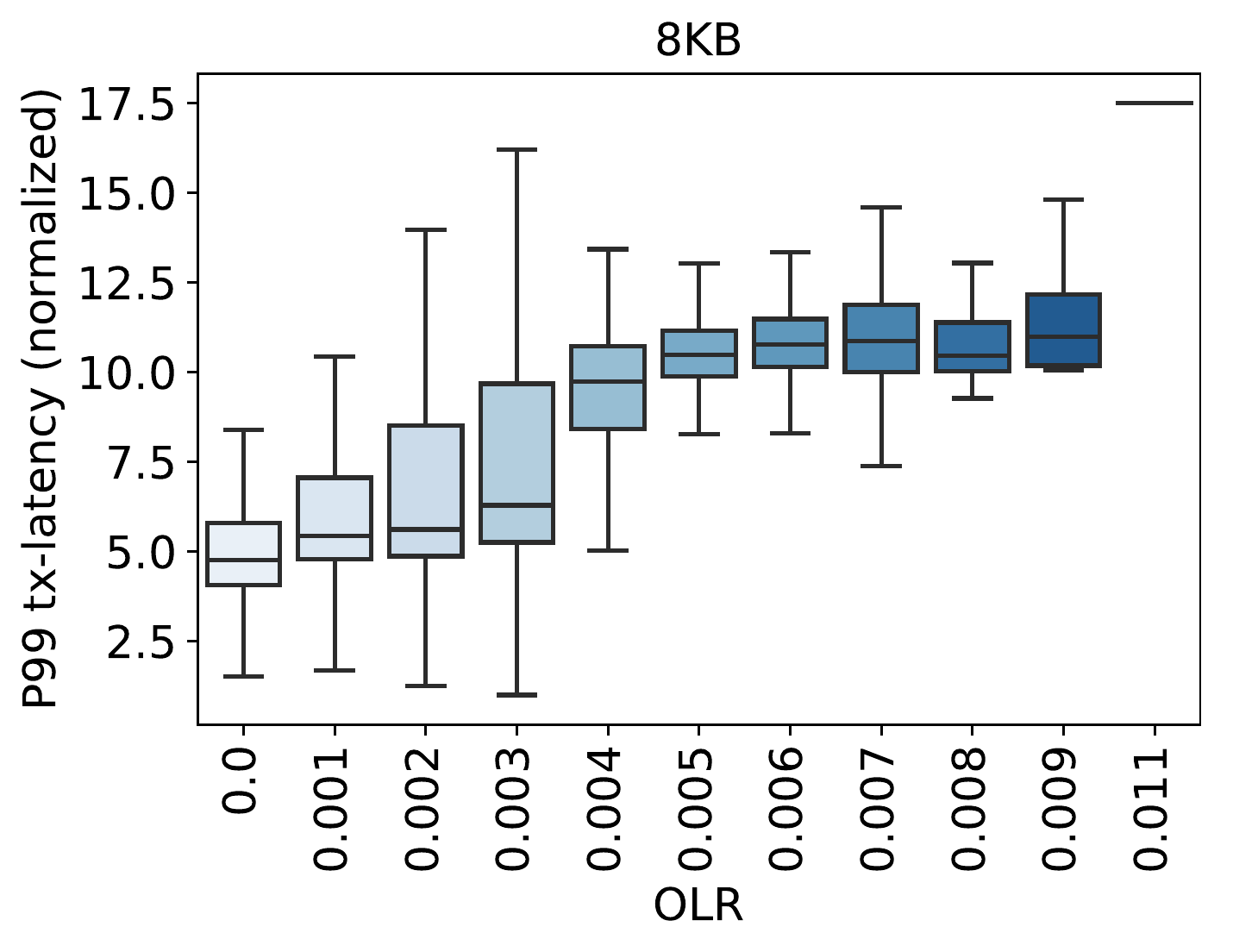}
\end{minipage}
\begin{minipage}{0.4\columnwidth}
  \centering
  \includegraphics[width=0.95\columnwidth]{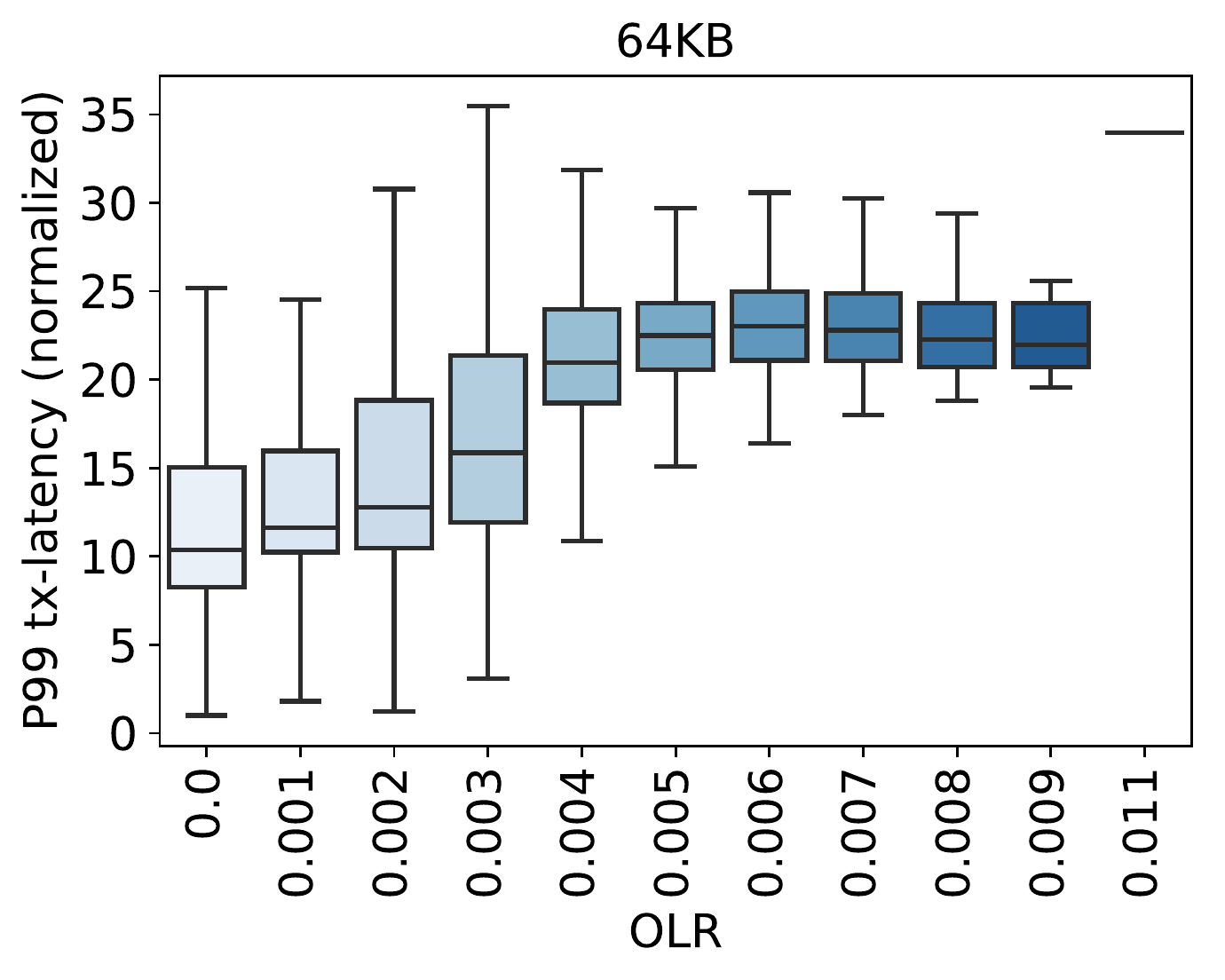}
\end{minipage}
\begin{minipage}{0.4\columnwidth}
  \centering
  \includegraphics[width=0.95\columnwidth]{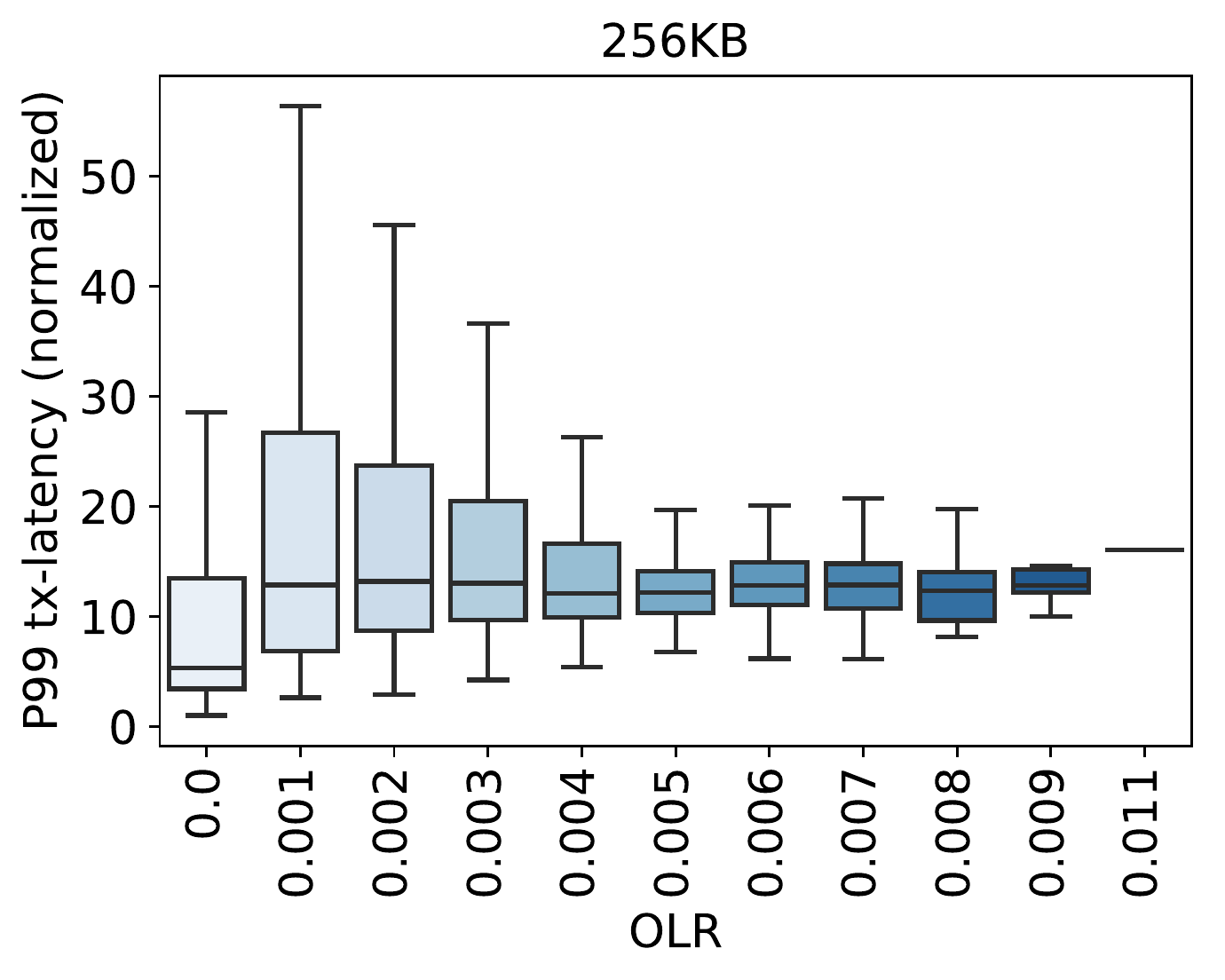}
\end{minipage}
\begin{minipage}{0.4\columnwidth}
  \centering
  \includegraphics[width=0.95\columnwidth]{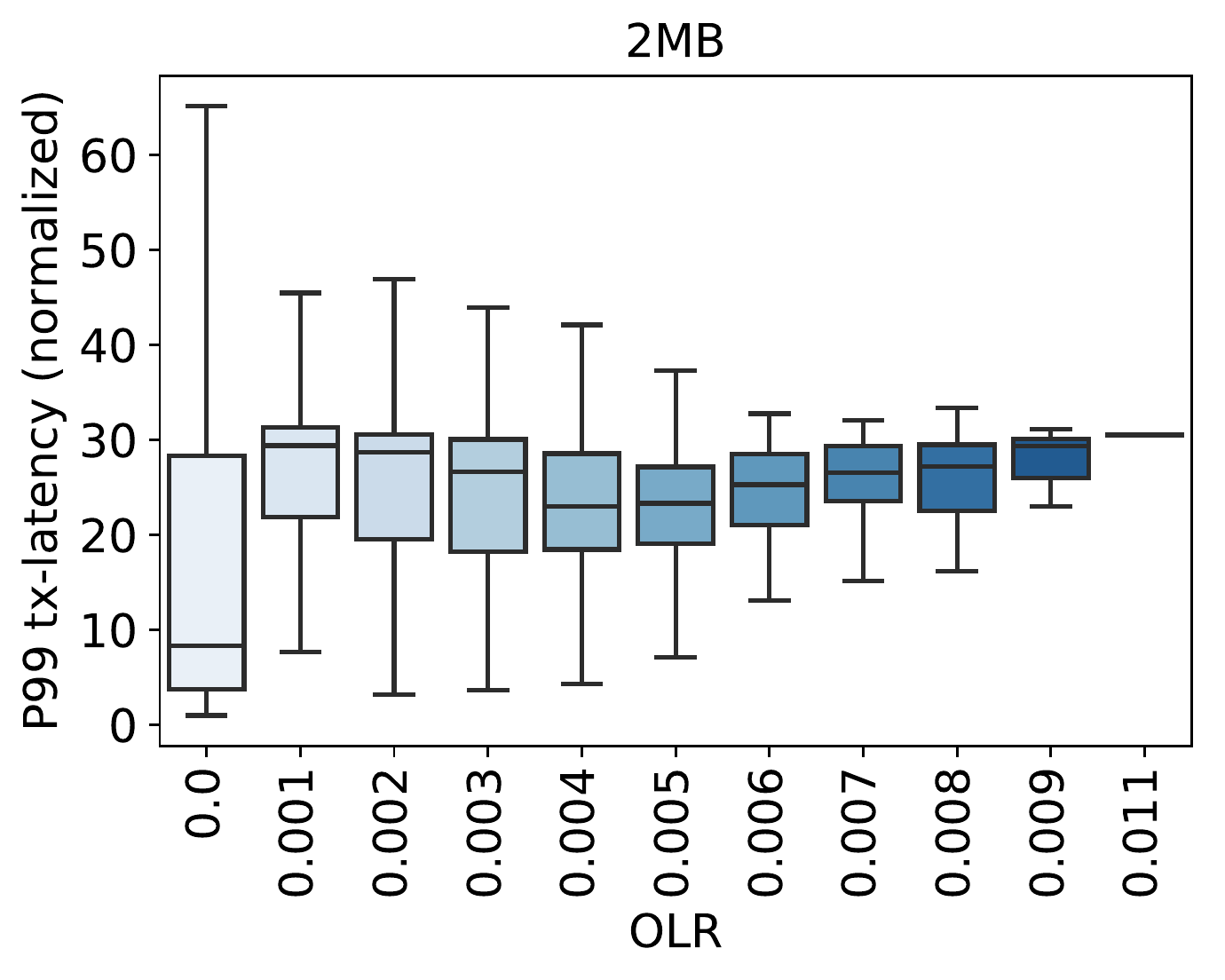}
\end{minipage}
\setlength{\abovecaptionskip}{-2pt}
\setlength{\belowcaptionskip}{-4pt}
\caption{FCTs (inter-pod flows) vs p99 OLRs (all links) on production fabrics}
\label{fig:fct_vs_olr_all_links}
\end{figure*}

In \secref{sec:metrics} we presented correlations between FCTs for inter-pod flows with DCNI-level link-utilization
metrics.  We also collected link-utilization data for all other links, including host-to-ToR links as well as pod-internal
links.   This data provides stronger evidence for correlations between link utilizations and FCTs, but is less indicative
of whether the DCNI-only simulated utilizations in \secref{sec:evaluation} would be predictive of FCT benefits.
\figref{fig:fct_vs_mlu_all_links} shows FCTs \vs p99 all-links MLUs; 
\figref{fig:fct_vs_alu_all_links} shows FCTs \vs p99 all-links ALUs;
\figref{fig:fct_vs_olr_all_links} shows FCTs \vs all-links overloaded link ratios (OLRs).  

As in \secref{sec:metrics}, FCT values are normalized to the best sample for each size,
and the message size shown at the top of each graph is the upper bound for the message-size bucket represented by that graph.

Note that with this dataset, there does
appear to be a correlation between FCTs and OLRs.

\section{FCTs in testbed experiments}
\label{sec:testbed_fcts}

Here we report some additional, inconclusive results regarding FCTs in the testbed experiments of \secref{sec:testbed}.

Note that the workload changed measurably between the baseline and best-predicted trials:
\squishlist
    \item The daily-average traffic volume increased by 17\%
    \item The fraction of well-bounded pod-pairs decreased from 0.93 to 0.825, indicating that the traffic became less predictable.
    \item The maximum DMR (demand-to-max ratio) increased considerably, from 1.67 to 5.49, also indicating a decrease in predictability.
\squishend
We currently lack access to this testbed that would allow us to repeat the experiments with less of a change in workload
between trials.

\parab{FCT metrics} We collected per-flow metrics: min RTT, message transmission latency, and delivery rate (for transfers that were network, not application, limited). For each of these, we report the median and 99-th percentile values. Transmission latencies are bucketed by transfer size into 5 buckets ranging from 1KB to 2MB and a sixth bucket with transfers larger than 2MB.
As in \secref{sec:metrics}, we report FCT values that are normalized to the best sample for each size.

\begin{figure}
\centering
\begin{minipage}{0.23\columnwidth}
  \centering
  \includegraphics[width=0.95\columnwidth]{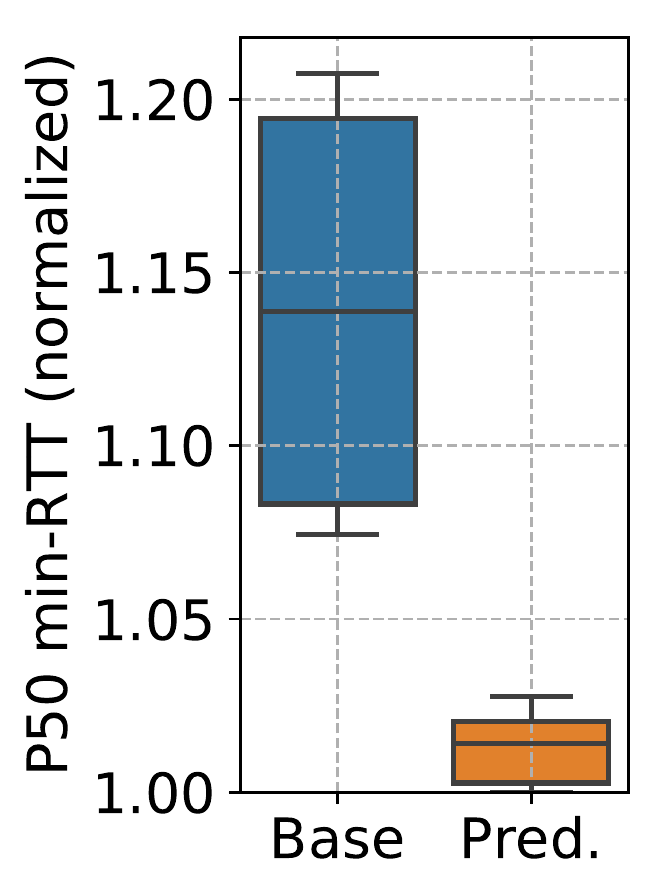}
  \label{fig:p50_min_rtt}
\end{minipage}
\begin{minipage}{0.23\columnwidth}
  \centering
  \includegraphics[width=0.95\columnwidth]{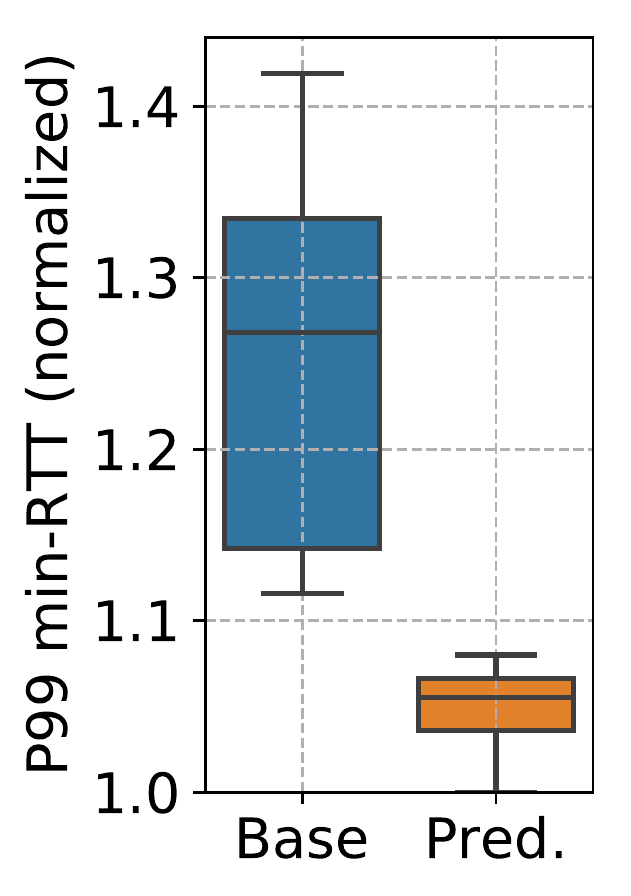}
  \label{fig:p99_min_rtt}
\end{minipage}
\begin{minipage}{0.23\columnwidth}
  \centering
  \includegraphics[width=0.95\columnwidth]{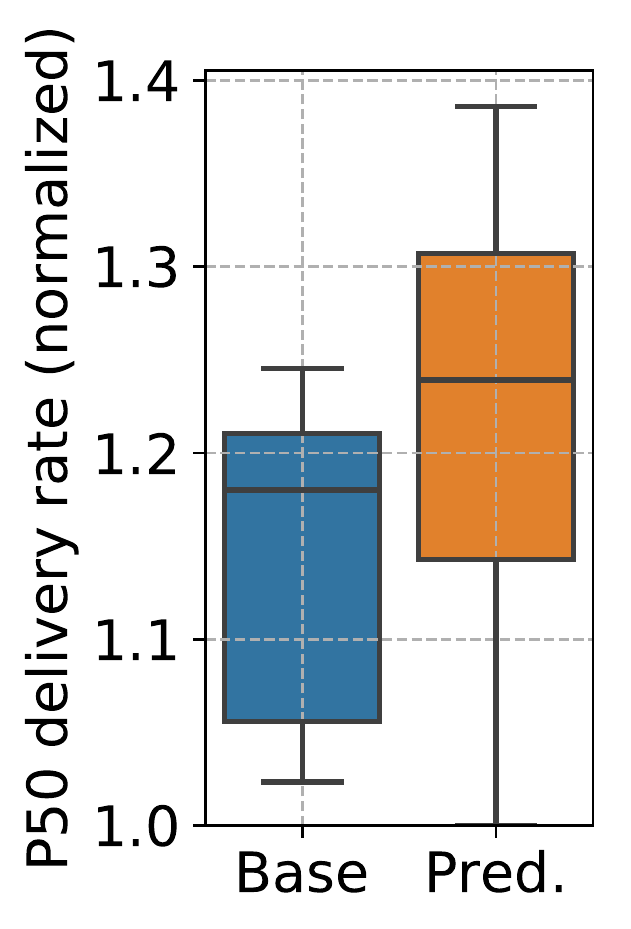}
  \label{fig:p50_delivery_rate}
\end{minipage}
\begin{minipage}{0.23\columnwidth}
  \centering
  \includegraphics[width=0.95\columnwidth]{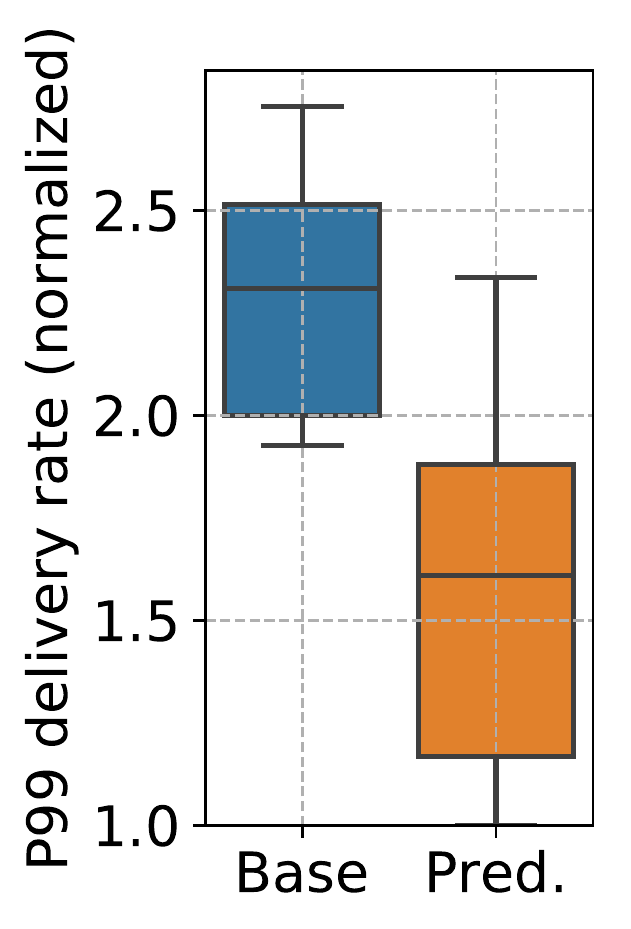}
  \label{fig:p99_delivery_rate}
\end{minipage}
{\small ``Base'' $=$ baseline; ``Pred.'' $=$ predicted-best}
\caption{\small Testbed experiments -- min-RTT, delivery rate}
\label{fig:rtt-throughput}
\end{figure}

\begin{figure*}
\centering
\begin{minipage}{0.20\columnwidth}
  \centering
  \includegraphics[width=0.95\columnwidth]{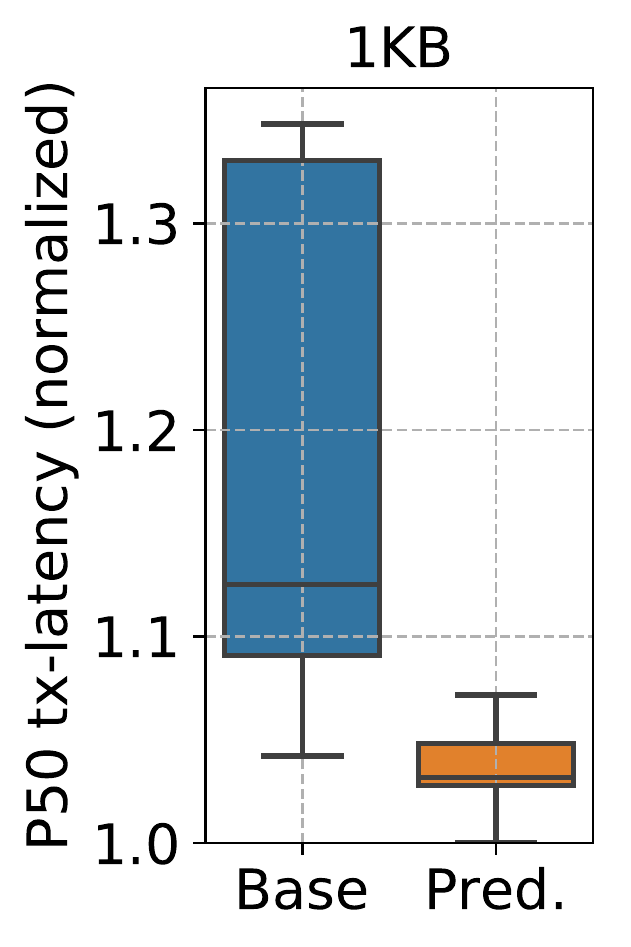}
  \label{fig:p50_tx_latency_1kb}
\end{minipage}
\begin{minipage}{0.20\columnwidth}
  \centering
  \includegraphics[width=0.95\columnwidth]{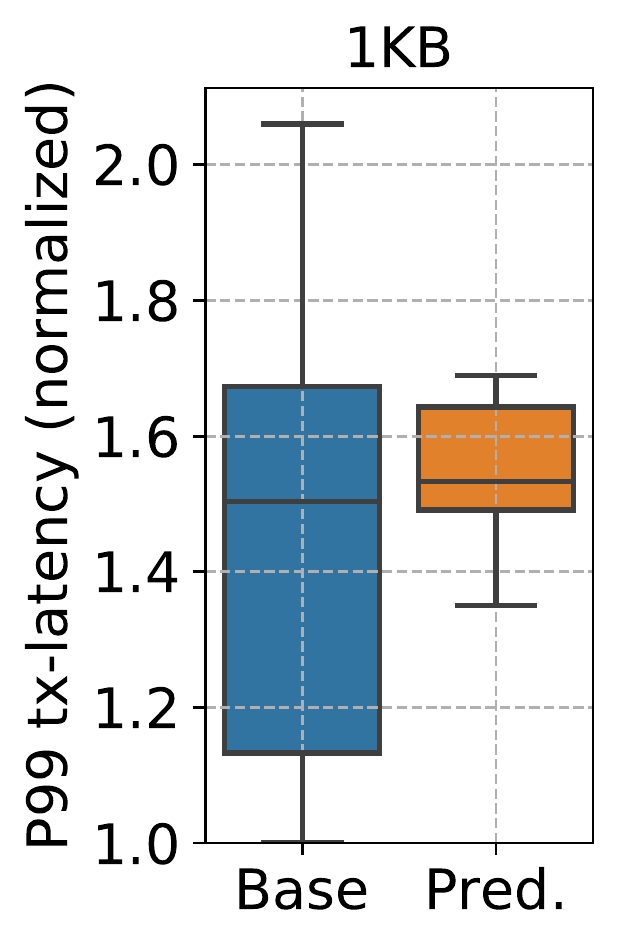}
  \label{fig:p99_tx_latency_1kb}
\end{minipage}
\begin{minipage}{0.20\columnwidth}
  \centering
  \includegraphics[width=0.95\columnwidth]{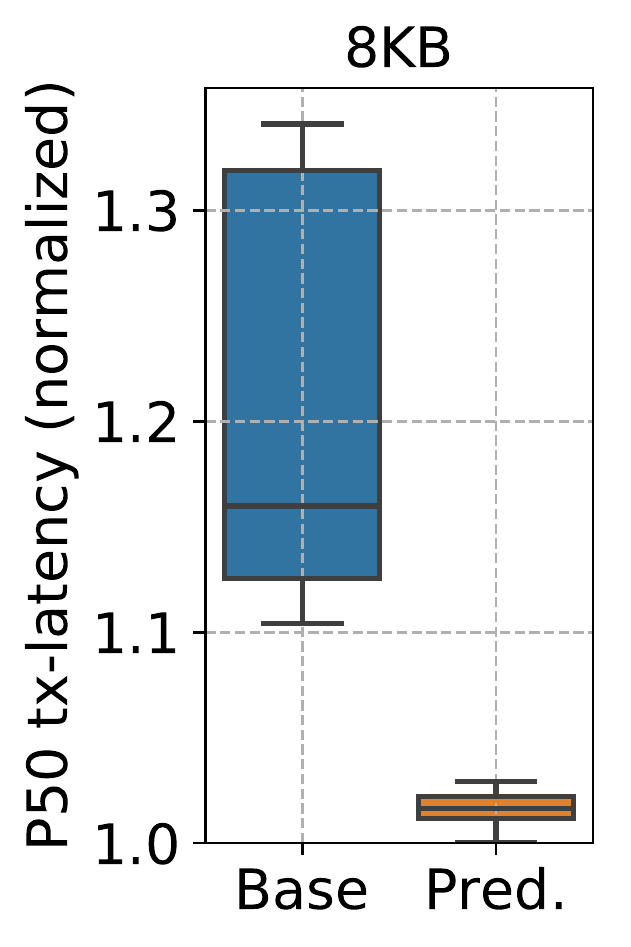}
  \label{fig:p50_tx_latency_8kb}
\end{minipage}
\begin{minipage}{0.2\columnwidth}
  \centering
  \includegraphics[width=0.95\columnwidth]{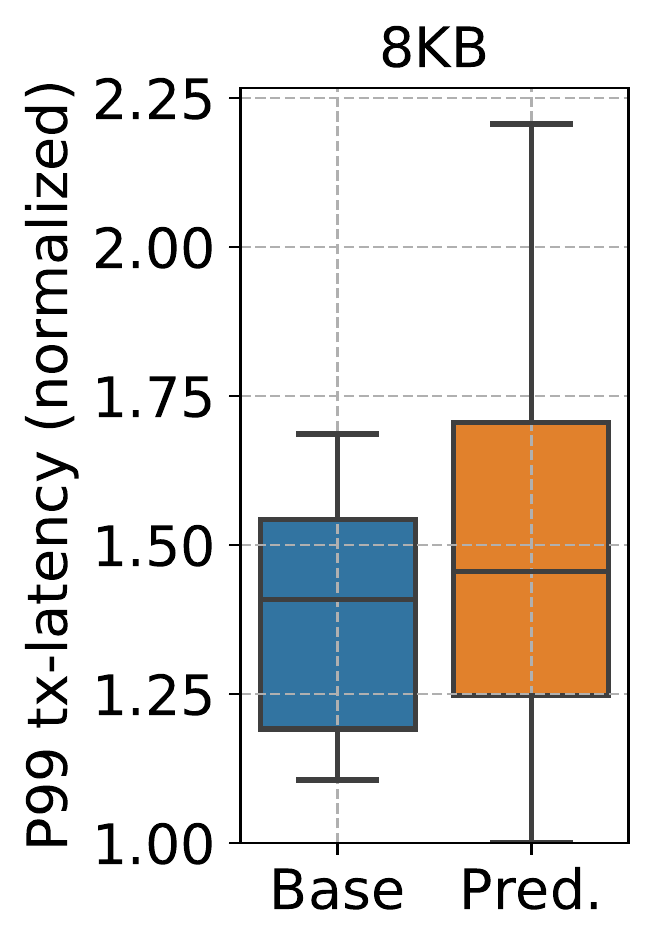}
  \label{fig:p99_tx_latency_8kb}
\end{minipage}
\begin{minipage}{0.20\columnwidth}
  \centering
  \includegraphics[width=0.95\columnwidth]{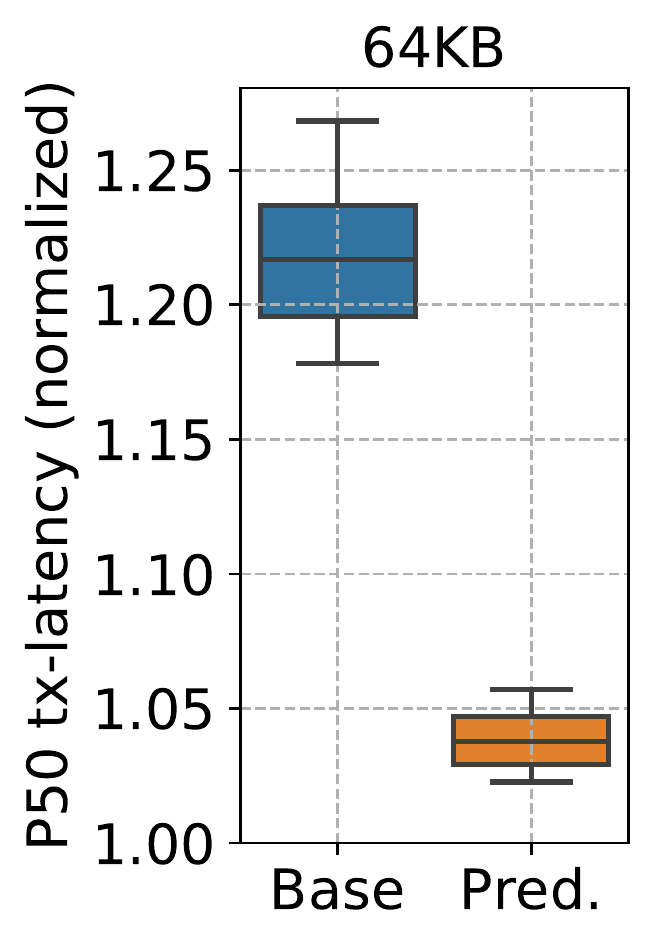}
  \label{fig:p50_tx_latency_64kb}
\end{minipage}
\begin{minipage}{0.19\columnwidth}
  \centering
  \includegraphics[width=0.95\columnwidth]{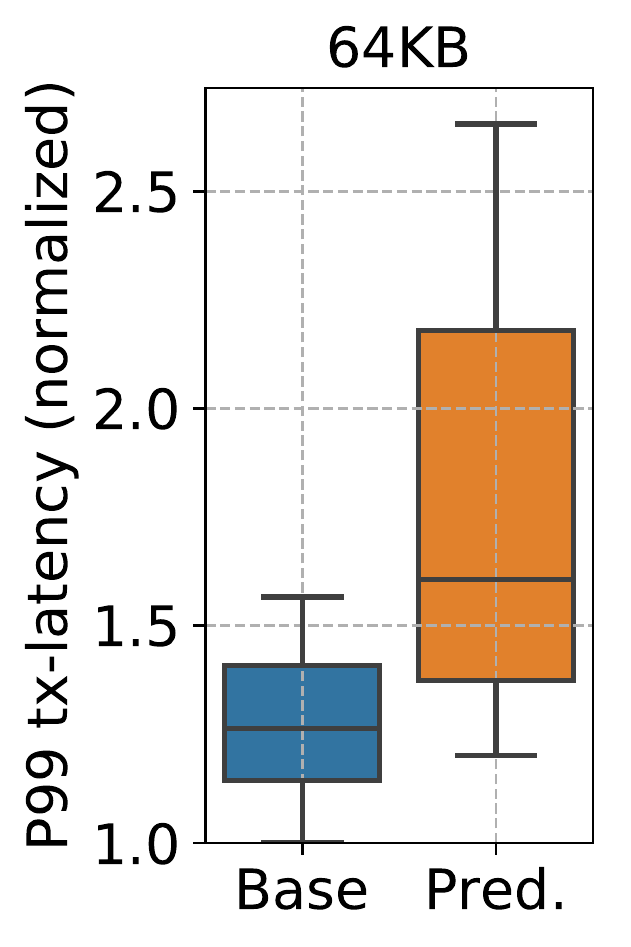}
  \label{fig:p99_tx_latency_64kb}
\end{minipage}
\begin{minipage}{0.21\columnwidth}
  \centering
  \includegraphics[width=0.95\columnwidth]{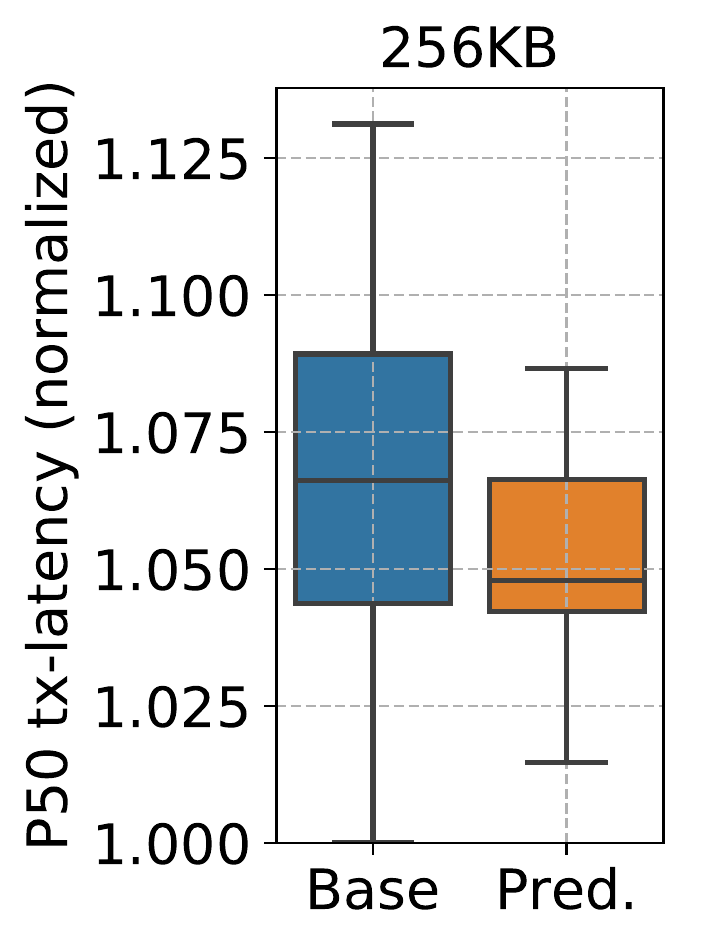}
  \label{fig:p50_tx_latency_256kb}
\end{minipage}
\begin{minipage}{0.19\columnwidth}
  \centering
  \includegraphics[width=0.95\columnwidth]{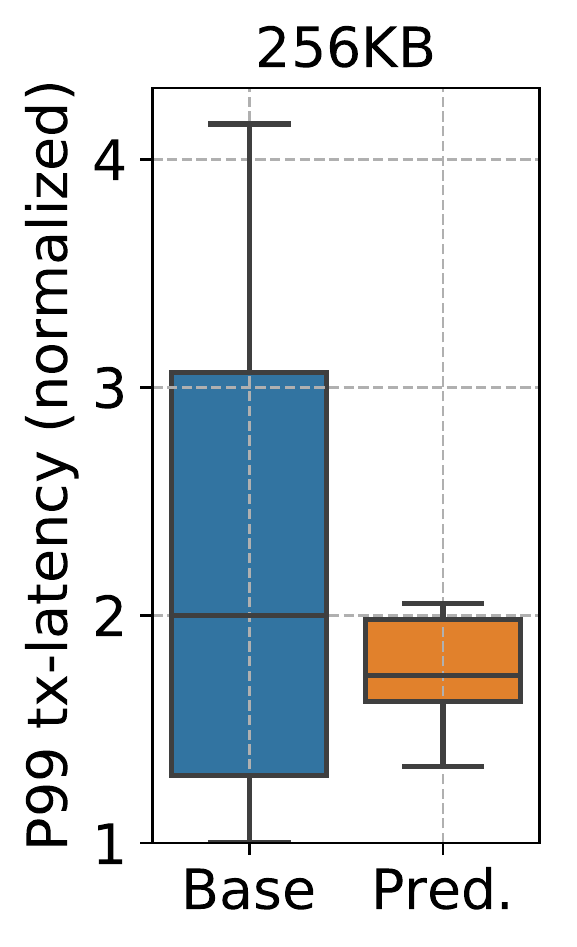}
  \label{fig:p99_tx_latency_256kb}
\end{minipage}
\begin{minipage}{0.2\columnwidth}
  \centering
  \includegraphics[width=0.95\columnwidth]{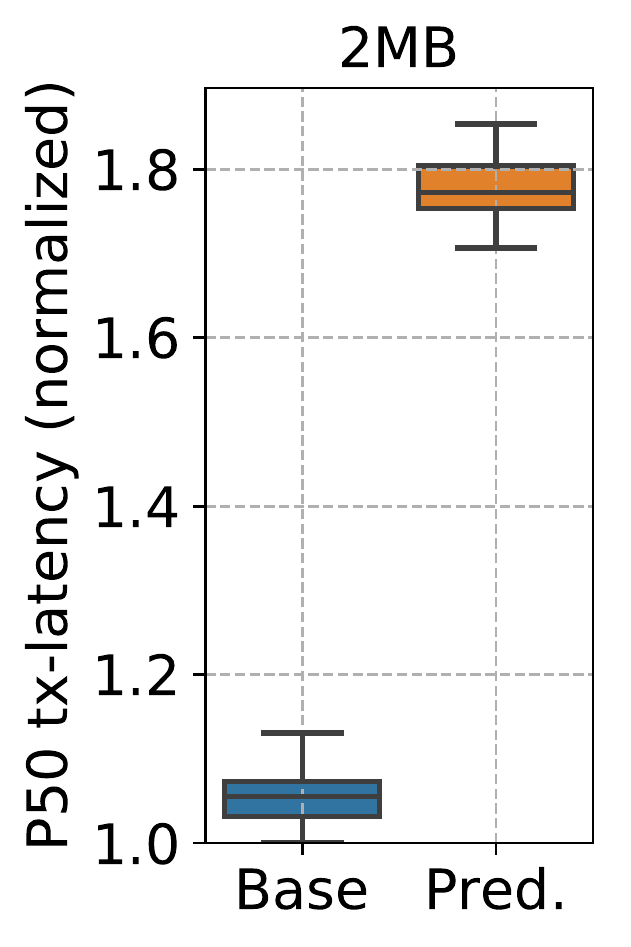}
  \label{fig:p50_tx_latency_2mb}
\end{minipage}
\begin{minipage}{0.19\columnwidth}
  \centering
  \includegraphics[width=0.95\columnwidth]{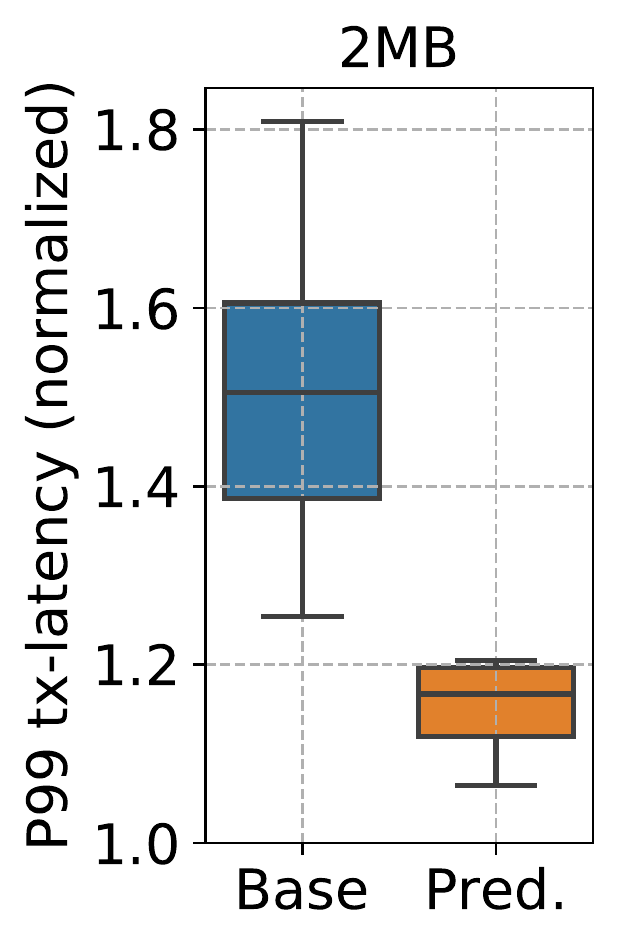}
  \label{fig:p99_tx_latency_2mb}
\end{minipage}
\setlength{\abovecaptionskip}{-2pt}
\setlength{\belowcaptionskip}{-8pt}
\caption{Testbed experiments -- message-transfer latency}
\label{fig:canary-fct}
\end{figure*}

\figref{fig:rtt-throughput} shows how \gemini's
suggested non-uniform topology and routing, in
our testbed experiments, affects the normalized min-RTT and delivery rates observed at the endpoints;
\figref{fig:canary-fct} shows the effects on FCTs.
In both figures, the \emph{predicted-best} topology appears to improve results in some cases, but worsens them
in others.   We lack sufficient information to conclude whether any of these changes are attributable to the
configuration or to the difference in workload (especially, the significant difference in variability).

The developers of the FCT-measurement system have warned us that the p99 delivery-rate results
could be unreliable, due to some practical difficulties in measuring these rates at the tail.   They are more confident in the other measurements (RTT, FCT, and p50 delivery rates).

\end{document}